\documentstyle[12pt,equation]{article}
\advance \hoffset by 0.5 cm
\begin{document}
\newcommand\be{\begin{equation}}
\newcommand\ee{\end{equation}}
\newcommand\bea{\begin{eqnarray}}
\newcommand\eea{\end{eqnarray}}
\newcommand{\een}{\end{subequations}}
\newcommand{\ben}{\begin{subequations}}
\newcommand{\beq}{\begin{eqalignno}}
\newcommand{\eeq}{\end{eqalignno}}
\def\ext{{\rm ext}}
\def\lsim{\:\raisebox{-0.5ex}{$\stackrel{\textstyle<}{\sim}$}\:}
\def\gsim{\:\raisebox{-0.5ex}{$\stackrel{\textstyle>}{\sim}$}\:}
\def\a{
\begin{picture}(50,1)
\put(0,1){\vector(1,0){50}}
\end{picture}}
\renewcommand{\theequation}{\arabic{section}.\arabic{equation}}
\pagestyle{empty}
\tableofcontents
\newpage
\pagestyle{empty}
\begin{flushright}
BU TH-94/04 \\
hep-ph/9407210 \\
\end{flushright}

\begin{center}
{\Large  Structure of proton and nuclei}
\footnote{Lectures given
at the S.E.R.C. school in Nuclear Physics, Goa University, Feb. 1993}\\
\vspace*{5mm}
Rohini M. Godbole\\
{\it Dept. of Physics, Univ. of Bombay,
Kalina Campus,\\ Santa Cruz(E), Bombay 400 098, India.\\}
\end{center}

\begin {abstract}
We discuss in this set of lectures the structure of
proton/neutron as revealed through a study of form-factors. This
is followed by a discussion of the structure functions of
proton/nuclear targets as measured in the deep inelastic
scattering (DIS) of leptons off these targets. We discuss the
parton model in DIS as well as outline the usage of parton model
in processes other than the DIS. We then go on to dicsuss the EMC effect:
the nuclear dependence of the structure functions. After
a discussion of  different models of the EMC effect we end by
pointing out the possibility of distinguishing between these
different models by studying the correlation between the
$A$--dependence of different hard processes and the EMC effect.
\end{abstract}
\newpage
\section*{Introduction}
Structure of hadrons, the strongly interacting particles, has been the
subject of theoretical and experimental investigations over the period of
past four decades and has played a crucial role in the birth of the
subject of particle physics.  Traditionally, information about the
constituents of a system has come from two sources: one is the study of
static properties such as mass, magnetic moments, spin, parity, etc., and
the other is scattering experiments.  The famous Rutherford scattering
experiment is the prototype of the latter.  In the case of hadrons, the clues
to quark structure came in the form of the quark model put forward by
Gell-Mann  \cite{qmodel}, but the final confirmation came from the Deep
Inelastic Scattering (DIS) experiments \cite{DIS} where the observed
phenomenon  of `scaling' was explained \cite{bjpaschos} in terms of
`partons'(pointlike  constituents of proton) \cite{feynman}.  The
discovery of `asymptotic freedom' \cite{gross} of `Quantum
Chromodynamics' (QCD) \cite{QCD}: the gauge theory of quarks and
gluons, coupled with the operator  product expansion  gave an explanation,
in the context of perturbative QCD(pQCD), why the parton model works so
well\cite{QCDper}.  By now QCD is the accepted theory of strong
interactions.  In the set of these six lectures an effort will be made
to review the phenomenological understanding of  DIS and how its
study led to the  parton model, along with some aspects of parton
structure of nuclei.  We begin with an introduction as to how one
probes the structure of a nucleus via electromagnetic interactions and
explain the concept of the electric and magnetic form factors of the
proton/neutron.  Following this, we will discuss the DIS kinematics and
introduce the idea of two structure functions in terms of which the
electromagnetic DIS cross-section can be parametrised.  This will be
followed by a discussion of scaling of the structure functions and
parton model.  The scaling violations predicted by QCD will be alluded
to very briefly.  In the case of the DIS processes mediated by the
electroweak gauge bosons, $W^\pm/Z^0$, three independent structure
functions are required to parametrise the DIS cross-section.  We
introduce these and briefly discuss the important role played by
neutrino induced DIS processes in providing conclusive evidence for the
parton model.  The momentum distribution of partons measured in both
the electromagnetic and the weak DIS processes is found to be affected
by the nuclear environment.  This is the so called EMC effect
\cite{oldemc,newreview}.  This will be discussed in the last lecture.
Various models \cite{newreview,myreview} have been put forward as a
theoretical explanation of this effect.  Some of the models will be
summarized along with a discussion of possible experimental tests to
distinguish among these models \cite{myreview}.

\setcounter{equation}{0}
\section{Electromagnetic Structure of Hadrons}

Historically, hadrons were classified as particles with large masses
$(m \simeq 1$ GeV) which participate in strong interactions.  The
wide variety of these hadrons
$(p,n,\Lambda,$ $\Sigma^0,\Sigma^\pm,\Xi^\pm$
$,\pi^\pm,\pi^0,K^\pm,K^0,\bar K^0,
\cdots)$ already indicated that these are not elementary.  Further, for an
elementary, spin ${1\over2}$ particle, Dirac theory predicts gyromagnetic
ratio $g$ to be 2.  Even the deviation of $g$ from 2 can be computed in
perturbation theory.  For example, for an $e^-$, theoretical computations
\cite{(g-2)cal} predict
\bea
\left({g-2 \over 2}\right)^{th}_e &=& {\alpha \over 2\pi} - 0.328478966
\left({\alpha \over \pi}\right)^2 + 1.1765(13) \left({\alpha \over
\pi}\right)^3 -.8(2.5)\left({\alpha \over
\pi}\right)^4 \nonumber\\[2mm]
&=& 0.001159652\underline{460}(192)
\label{one.1}
\eea
where $\alpha$ is the fine structure constant given by
$e^2/4\pi$.\footnote{I will use units $\hbar = c = 1$ throughout these
lectures.}  The experimentally measured value \cite{pdg} is
\be
\left({g-2 \over 2}\right)^{expt.}_e = 0.0011596521\underline{93}(10)
\label{one.2}
\ee
In eqs. (\ref{one.1}) and (\ref{one.2}) the numbers in the bracket
indicate the theoretical and
experimental uncertainty respectively.  The excellent agreement between
the theoretical prediction and the experimentally measured value
indicates that the electron is indeed elementary.  However, for a proton
and neutron, the measured values \cite{pdg} of magnetic moments are
\bea
\mu_p = 2.79 ~\mu_N = 2.79 {{|e|} \over 2M_p}\;\;\; {\huge {^,}}
\nonumber\\[2mm]
\mu_n = -1.91~\mu_N = -1.91 {{|e|} \over 2M_p}\;\; {\huge ^{.}}
\label{one.3}
\eea
These correspond to values of $1.79~\mu_N$ and $-1.91~\mu_N$ for anomalous
magnetic moments of the proton and neutron respectively.  This clearly
indicates that they are not elementary particles {\it i.e.} they have spatial
extension.  They have a structure.  But this is, strictly speaking, an
energy dependent statement.  At low energies, the scattering experiments
are not able to `resolve' this structure.  After all, Rutherford concluded
from his scattering experiments  \cite{rutherford} with $\alpha$ particles
that all the positive charge in an atom is concentrated in a
{\it point} nucleus and the electrons occupy the rest of the atomic
volume.  But what this conclusion really meant was that the nuclear radius,
$r_N$, is much smaller than atomic sizes $(\sim {\cal O}$
(few $\buildrel \circ \over A \; \approx 10^{-10} ~m))$ .
One had to perform the scattering
experiments \cite{hofstadter} with higher energy electrons $(\sim 100 -
500 ~{\rm MeV})$ to reveal the structure of nuclei at a distance scale
$\sim$ few fm $= 10^{-15}~ m$.



\subsection{Effect of structure of the scattering centers on
scattering amplitudes}
To understand this let us consider a general scattering process
\be
e^- + A \rightarrow e^- + A \; .
\label{one.4}
\ee
The process is represented digramatically in fig. \ref{fig1}.  Let $P_1$ and
$P_2$\footnote{I use Pauli metric such that $P^2_i = -m^2_i$.  Details of
notation are given in Appendix A.} denote the four momentum of the target
nucleus and the incoming electron respectively and $P_3,P_4$ be the four
momenta of the  $e^-$ and nucleus in the final state, respectively.  Let
\bea
P_1 \equiv (\vec 0,iM_A), ~~~P_2 = (\vec p_0,iE_0) \;, \nonumber\\[2mm]
P_3 \equiv (\vec p,iE), ~~~~~P_4 = (\vec p^{~\prime},iW) \; \; .
\label{one.5}
\eea
As shown in the figure, the electromagnetic scattering takes place via the
exchange of a photon.  Application of four-momentum conservation at each
vertex implies that the four momentum of the photon is given by
\be
q = P_2 - P_3 = P_4 - P_1 \;\; .
\label{one.6}
\ee
In the notation of eq. (\ref{one.5}) we have,
\bea
q \equiv (\vec q,iq_0) &=& (\vec p^0 - \vec p,\; i(E_0 - E)) \nonumber\\[2mm]
&=& (\vec p^{~\prime},\;i(W-M_A)) \;.
\label{one.6p}
\eea
Using eqs. (\ref{one.5}) and (\ref{one.6}), we get for the invariant mass
of the photon,
\be
q^2 = (P_2-P_3)^2 = -2m^2_e - 2|\vec p_2|~|\vec p_3| \cos\theta +
2E_2E_3 \;\; .
\label{one.7}
\ee
Neglecting the electron mass $m_e$ (which also implies $|\vec p_2| = E_0 =
|\vec p_0| = p_0$ and $|\vec p_3| = E = p$) we get,
\be
q^2 = 2pp_0 (1 - \cos\theta) = 4pp_0 \sin^2 {\theta\over2} \; \;\; {\huge^{.}}
\label{one.8}
\ee
Thus for a scattering process one always has $q^2 > 0$.  This means that the
exchanged photon is not a real photon but a virtual one.

It can be shown, on very general grounds, that the net effect of the
presence of a structure in the scattering center on the scattering
amplitude, calculated from a diagram such as shown in fig. \ref{fig1},
is to multiply it by a form factor $F(q^2)$.  This form factor is given by,
\be
F(q^2) = \int e^{iq \cdot y}\; f(y)\; d^4 y
\label{one.9}
\ee
where $f(y)$ is the distribution function describing the target.  In the
limit of recoil-less, elastic scattering $(W \approx M_A)$ eq.
(\ref{one.6}) gives
$\vec q \equiv (\vec q,i0)$.  In this case the four-fourier transform
reduces to the spatial Fourier transform.

%

\subsection{Relation of spatial charge distribution to form factor}
Let us consider scattering of a spinless electron from a spinless charge
distribution and derive an expression for the form factor in this case.
Consider an $e^-$ of energy $E_0$ and momentum $\vec p_0$ incident on a
target nucleus of charge $Z|e|$ at rest.  The four momenta of various
particles are then given by eq. (\ref{one.5}).  For a point nucleus, the
classical Rutherford scattering formula for the differential
cross-section\footnote{Note that this formula can be derived classically
as well as in non-relativistic quantum mechanics using Born approximation
for the scattering of an $e^-$ in the screened, Coulomb field of
a nucleus.} is given by
\be
{d\sigma\over d\Omega}\Bigg|_{\rm Rutherford} = {m^2_e Z^2(e^2/4\pi)^2
\over 4p^4_0 \sin^4 {\theta\over2}} = {Z^2\alpha^2m^2_e \over 4p^4_0 \sin^4
{\theta\over2}} \;\;\;\; {\huge^{,}}
\label{ruther}
\ee
where $\Omega$ is the solid angle and $d\Omega = \sin\theta d\theta
d\phi$, $\theta$ being the scattering angle.  For a spinless, relativistic
$e^-$ scattered by the screened, Coulomb potential of a point nucleus,
this expression is,
\be
{d\sigma \over d\Omega} = {Z^2\alpha^2 \over 4E^2_0 \sin^4
{\theta\over2}} \; \; \; {^{.}}
\label{one.10}
\ee
If we assume the initial electron to be incident along the $z$ axis, the
kinematics is as depicted in fig. \ref{fig2}.  Now using eqs. (\ref{one.5}) and
(\ref{one.6}), we see that
\begin{subequations}
\label{one.11}
\begin{eqalignno}
\vec p_0 &= \vec p + \vec p^{~\prime} \;\; ,\label{one.11a}\\
M_A + E_0 &= E + W \;\; . \label{one.11b}
\end{eqalignno}
\end{subequations}
If $E_i$ and $E_f$ are used to indicate total initial and final state
energies, we have
\be
E_i = M_A + E_0 = M_A + p_0 = E_f = W + E = W + p \; \; .
\nonumber
\ee
Let us suppose that the nuclear charge $Z|e|$ is distributed with a
distribution function $\rho(R)$, normalised to unity by,
\be
\int \rho(\vec R) d^3 R = 1 \; \;.
\label{norm}
\ee
Hence the total electric charge in a volume element $d^3 R$ is given
by $Z|e| \rho(\vec R) d^3 R$.  The scattering cross-section is
\be
d\sigma = {{\cal W} \over v} \; \;,
\label{one.12}
\ee
where ${\cal W}$ is the transition probability and $v$ is the velocity of
the incident $e^-$ w.r.t. the center (which for relativistic electrons is
$1$ in our units).\footnote{Note here that I am normalising with one
particle per unit volume.}  The transition probability ${\cal W}$ is
calculated using Fermi's golden rule
\be
{\cal W} = 2\pi \rho_f |{\cal M}_{if}|^2 \; \; ,
\label{one.13}
\ee
where $\rho_f$ is the density of states in the final state and
${\cal M}_{if}$ is
the scattering amplitude, computed in Born approximation for the screened,
Coulomb nuclear potential.  For the two particle final states, the density
of final states $\rho_f$ is given by (with $\hbar = 1$),
\be
\rho_f = {dN \over dE_f} = p^2 d\Omega {dp \over dE_f}
\label{one.14}
\ee
and ${\cal M}_{if}$ is given in the Born approximation as
\be
{\cal M}_{if} = \int \psi^\star_f (\vec r) V(\vec r) \psi_i (\vec r)
d^3 r \;\;\; .
\label{one.15}
\ee
$dp /dE_f$ in eq. (\ref{one.14}) can be computed using eq.
(\ref{one.11a}) as follows. Eq. (\ref{one.11a}) gives
\bea
E_f = p + W &=& \sqrt{|\vec p^{~\prime} |^2 + m^2_A} + p \nonumber \\[2mm]
&=& \sqrt{M^2_A + p^2_0 + p^2 - 2pp_0\cos\theta} + p  \;\;\; .\nonumber
\eea
This in turn gives,
\be
{dp \over dE_f} = {W \over E_f - p_0 \cos\theta} \;\;\; {^{.}}
\label{one.16}
\ee
The energy and three momentum conservation of eqs. (\ref{one.11a}) ,
(\ref{one.11b})  gives
\be
E_f = p_0 + M_A = p + W = p + \sqrt{p^2 + p^2_0 - 2pp_0 \cos\theta +
m^2_A} \;\; .
\ee
This gives us,
\be
{p_0 \over p} M_A = E_f - p_0 \cos\theta \;\; \; .
\label{one.17}
\ee
This can also be rewritten as ,

\be
{p \over p_0} = {{1}\over {1 + 2p_0 /M_A} \sin^2
{\theta\over2}} \;\; {\huge ^{.}}
\label{one.18p}
\ee
Combining eqs. (\ref{one.16}) and (\ref{one.17}) we have
\be
{dp \over dE_f} = {W \over p_0}  {p \over M_A} \;\; .\label{one.18}
\ee

To calculate ${\cal M}_{if}$ in eq. (\ref{one.15}) we need $V(\vec r)$.  If we
take into
account the screening of the nuclear charge due to atomic electrons, the
potential $V(\vec r)$ felt by the electron at $\vec r$, due to the nuclear
charge distribution shown in fig. \ref{fig3} will be,
\be
V(\vec r) = - {Ze^2 \over 4\pi} \int {\rho(\vec R) d^3 R \over |\vec r -
\vec R|} e^{-|\vec r - \vec R|/a}\;\; ,
\label{one.19}
\ee
where $a$ is a damping factor $\sim {\cal O} $ (atomic radius)
which arises from the screening of the nuclear charge by the
atomic electrons and is called
the screening radius.  Clearly $a \gg R$.  If one uses the plane wave
approximation for the wave functions of the incident and scattered
electrons, with momenta $\vec p_0$ and $\vec p$ respectively, we have
\be
\psi^\star_f (\vec r) = e^{-i\vec p \cdot \vec r}, ~~\psi_i (\vec r) =
e^{i\vec p_0 \cdot \vec r} ~~~~ (\hbar = c = 1) \;\; .
\ee
This then gives for ${\cal M}_{if}$~,
\bea
{\cal M}_{if} &=& {-Ze^2 \over 4\pi} \int e^{i\vec q \cdot \vec r} d^3 r \int
{\rho (\vec R) e^{-|\vec r - \vec R|/a} \over |\vec r - \vec R|} d^3 R
\nonumber \\[2mm] &=& {-Ze^2 \over 4\pi} \int e^{i\vec q \cdot \vec R}
\rho (\vec R) d^3R \int {e^{i\vec q(\vec r - \vec R) - |\vec r - \vec
R|/a} \over |\vec r - \vec R|} d^3 R \;\; .\nonumber
\eea
Choosing the $z$ axis along $\vec q$ and $\vec s = \vec r - \vec R$, we get
\be
{\cal M}_{if} = {-Ze^2 \over 4\pi} \int e^{i\vec q \cdot \vec R} \rho (\vec R)
d^3R \int^\infty_0 2\;\pi~s ds \int^1_{-1} d\cos \alpha\
e^{iqs\cos\alpha} \ e^{-s/a} \;\;.
\ee
The first integral is a function only of $ Q^2 = |\vec q |^2$.  Then we have
\be
{\cal M}_{if} = -  F(Q^2) {Ze^2 \over 4\pi} {4\pi \over Q^2 + 1/a^2}
\;\; ,
\ee
where
\be
F(Q^2) = \int e^{i\vec q \cdot \vec R} \rho(R) d^3R \;\; .
\label{one.19p}
\ee
$F(Q^2)$ is called the form factor.  Note here that the screening
factor $e^{-s/a}$ in the integrand is essential to make this integral
convergent.  $Q^2 = |\vec q|^2 = p^2_0 + p^2 - 2pp_0 \cos\theta$ is
$\sim {\cal O}({\rm MeV}^2)$ for the energies under consideration
whereas $a \sim {\cal O} (10^{-8} {\rm cm})$ ; {\it i.e.} ${1}/{a^2}
\simeq 4({\rm KeV}^2)$. \footnote{Use $\hbar c = 197$ MeV fm to arrive
at this result.}  Hence for $E_0 \gsim O $ (MeV) , ${1}/{a^2} \ll Q^2$.
Therefore the expression for ${\cal M}_{if}$ can be approximated as
\be
{\cal M}_{if} \simeq \frac{-Ze^2}{q^2} \simeq \frac{-Ze^2}{Q^2} \;\; .
\label{one.20}
\ee
Substituting eqs. (\ref{one.20}), (\ref{one.16}) and (\ref{one.14}) in
(\ref{one.13}), we get
\be
d\sigma = 2\pi \frac{p^2d\Omega}{(2\pi)^3}  \frac{W}{E_f - p_0
\cos\theta}  \left(\frac{Ze^2}{4\pi}\right)^2
\frac{(4\pi)^2}{Q^4} |F(Q^2)|^2 \; \; .
\ee
Using eq. (\ref{one.18p}) we get finally,
\be
\frac{d\sigma}{d\Omega} = 4p^2 \frac{W}{M_A}  \frac{p}{p_0}
\left(\frac{Ze^2} {4\pi}\right)^2\;  \frac{|F(Q^2)|^2}{Q^4} \; \; .
\ee
Since the nuclear mass  $M_A \sim {\cal O} $ (GeV) $\sim {\cal O} (1000$ MeV),
for electron beam energies of $\sim {\cal O} $ (MeV) , $W \simeq M_A$ .  Hence
eq. (\ref{one.11b})  tells  us $p \approx p_0$ and hence
$Q^2 \simeq
4p^2_0 \sin^2 {\theta / 2} = 4E^2_0 \sin^2 {\theta/ 2}$.  Using
all this the expression for the differential cross-section becomes
\be
\frac{d\sigma}{d\Omega} = \frac{Z^2  \alpha^2}{4E^2_0\sin^4
\frac{\theta}{2}} \; |F(Q^2)|^2 \;\; .
\label{one.21}
\ee
Comparing eqs. (\ref{one.21}) and (\ref{one.10}) we see that the net
effect of the distribution of the nuclear charge  over a spatial volume
according to a distribution function $\rho(\vec R)$ is to multiply the
cross-section by the square of the form factor $|F(Q^2)|^2$.  $F(Q^2)$
is nothing but the three dimensional Fourier transform of the charge
distribution $\rho (\vec R)$.  If $|F(Q^2)| = 1$ for all $Q^2$, then
eq. (\ref{one.21}) reduces to eq.  (\ref{one.10}) which is the
Rutherford scattering cross-section formula for the scattering of a
relativistic electron from a spinless point nucleus; {\it i.e.},
\be
\left(\frac{d\sigma}{d\Omega}\right)_{\rm charge~distn.} = |F(Q^2)|^2
\left(\frac{d\sigma} {d\Omega}\right)_{\rm point~nucleus}\;\;\;,
\label{one.22}
\ee
where the form factor is given by eq. (\ref{one.19p}).  For a spherically
symmetric charge distribution, choosing the $z$ axis along $\vec q $,
the equation for $F(Q^2)$ becomes
\be
F(Q^2) = 4\pi \int^\infty_0 \rho (R) \frac{\sin QR} {QR} R^2 dR \;\;\; .
\label{one.23}
\ee
The charge distribution $\rho (R)$ is given in terms of $F (Q^2)$ by the
inverse Fourier transform,
\be
\rho (R) = \frac{1} {2\pi^2} \int F(Q^2)  \frac{\sin QR}{QR} Q^2 dQ
\;\; .
\label{one.24}
\ee
At values of $Q = |\vec q|$ such that $QR \ll 1$ over all the region where
$\rho (R)$ is appreciable, one can expand $\sin QR / QR $
in the integrand in eq. (\ref{one.23}) and we get
\be
F(Q^2) = 4\pi \int^\infty_0 R^2 \rho(R) dR - \frac{Q^2}{6} \int^\infty_0
4\pi R^2 \rho (R) R^2 dR + O(Q^4 R^4) \;\; .
\ee
Hence for a charge distribution normalised as given by  eq.
(\ref{norm}) we have,
\be
F(Q^2) = 1 - \frac{Q^2 \langle R^2\rangle} {6} \;\;  .
\label{one.25}
\ee
Thus a measurement of $F(Q^2)$ at small $Q^2$ such that $Q^2 \langle
R^2\rangle \ll 1$ gives us $\langle R^2 \rangle$ for different charge
distributions. Table \ref{tabform} summarises the form factors $F(Q^2)$
and $ \langle R^2 \rangle $ for different charge distributions where
these can be calculated analytically. As can be seen, all of them satisfy eq.
(\ref{one.25}).

\begin{table}
\caption{
The form factors and the Root Mean Square (r.m.s.) radii
$ \langle R^2 \rangle ^\protect{1/2}$ for different charge distributions.}
\vspace{0.2cm}
\begin{center}
\begin{tabular}{|c|c|c|c|}
\hline
& & & \\
$\rho(R)$ & $F(Q^2)$ &$ \langle R^2 \rangle ^{1/2}$&  \\[2mm]
\hline
& &  &\\
$\delta^3(\vec R)$ & 1 & 0.0 &\\[4mm]
\hline
& & & \\
$\frac{e^{-mR} m^2}{4 \pi R}$ &
$\frac{1}{(1 + \frac{Q^2}{m^2})}$ &
$\frac{\sqrt{6}} {m}$& {\rm Monopole}\\[4mm]
\hline
& & & \\
${e^{-mR} m^3 \over 8 \pi}$&
${1 \over (1 + \frac{Q^2}{m^2})^2}$&
$ {\sqrt{12} \over m}$& {\rm Dipole}\\[4mm]
\hline
\end{tabular}
\label{tabform}
\end{center}
\end{table}

Recall now from eq. (\ref{one.6p}) that the square of the four-momentum
transfer $q^2$ is given by
\be
q^2 = |\vec p^{~\prime}|^2 - (W - M_A)^2 = -2M^2_A + 2WM_A \;\; .
\ee
This gives us
\be
\frac{q^2}{2M^2_A} + 1 = {\frac{W}{M_A}} .
\label{one.26}
\ee
Further for $W \simeq M_A$ ({\it i.e.} $q^2 \ll 2m^2_A$; small nuclear recoil),
we also have $q^2 \simeq Q^2$.  This means that in this approximation
we can replace $Q^2$ by $q^2$ in all the earlier equations.

If we repeat the analysis above for a point $e^-$ incident on a point
spinless charge $Z|e|$ but keeping the effect of $e^-$ spin we
get (see problem 4, Appendix B)
\be
\left(\frac{d\sigma}{d\Omega}\right)_{{{\rm point}~e^-} \atop {\rm
point,~spinless~nucleus}} = \frac{Z^2 \alpha^2 \cos^2 \frac{\theta}{2}}
{4E^2_0 \sin^4 \frac{\theta}{2} \left[1 + {2p_0 \over M_A} \sin^2
\frac{\theta}{2}\right]} \;\;\; .
\ee
Neglecting the recoil, for relativistic $e^-$-nucleus scattering this
becomes,
\be
\left(\frac{d\sigma}{d\Omega}\right)_{\rm Mott.} = \frac{Z^2 \alpha^2
\cos^2 \frac{\theta}{2}} {4 E^2_0 \sin^4 \frac{\theta}{2}} = \frac{4Z^2
\alpha^2 \cos^2 \frac{\theta}{2}\; E^2} {q^4} \;\; .
\label{one.27}
\ee
The factor of $\cos^2 \frac{\theta}{2}$ in the numerator in eq. (\ref{one.27})
indicates the impossibility of 180$^\circ$ scattering for longitudinally
polarised, spin $\frac{1}{2}$ electrons (see problem 3, Appendix B for a
further discussion of this point).  Once again, repeating the
exercise for a nuclear charge distribution, we get
\be
\left(\frac{d\sigma}{d\Omega}\right)_{{{\rm spin}~\frac{1}{2} e^-} \atop
{\rm spinless,~charge~distn.}} = \left(\frac{d\sigma}{d\Omega}\right)_{\rm
Mott} |F(q^2)|^2 \;\; ,
\ee
where $F(q^2)$ is the form factor given by eq. (\ref{one.19p}).  Hence for
a spinless nucleus the form factor can be measured by the ratio:
\be
|F(q^2)|^2 = \frac{(d\sigma/d\Omega)^{eA \rightarrow eA}}
{(d\sigma/d\Omega)_{\rm Mott}} \;\; .
\label{one.28}
\ee
The charge distribution of the nucleus is then given by the inverse
Fourier transform of $F(q^2)$.  Hofstadter studied \cite{hofstadter} the
$e^-$-nucleus scattering for $E_{e^-} \lsim 600$ MeV.  Hence the electrons
were relativistic.  Since $M_A \simeq 1000 $ A MeV, the slow recoil
assumption is also justified in this case.  Hence, the nuclear form
factors could be determined in this case using eq. (\ref{one.28}).  Thus a
deviation of the angular dependence of the cross-section from the one
expected for a point target, measures the form-factor.  Therefore to get
information about a possible structure of the scattering center, we must
know the theoretical predictions for the results expected for a point
target.  These, along with the experimental information on the
form-factor, for the case of proton/neutron will be discussed in the next
lecture.  It must be noted here that the interpretation of the form-factor
as a three dimensional Fourier transform of the charge distribution is
strictly true only in the case of recoilless scattering.  At higher
energies, when this assumption is not justified, the form-factor $F(q^2)$
can be looked upon as the ratio of the scattering amplitudes from an
extended and point target.

\setcounter{equation}{0}
\section{Form Factors of Proton and Neutron}

\subsection {$e^-p \rightarrow e^-p$ for pointlike ``Dirac''
proton}

The discussion in the last section and eq. (\ref{one.3}) in particular
indicates that the proton and neutron are not pointlike and
elementary. It also makes it clear that one can obtain information
about the charge distribution in a proton (neutron) by studying the
electromagnetic scattering process
\be
e^- + p\; (n) \rightarrow e^- + p \; (n) \;\; .
\label{two.1}
\ee
Eq. (\ref{one.28}) indicates that to be able to do this it  is
essential to know the expected cross-section ${d\sigma}/{d\Omega}$ for
a point proton.  In the discussion below we will try to indicate how
such a calculation is done.

At the energies which we are considering the proton and neutron also need
to be treated relativistically.  For a Dirac ({\it i.e.} point) $e^- (p)$, the
electromagnetic current is given by
\bea
J^{e(p)}_\mu &=& iq_{e(p)} \overline \psi \gamma_\mu \psi \nonumber \\[2mm]
&=& i \frac{q_{e(p)}} {2m_e(M_p)} \left(\frac{\partial\overline\psi}{\partial
x_\mu} \psi - \overline\psi \frac{\partial\psi}{\partial x_\mu}\right) -
\frac{q_{e(p)}} {2m_e(M_p)} \frac{\partial}{\partial x_\nu} (\overline\psi
\sigma_{\nu \mu} \psi) \;\;\; ,
\label{two.2}
\eea
where $\gamma_\mu$ are the usual Dirac matrices and $\sigma_{\mu\nu} =
\frac{1}{2i} [\gamma_\mu,\gamma_\nu]$; $q_{e(p)}$ is the charge of the
$e^-(p)$ and $m_e(M_p)$ is the mass of the $e^-(p)$.  The fermion index
$(e^-$ or $p$) on the spinor is suppressed.  It can be shown easily that
this expression for the electromagnetic current means $g_e = 2$, $\mu_e =
1$ B.m., $g_p = 1$ and $\mu_p = 1$ n.m.  To calculate $d\sigma/d\Omega$
one needs the matrix element for the scattering process $e^- p \rightarrow
e^- p$.  This is given by
\be
{\cal M} \sim \frac{1} {q^2}  J^e_\mu J^p_\mu \;\; ,
\label{two.3}
\ee
where $q^2$ is the square of the four momentum transfer in this process.
The Feyman diagram for the scattering process and the four momentum
assignments of various particles involved in it are shown in fig.
\ref{fig4}.  The kinematics of the process is the same as given by eqs.
(\ref{one.5}) - (\ref{one.8}), (\ref{one.11a}, \ref{one.11b})
and (\ref{one.17}) - (\ref{one.18}), after replacing $M_A$ by $M_p$.
Again the square of invariant mass of exchanged photon $q^2 \not
= 0$, and hence it is virtual.

To calculate ${d\sigma}/{d\Omega}$ now one has to use the rules of
quantum field theory.  The cross-section for a process $A + B \rightarrow
C + D$ is given by
\be
d\sigma = \frac{1} {2E_A} \frac{1} {2E_B} \frac{(2\pi)^4 \delta^4 (P_A +
P_B - P_C - P_D)} {(2\pi)^6} \frac{d^3 p_C} {2E_C} \; \frac{d^3 p_D}
{2E_D}\; \overline{|{\cal M}|^2}
\label{two.4}
\ee
where $P_A,P_B,P_C$ and $P_D$ are the four momenta of the four particles
involved and $\overline{|{\cal M}|^2 } $ is the square of the matrix element
averaged over initial state spins and summed over the final state spins.
The matrix element ${\cal M}$ itself for the scattering process is given by eq.
(\ref{two.3}).  Using $J^{e(p)}_\mu$ given by eq. (\ref{two.2})
and the trace rules given in  Appendix A , we get for a point
proton (see problem 4, Appendix B) ,
\begin{eqnarray}
\frac{d\sigma} {d\Omega} (e^- p \rightarrow e^-p)\Big|_{\rm
Point~e^-/p} &=& {\alpha^2  \cos^2 {\theta \over 2} \over
4E^2_0 \sin^4 {\theta \over 2}} {1 \over 1 + {2p_0 \over M_p} \sin^2
{\theta \over 2}} \left[1 + {q^2 \over 2M^2_p} \tan^2
{\theta \over 2}\right]  \nonumber \\[2mm]
&=& \left({d\sigma \over d\Omega}\right)_{{\rm Mott} \atop (Z=1)}
{p \over p_0} \left[1 + {q^2 \over 2M^2_p} \tan^2
{\theta \over 2}\right]  \;\; .
\label{two.5}
\end{eqnarray}
The first term is just the Mott electrostatic scattering cross-section of eq.
(\ref{one.27}) with the nuclear charge $Z=1$, the factor of $p/p_0$ is the
recoil factor which goes to unity for recoil-less scattering
(recall  eq. (\ref{one.11b})) and the last factor is due to the
magnetic moment of the proton and electron which reduces to unity if
proton were to be replaced by a spinless point charge. Thus
for a scalar, point proton we can write (see problem 4, Appendix B),
\begin{eqnarray}
\frac{d\sigma} {d\Omega} (e^- p \rightarrow e^-p)\Big|_{\rm
Point~e^-/p} &=& {\alpha^2  \cos^2 {\theta \over 2} \over
4E^2_0 \sin^4 {\theta \over 2}} {1 \over 1 + {2p_0 \over M_p} \sin^2
{\theta \over 2}} \left[1 + {q^2 \over 2M^2_p} \tan^2
{\theta \over 2}\right]  \nonumber \\[2mm]
&=& \left({d\sigma \over d\Omega}\right)_{{\rm Mott} \atop (Z=1)}
{p \over p_0}  \; \; .
\label{two.5p}
\end{eqnarray}
Note also that we are no longer
restricted to $W \simeq M_p$ and hence $q^2 \neq |\vec q|^2 = Q^2$.

%
\subsection{Effect of anomalous magnetic moment of proton}
The above discussion assumes that the proton is a Dirac particle with
$g_p = 2 $.  In reality, even if we were to neglect the possibility
that the proton is an extended object, the anomalous magnetic moment
of the proton cannot be neglected, {\it i.e.}, eq.  (\ref{two.2}) cannot give
the electromagnetic current for the physical proton even if were to
be point like.  The anomalous magnetic moment gives an additional
contribution to the electromagnetic current of a proton and it can be
written as\footnote{see for example Advanced Quantum Mechanics, J.J.
Sakurai, (Addison Wesley, Reading, Mass., USA)}
\be
``{J_{\mu}^{p}}\; "= {iq_p \over 2M_p}
\left[{\partial \overline\psi_p \over \partial x_\mu} \psi_p - \overline\psi_p
{\partial \psi_p \over \partial x_\mu}\right] - {q_p \over 2M_p} (1 +
\kappa_p) {\partial \over \partial x_\nu} \overline \psi_p \sigma_{\nu \mu}
\psi_p \; \; .
\label{two.6}
\ee
Note here that the effects of a possible spatial structure of the proton
are not taken into account  apart from the nonzero value of the
anomalous magnetic moment.  The interaction Hamiltonian ${\cal H}^{int}$
is given by $\sim `` {J^p_{\mu}}\; " A_\mu$ and
hence the scattering amplitude will be determined by this.  Recall also
$\psi (x) \sim u(P) \exp(iP \cdot x)$.  Since $P_1$ and $P_4$ are the four
momenta of the initial and final state respectively, the expression for
the current of eq. (\ref{two.6}), in momentum space, becomes
\be
`` {J^p_{\mu}}\; " = J^{Dirac}_\mu + {iq_p \over 2M_p}
\bar u_p (P_4) \sigma_{\nu \mu} u_p (P_1) q_\nu \; \; ,
\label{two.7}
\ee
where $J^{Dirac}_\mu$ represents the current of eq. (\ref{two.2}), in momentum
space, given by
\bea
J^{Dirac}_\mu &=& iq_p \bar u_p (P_4) \gamma_\mu u_p (P_1)
\nonumber \\[2mm]
&=& {q_p \over 2M_p} (P_{4\mu} + P_{1\mu}) \bar u_p (P_4) u_p(P_1)
+ {iq_p \over 2M_p} q_\nu \bar u_p (P_4) \sigma_{\nu \mu} u_p (P_1).
\label{two.8}
\eea
This is called the Gordon decomposition of the electromagnetic
current of a spin 1/2 particle.
%
\subsection{Effect of the spatial extension of proton}
The discussion upto now has neglected a possible spatial extension for
the proton.  However, the large value of $\kappa_p$ given by eq.
(\ref{one.3}) implies that the proton is not a spin 1/2, point
particle.  Our discussions on nuclear form factor indicate that the
proton structure will be reflected in a multiplicative form factor
for the scattering amplitude.  Following this, we can try to write
down the most general expression for the electromagnetic current of a
spin 1/2 object by constructing a four vector using the various
bilinear covariants made up of $\overline \psi_p (P_4)$ and $\psi_p
(P_1)$ along with the independent four momenta available in the
scattering process of eq. (\ref{two.1}).  The requirement that parity
be conserved, {\it i.e.}, the electromagnetic current be a vector and
not a pseudovector, rules out all the bilinears containing
$\gamma_5$.  The possible 4-vectors that can be constructed are then:
\begin{enumerate}
\item $\bar u_p (P_4) \gamma_\mu u_p (P_1)$
\item $q_\nu \bar u_p (P_4) \sigma_{\nu \mu} u_p (P_1)$
\item $(P_{1\nu} + P_{4\nu}) \bar u_p (P_4) \sigma_{\nu \mu} u_p
(P_1)$
\item
\subitem[a] $(P_{1\mu} + P_{4\mu}) \bar u_p(P_4) u_p (P_1)$
\subitem[b] $(P_{4\mu} - P_{1\mu}) \bar u_p (P_4) u_p(P_1)$
\end {enumerate}
Out of these, (3) can be shown proportional to (4.b) by
using Dirac equation, \linebreak
$\not\!\!P u_p(P) = iM_p u_p(P)$ as well as the usual
$\gamma$-matrix algebra.  It is also clear from eq. (\ref{two.8}) that (1),
(2) and (4.a) are linearly dependent.  Hence (1), (2) and (4.b)
exhaust the set of linearly independent fourvectors that can be
constructed out of the spinors $\bar u_p(P_4)$, $u_p(P_1)$ and the
fourmomenta $P_1,P_4$ of the incoming and outgoing proton respectively.
Hence the most general expression for the electromagnetic current of a
spin 1/2 particle, which has a structure, can be written as,
\bea
J^{p,\ext}_\mu &=& iq_p \Bigg[{\cal F}^p_1 (q^2) \;
\bar u_p (P_4) \gamma_\mu u_p(P_1)
 + {\cal F}^p_2 (q^2) \; {\kappa_p \over 2M_p} q_\nu \bar u_p (P_4)
\sigma_{\nu \mu} u_p (P_1) \nonumber \\[2mm]
& & + {\cal F}^p_3 (q^2)\; \bar u_p (P_4) q_\mu u_p (P_1)\Bigg] \;\; ,
\label{two.9}
\eea
where due to Lorentz invariance,  ${\cal F}^p_1$, ${\cal F}^p_2$ and
${\cal F}^p_3$ are three arbitrary functions of the scalars that can be
constructed out of the available linearly independent fourvectors.  Since the
incoming and outgoing proton are on mass-shell, this means that ${\cal
F}^p_i$ $(i=1,2,3)$ can be functions of $q^2$ alone.  Eq.
(\ref{two.9})  means that the structure of the proton can be
parametrised by three arbitrary functions.

We know, however, that the electromagnetic current is a conserved
quantity, {\it i.e.}, $\partial_\mu J_\mu^{p,\ext} = 0$.  This
translates in momentum space into
\be
q_\mu J^{p,\ext}_\mu = 0 ~~~{\rm where}~~~ q_\mu = P_{4\mu} - P_{1\mu}.
\ee
Applying this condition to (\ref{two.9}) we get,
\bea
0 = q_\mu J^{p,\ext}_\mu &=& iq_p \Bigg[{\cal F}^p_1 (q^2) \bar u_p (P_4)
\gamma_\mu q_\mu u_p (P_1)+ {\cal F}^p_2 (q^2) {\kappa_p \over 2M_p}
q_\mu q_\nu \bar u_p (P_4)\sigma_{\nu \mu} u_p (P_1)\nonumber\\
&+& {\cal F}^p_3 (q^2) \bar u_p (P_4) q^2 u_p (P_1)\Bigg]\;\; .
\label{two.10}
\eea
The first term in the square bracket vanishes identically as can be seen
by using Dirac equation for proton.  The second term is identically zero
since $q_\mu q_\nu$ is a symmetric tensor in $(\mu,\nu) $ and it is
contracted with $\sigma_{\mu\nu}$ which is antisymmetric under $ (\mu
\leftrightarrow \nu )$.  Hence eq. (\ref{two.10}) becomes
\be
{\cal F}^p_3 (q^2)  q^2  \bar u_p (P_4) u_p (P_1) = 0 \;\; .
\ee
This then tells us that ${\cal F}^p_3 (q^2) \equiv 0$.  Hence the most
general expression for the electromagnetic current of a spin 1/2
proton with structure, which is consistent with Lorentz and gauge
invariance is given by
\bea
J^{p,\ext}_\mu &=& iq_p \Bigg[{\cal F}^p_1 (q^2) \bar u_p (P_4) \gamma_\mu
u_p (P_1)  +  {\cal F}^p_2 (q^2) {\kappa_p \over 2M_p} q_\nu \bar u_p (P_4)
\sigma_{\mu_\nu} u_p (P_1)\Bigg]\nonumber \\[2mm]
&\equiv& \bar u_p (P_4) \Gamma_\mu u_p (P_1)  \;\; .
\label{two.11}
\eea

Thus we see that the most general expression for the electromagnetic
current of an extended, spin 1/2 object is specified completely in terms
of just two arbitrary functions ${\cal F}_1 (q^2)$ and ${\cal F}_2 (q^2)$,
{\it i.e.}, the effect of the spatial extension of the scattering centre on the
scattering amplitude is simply parametrised in terms of these two
functions.  Comparing eq. (\ref{two.11}) with eq. (\ref{two.7}) we see that we
recover the case for a proton with an anomalous magnetic moment $\kappa_p$
but without a structure if ${\cal F}^p_2 (q^2)$ and ${\cal F}^p_1 (q^2)$
both are unity.  Our discussions on nuclear form factors tell us that for
$q^2 \ll {1 / \langle R^2\rangle_p}$, any form factor associated with
the
proton will be close to unity (recall eq. (\ref{one.25})).  In
particular eq. (\ref{two.7}) gives us,
\be
{\cal F}^p_1 (0) = 1, ~~~{\cal F}^p_2 (0) = 1 \;\; .
\label{two.12}
\ee
The two functions ${\cal F}^p_i (q^2)$ are called the two form factors
of the proton and are associated with two linearly independent
fourvectors in terms of which the electromagnetic current can be
decomposed.  For a neutron which has no electric charge and
only an anomalous magnetic moment we will have,
\be
{\cal F}^n_1 (0) = 0; ~~~{\cal F}^n_2 (0) = 1 ,
\label{two.13}
\ee
and $\kappa_p$ in eq. (\ref{two.11}) will be replaced by $\kappa_n$.
%
%
\subsection{Electric and Magnetic Form Factors}
Having established the most general expression for the electromagnetic
current of an extended object with spin 1/2, it is now necessary
to calculate ${d\sigma/d\Omega}$ for the scattering process of
eq. (\ref{two.1}) in terms of these form factors.  We can calculate
$d\sigma/d\Omega$ by using eq. (\ref{two.4}) for $d\sigma$, with
${\cal M}$ given by eq. (\ref{two.3}), where $J^p_\mu$ is given by eq.
(\ref{two.11}) and $J^e_\mu$ by eq. \ref{two.8} with a replacement of $q_p
\rightarrow q_e$, $M_p \rightarrow m_e$, $P_4 \rightarrow P_3$
and $P_1 \rightarrow P_2$.  We then get (see problems 4, Appendix B)
\bea
{d\sigma \over d\Omega} (e^-p \rightarrow e^- p)\Bigg|_{p,\ext}
&=& \left({d\sigma \over d\Omega}\right)_{\rm Mott}
\left({p \over p_0}\right) \Bigg\{\left(({\cal F}^p_1)^2 + {\kappa^2_p q^2
\over 4M^2_p} ({\cal F}^p_2)^2\right)  \nonumber \\[2mm]
& & \;\;\;+ {q^2 \over 2M^2_p} ({\cal F}^p_1 + \kappa_p {\cal F}^p_2)^2 \tan^2
{\theta \over 2}\Bigg\}  .
\label{two.14}
\eea
In the limit $q^2 \ll {1 / \langle R^2\rangle_p}$ and with $\kappa_p =
0$ this reduces to eq. (\ref{two.5}) as it should.  Since the term $(q^2/
2M^2_p)\tan^2 {\theta \over 2}$ in eq. (\ref{two.5}) is known to arise from
the magnetic moment of the ``Dirac'' proton and electron, on
comparing eqs. (\ref{two.5}) and  (\ref{two.14}) we see that
while the form factors ${\cal F}^p_1$ and ${\cal F}^p_2$ were
the natural ones while considering the tensor decomposition of
$J^{p,\ext}_\mu$, the combination $({\cal F}^p_1 + \kappa_p {\cal F}^p_2)$
has the more natural interpretation as the magnetic form factor.  If
we define the electric and magnetic form factors of the proton by
\bea
G^p_M (q^2) &\equiv& {\cal F}^p_1 (q^2) + \kappa_p {\cal F}^p_2 (q^2) ,
\nonumber \\[2mm]
G^p_E (q^2) &\equiv& {\cal F}^p_1 (q^2) - \kappa_p {q^2 \over 4M^2_p}
{\cal F}^p_2 (q^2)  .
\label{two.15}
\eea
the expression for the differential cross--section becomes,
\be
{d\sigma \over d\Omega} (e^-p \rightarrow e^-p) = \left({d\sigma \over
d\Omega}\right)_{\rm Mott} {p \over p_0} \left[{{G^p_E}^2 (q^2) + {q^2 \over
4M^2_p} {G^p_M}^2 (q^2) \over 1 + q^2/4M^2_p} + {q^2 \over 2M^2_p}
{G^p_M}^2 (q^2) \tan^2 {\theta \over 2}\right] .
\label{two.16}
\ee
The physical significance of $G^p_M (q^2)$ and $G^n_M (q^2)$ can be
understood in terms of electromagnetic scattering of protons with definite
helicity in the Breit frame.  The latter choice means that the scattering
involves no change in energy but only in the sign of the three momentum.
To see this let us analyse the expression for $J^{p,\ext}_\mu$ of
(\ref{two.11}), but for proton states of definite helicity (spin projection
along direction of motion).  Using eq. (\ref{two.8}) we can rewrite this as
\bea
&J&^{p,\ext}_{\mu,\lambda,\lambda'} = \bar u(P_4,\lambda') \Gamma_\mu
u(P_1,\lambda)  \nonumber \\[2mm]
&=& iq_p \bar u(P_4,\lambda') \left[\gamma_\mu G^p_M (q^2) + {i(P_{4\mu} +
P_{1\mu}) \over 2M_p} {(G^p_M (q^2) - G^p_E(q^2)) \over (1 +
q^2/4M^2_p)}\right] u(P_4,\lambda) .
\eea
The Dirac spinor for a particle with helicity $\lambda$ when its
momentum is along the $z$ axis $\vec p_1 = \hat z|\vec p_1|$ is given
by
\be
u(\vec p_1,\lambda) = N \left(\matrix {\chi_\lambda  \cr
\displaystyle{|\vec p_1| \sigma_3 \over E + M_p}  \chi_\lambda}\right) \; ,
\ee
where
\be
N = \sqrt{{E + M_p \over 2M_p}}, ~~~ \chi_{+1/2} = \left(\matrix{1 \cr
0}\right), ~~{\rm and}~~ \chi_{-1/2} = \left(\matrix{0 \cr 1}\right)\; .
\ee
$\chi_{+1/2}$ and $\chi_{-1/2}$ correspond to states with helicity +1 and -1
respectively.  The Dirac spinor for a particle with helicity
$\lambda$, but moving along the
negative $z$ direction will be obtained by considering the transformation
of the above spinor under rotation through $180^o$ about $y$ axis.
This transformation is given by $\exp(-i\pi \sigma_y)$, {\it i.e.},
$-i\sigma_y$.  (The negative sign comes from the fact that we are
rotating the physical state and not the co-ordinates.)  Hence we get,
\be
u(-\vec p_1,\lambda) = N  \left(\matrix{-i\sigma_y  \chi_\lambda \cr
\displaystyle {|\vec p_1| \sigma_x \over E+M_p}  \chi_\lambda}\right) .
\ee
In the Breit frame $\vec p_4 = -\vec p_1$ and therefore we can use the
above expressions for the spinors for a fixed helicity.  We then get,
\bea
J^{p,\ext}_{\mu,+1/2,+1/2} &=& G^p_M (q^2) {|\vec p_1| \over 2M_p} q_p
(1,i,0,0) , \nonumber \\[2mm]
J^{p,\ext}_{\mu,+1/2,-1/2} &=& q_p G^p_E (q^2) (\vec 0,-i)  .
\label{two.16p}
\eea
The first amplitude in eq. (\ref{two.16p}) corresponds to no helicity
flip (since the three momenta of the scattered and incident electron
are opposite to each other, this means spin flip) scattering and this
is proportional to the magnetic form factor, whereas the second
amplitude represents the contribution  with  helicity flip
(hence no spin flip) which corresponds to
the electric form factor.  Also note that the spin-flip amplitude due
to the magnetic scattering vanishes in the non-relativistic (NR)
limit.  This observation also explains clearly the terminology used,
viz., the nomenclature of electric/magnetic form factors.  Thus
electrostatic scattering cannot flip the spin of the electron whereas
magnetic scattering does in this kinematical configuration (see
discussion in problems).

The discussion for ${d\sigma \over d\Omega} (e^-p \rightarrow e^-p)$ will
be applicable to the case of neutrons equally well.  An expression similar
to eq. (\ref{two.16}) can be written for ${d\sigma \over d\Omega} (e^-n
\rightarrow e^-n)$ as well.  Only the boundary conditions on $G_{M,E} (q^2
= 0)$ will change appropriately.  Eqs. (\ref{two.12}), (\ref{two.13}) and
(\ref{two.14}) imply,
\bea
G^p_E (0) &=& 1, ~~~ G^p_M (0) = 1 + \kappa_p \equiv \mu_p ~({\rm in
{}~n.m.}) \; ,  \nonumber \\[2mm]
G^n_E (0) &=& 0, ~~~ G^n_M (0) = \kappa_n \equiv \mu_n ~({\rm
in~n.m.}) \; .
\label{two.17}
\eea
Experimental information on the form factors $G^{p(n)}_{M,E} (q^2)$ was
obtained  by studying ${d\sigma \over d\Omega} (ed
\rightarrow ed)$ and ${d\sigma \over d\Omega} (ep \rightarrow ep)
\equiv \left({d\sigma \over d\Omega}\right)_{ep}$. The
differential cross--section ${d\sigma \over d\Omega} (en \rightarrow
en)$ is then obtained from the first two measurements applying
correction factor for nuclear physics effects. It is clearly not
possible to do experiments with neutron targets as the neutron is not
stable. Since we are discussing elastic scattering, eq. (\ref{one.18p})
ensures that there is only one independent variable characterising the
final state.  This can be chosen to be either $\cos\theta$ or
$\displaystyle q^2 = (4 p^2_0 \sin^2 \theta/2)  \biggm /
\left( 1 + {2p_0 \over M} \sin^2 {\theta \over 2} \right)$ .
Hence eq. (\ref{two.16}) implies that
\be
{\left({d\sigma \over d\Omega}\right)_{ep}
\over\left({d\sigma \over d\Omega}\right)_{\rm Mott}}
= {p \over p_0}\left (A^p(q^2) + B^p(q^2) \tan^2 {\theta \over
2}\right) \;\; ,
\label{two.18p}
\ee
where
\bea
A^p(q^2) &=&  {1 \over  (1 + q^2/4M^2_p)} \;\;
\left({G^p_E}^2(q^2) + {q^2 \over 4M^2_p} {G^p_M}^2(q^2) \right)
 \nonumber \\
& = &\; \left(({\cal F}^p_1)^2 + {\kappa^2_p q^2
\over 4M^2_p} ({\cal F}^p_2)^2\right) ;\nonumber\\[2mm]
B^p(q^2) &=& {q^2 \over 2M^2_p} \; {G^p_M}^2(q^2)\nonumber\\
& =& \;\;{q^2 \over 2M^2_p} ({\cal F}^p_1 + \kappa_p {\cal F}^p_2)^2 .
\label{three.27}
\eea
The above formula is called Rosenbluth formula.  This also shows
clearly how one can determine the two functions $G^p_E (q^2)$ and
$G^p_M (q^2)$ from a measurement of $\left(d\sigma \over
d\Omega\right)_{ep}$.  From the kinematic considerations of
eq. (\ref{one.18p}), one sees that it is possible to obtain data at a
fixed $q^2$ and fixed angle $\theta$, by changing the incident
electron beam energy.

The form factors for both the proton and neutron have been measured now
over a wide range of $q^2$ $(q^2 \lsim 30$ GeV$^2$ for the proton and $q^2
\lsim 5$ GeV$^2$ for the neutron).
The experimentally measured form-factors for proton and
neutron seem to obey a scaling law in the following sense:
\be
G^p_E (q^2) = {G^p_M (q^2) \over |\mu_p|} = {G^n_M (q^2) \over |\mu_n|}
\equiv G (q^2) = {1 \over (1 + q^2/M^2_V)^2} \;\; {\huge ^{,}}
\label{two.18}
\ee
with $M^2_V = (0.84)^2$ GeV$^2$.  This is the dipole form factor of
Table \ref{tabform}. $G^n_E (q^2)$ is identically zero as is
to be expected.  If we again compare (\ref{two.16}) and (\ref{two.5}), we
see that to leading order in $q^2$,
\be
\displaystyle{{\left({d\sigma \over d\Omega}\right)_{ep}
\over\left({d\sigma \over d\Omega} \right)_{\rm ep,point}}
\approx G^2 (q^2) \simeq{1 \over \left(1 + {q^2/ M^2_V}\right)^4}}\;\;
{\huge ^{.}}
\label{two.19}
\ee
This means that the elastic cross-section falls off very sharply with
increasing values of $q^2$ for scattering off an extended object.  This
behaviour then tells us that the probability of elastic scattering falls
off very sharply with increasing $q^2$ ({\it i.e.}, increasing energy of the
incident $e^-$ beam).  Some of the SLAC data on $G^p_M(q^2)$ taken
from \cite{kirk} are shown in fig. \ref{fig5}.

The experimentally observed $q^2$-dependence of the form factors ,
eq. (\ref{two.19}), along with eq. (\ref{one.25}) implies that the r.m.s.
radius of the electric charge/magnetic moment distribution in a
proton or neutron is
\be
\langle R^2\rangle^{1/2} = {\sqrt{12} \over 0.84~{\rm GeV}} \simeq 0.81
{}~{\rm fm} \;\;.
\label{two.20}
\ee
The dipole form factor of eq. (\ref{two.19}) can also be interpreted (see
Table \ref{tabform}) as an exponential charge/magnetic moment
distribution given by
\be
\rho (R) = {M^3_V \over 8\pi} \exp(-M_V R) \;\; .
\label{two.21}
\ee
While it is true that the electric and magnetic form factors provide a
neat way to parametrise the effects of spatial extension of the target on
a scattering process, attempts to calculate these form factors from first
principles, in some model, did not add  to our knowledge of strong
interaction dynamics or that of the structure of proton, beyond the
information provided by eq. (\ref{two.21}).  Real progress in the information
on proton/neutron structure as well as strong interaction dynamics was
made by increasing the energy of the electron beams incident on the
proton/neutron target and studying the scattering process, which  will
be discussed in the next section.

%
%
\subsection{Summary of the dependence of the
cross--sections for elastic scatterring on the nature of the target}
In this section we saw how  the differential cross--section for elastic
scattering  depends on the nature of the scatterer and how this
can be parametrised in terms of one (two) form factors for a
scalar (spin 1/2) target. This information can be summarised in
a compact form as in Table \ref{tabcomp}.

\begin{table}[hbt]
\caption{
Dependence of the elastic differential cross--section on the
nature of the target and projectile (Y: effect included, N : Not included).}
\begin{center}
\begin{tabular}{|c|c|c|c|c|c|c|}
\hline
&\multicolumn{2}{c|}{\rm Projectile e} &\multicolumn{4}{c|}
{{\rm Target} $Z\vert e \vert$} \\
\cline {2-3} \cline{3-7}
{\rm Formula}& {\rm Spin} &{\rm Energy} &{\rm Spin}&{\rm Anom.} $\mu$
&{\rm Size} &{\rm Recoil} \\
& & &  & $\kappa$& &(Target mass)\\
\hline
Rutherford         &    &Non Rel.&   &   &  & \\
eq. (\ref{ruther})& N  & (NR)   & N & N & N & N \\
\hline
Mott   &  & Rel.                            &   & &  & \\
eq. (\ref{one.27})& Y & $ E/m \rightarrow \infty$ & N & N & N &N \\
\hline
eq. (\ref{two.5p})&   &   &   &   &   &  \\
{\rm spinless}  & Y & ''& N & N & N & Y  \\
{\rm point p}   &   &    &   &   &   &   \\
\hline
eq. (\ref{two.5}) &   &    &   &   &   &  \\
{\rm point}  & Y & '' & Y & N & N & Y \\
{\rm spin 1/2 p}&&&&&&\\
\hline
Rosenbluth   &   &   &   &   &   & \\
eq. (\ref{two.14}) & Y & ''& Y & Y & Y & Y \\
\hline
\end{tabular}
\end{center}
\label{tabcomp}
\end{table}
As we can see, in the limit of the point proton, the form factors reduce
to 1 and the Rosenbluth formula reduces to eq. (\ref{two.5}) as mentioned
in the line 4 of the table. This does not incorporate the anomalous
magnetic moment either.  The term proportional to $\tan ^2 \theta/2 $
in both these equations is due to the magnetic scattering and is
present when the spin 1/2 nature of {\it both} the target and the
projectile  is taken into account.  As we have seen in our discussions
at lower energies, the electrostatic scattering dominates and hence the
second term can be dropped.  If we drop this term from eq. (\ref{two.5})
we recover the expression for the  cross--section for the
scattering of a spin 1/2 electron from a spinless, pointlike particle.
For a spinles non--pointlike particle  this would be multiplied by an
appropriate form factor which at low energies of the projectile will
reduce to one.  If the target has no spin and has infinite mass (as
compared with the projectile energy) then the target recoil in the
scattering process can be neglected and $E \simeq E_0$ for the
projectile. In this limit this formula reduces to the Mott scattering
cross--section of eq. (\ref{one.27}). At lower projectile energies, the
effects of the spin 1/2 nature of the projectile also become negligible
and the relativistic Mott scattering cross--section reduces to the
Rutherford scattering cross--section. Again if the target is not a
pointlike charge but a charge  distribution instead, then this formula
gets multiplied by a form factor  as we have seen before. Thus we see
that as we go to lower and lower projectile energies, various factors
which represent effects of the projectile spin (e.g. the factor $\cos
^2 (\theta/2) $ in the Mott scattering cross--section of the second
line in the table) or the target spin, size and mass ( the term with $
\tan ^2 \theta/2$ factor, the form factor or the recoil factor
respectively) go either to one or zero, giving us in the end the
simple Rutherford scattering cross--section in the complete
non--relativistic (NR) limit.

\setcounter{equation}{0}
\section{Deep Inelastic Electron-Nucleon Scattering}
As we saw in the last section, study of elastic scattering of an electron
off a proton target gives information about the spatial charge/magnetic
moment distribution for a proton.  However, eq. (\ref{two.19}) and fig.
\ref{fig5} tell us that the elastic cross-section falls through four
orders of magnitude as $q^2$ changes from $2 \rightarrow 25\; {\rm
GeV}^2$.  These larger values of $q^2$ are reached using higher energy
electron beams.  With increasing $q^2$, quasi--elastic scattering with
excitation of baryon resonances becomes possible (recall the case of
nuclei in nuclear reactions), and at still higher energies (and hence at
higher $q^2$ values), the scattering is dominated by the inelastic process:
the so called Deep Inelastic Scattering (DIS).
%
%
\subsection{Kinematics of inelastic scattering}
The kinematics of the inelastic $e^-$ -- $p$ scattering is shown in fig.
\ref{fig6}, which is almost the same as fig. \ref{fig4}, except that the
hadronic final state is no longer a proton.  The reaction now is
\be
e^- + p \rightarrow e^- + X \;\; .
\label{three.1}
\ee
$X$ stands for the hadronic final state.  The four momenta of the various
particles involved in scattering are as shown in fig. \ref{fig6}.  The
invariant mass of the hadronic final state is now,
\be
M^2_X = -P^2_4 = -|\vec p~^\prime|^2 + W^2 \;\; .
\label{three.2}
\ee
Energy momentum conservation of eq. (\ref{one.6}) gives us
\be
q^2 = (P_2 - P_3)^2 = (P_4 - P_1)^2 = |\vec p~^\prime|^2 - (W - M_p)^2
= -M^2_X + M^2_p + 2M_p \nu \;\; ,
\label{three.3}
\ee
where $\nu = E_0 - E$ is the energy transfer from the electron to proton
in the laboratory frame.  Using eq. (\ref{one.8}) for $q^2$ we then get,
\be
q^2 = 4EE_0 \sin^2 {\theta \over 2} = -M^2_X + M^2_p + 2M_p (E_0 - E)
\;\; .
\label{three.4}
\ee
As we can see, if $M^2_X = M^2_p$, {\it i.e.}, the scattering is
elastic, then eq.  (\ref{three.4}) reduces to
\be
q^2 = 2M_p \nu = 2M_p (E_0 - E) \;\; .
\label{three.5}
\ee
Using the expression for $q^2$ given in eq. (\ref{three.4}), we can see that
eq. (\ref{three.5}) leads to eq.~(\ref{one.18p}) with $M_A \rightarrow M_p$.
Eq. (\ref{three.5}) means that $q^2$ and $\nu$ are not independent variables
and the reaction is characterised by only one independent variable.

For quasi--inelastic scattering, {\it i.e.},  excitation of a baryon
resonance (say $N^{\star +}$ in the reaction $e^-p \rightarrow
e^-N^{\star +} \rightarrow e^- p \pi^0$), $M_X = M^\star$.
Eq. (\ref{three.3}) then becomes,
\be
q^2 = 4EE_0 \sin^2 {\theta \over 2} = -M^{\star 2} + M^2_p + 2M_p \nu
\;\; .
\label{three.6}
\ee
Again there is only one independent variable except that the relation
between $q^2$ and $\nu$ is different from that for elastic
scattering (cf. eq. (\ref{three.3})).

If $M^2_X$ does not have a fixed value but changes continuously, then both
$q^2$ and $\nu$ are independent variables.  Clearly the proton no longer
remains intact.  This region is called the continuum region.  This
completely inelastic scattering is characterised by two independent
variables.  From eq. (\ref{one.18p}) and the definition of $\nu$ it is
clear that both of these are completely specified once the energy of the
scattered electron $E$ and its angle $\theta$ are measured; {\it i.e.},
{\it the kinematics of the event is independent of the precise details of the
hadronic final state $X$}. It then  makes sense to think of a measurement
 of the cross-section where one sums over all possible hadronic
final states $X$. Such a measurement is called an inclusive measurement.

The kinematically allowed region in the $q^2$ -- $\nu$ plane for the elastic,
quasi-elastic and inclusive inelastic scattering is shown in fig.
\ref{fig7}.  For elastic scattering the allowed region is the straight line
given by eq. (\ref{three.5}).  For quasi-elastic scattering, again the
allowed region is a straight line but now with an intercept
$(M^{\star 2} - M^2_p) / {2M_p}$, as given by eq. (\ref{three.6}).
The discussions of last section, particularly eqs. (\ref{two.16})
and (\ref{two.19}), make it quite clear that the probability of
elastic scattering goes down with increasing $q^2$ as more
inelastic channels open up.  When both $q^2$ and $\nu$ are large
then the cross-section is dominated by continuum excitation.
The allowed region in the $q^2$ -- $\nu$ plane in this case is the
entire region to the right of the straight line given by eq.
(\ref{three.5}).

The kinematics can be described in terms of any of the pairs of variables:
$(E,\theta)$ or $(q^2,\nu)$.  Equivalently, one can also define two
dimensionless variables
\begin{subequations}
\label{three.7}
\begin{eqalignno}
x &= {q^2 \over 2M_p \nu} = {-q^2 \over 2P_1 \cdot q}\;\; ; \label{three.7a} \\
y &= {\nu \over E_0} = {E_0 - E \over E_0}\;\; . \label{thrre.7b}
\end{eqalignno}
\end{subequations}
For elastic scattering $q^2$ and $\nu$ are related to each other via
eq. (\ref{three.5}). Hence  the variables $x$ and $y$ become,
\bea
x &=& 1 \;\; ; \nonumber \\[2mm]
y &=& {2E_0/M_p \; \sin^2 {\theta \over 2} \over 1 + {2E_0 \over M_p}
\sin^2 {\theta \over 2}}\;\; .
\label{three.8}
\eea
For inclusive, inelastic scattering both $q^2$ and $\nu$ (or equivalently
$x$ and $y$) can vary independently.  We notice from eqs. (\ref{three.4}),
(\ref{three.7}) that
\bea
q^2 = 2M_p \nu x = 2M_p E_0 xy &=& -M^2_X + M^2_p + 2M_p \nu \nonumber \\[2mm]
&=& 4EE_0 \sin^2 \theta/2 \;\; .
\label{three.8p}
\eea
Since $M_X > M_p$, the equations above trivially yield the kinematically
allowed region for variables $(x,y)$ as
\be
0 < x < 1, ~~~ 0 < y < {2E_0 \over 2E_0 + xM_p}\;\; .
\label{three.8pp}
\ee
Using eq. (\ref{three.8p}) we can also derive the allowed region in the
$(q^2,\nu)$ plane and it is given by
\be
0 < q^2 < {4M_p E^2_0 \over 2E_0 + M_p} \approx 2M_p E_0~~, ~~~{q^2 \over
2M_p} < \nu < E_0 - {q^2 \over 4E_0} \;\; .
\label{three.8ppp}
\ee
The early DIS experiments at SLAC \cite{DIS}  used electron beams
with energy $E_0 = 20$ GeV which corresponds to $q^2 \leq 40$
GeV$^2$.  The current experiments at Fermilab use $\nu$ beams with
energy 500 GeV and hence can reach $q^2$ values upto 1000 GeV$^2$,
whereas the $e^-$ -- $p$ collider HERA at DESY operating with an electron
beam of 30 GeV and proton beam of 800 GeV is capable of measuring DIS
cross-sections upto $q^2 \leq 10^5 \;$ GeV$^2$.
\subsection{Inelastic cross-section and structure functions}
We saw in the first lecture that for elastic scattering of eq.
(\ref{two.1}), the most general expression for the current
$J^{p,\ext}_\mu$ and hence that for the cross-section could be written in
terms of two arbitrary functions ${\cal F}^p_1 (q^2)$ and ${\cal F}^p_2
(q^2)$.  In case of the inclusive, inelastic measurement described above,
again the cross-section can be parametrised in terms of two arbitrary
functions $W_1 (q^2,\nu)$ and $W_2 (q^2,\nu)$ which have to be determined
experimentally.  First let us see how the number of these arbitrary
functions, which parametrise the effect of the structure of a proton on
the DIS process, can be restricted to two using gauge invariance, Lorentz
invariance, parity invariance and time reversal invariance.

Let us start by recalling that in case of elastic scattering we started
off with ${\cal M} \sim  (J^e_\mu J^p_\mu) / q^2 $,
where $J^e_\mu$ was given by eq. (\ref{two.8}) and $J^p_\mu$ for a
point Dirac proton was given by eq. (\ref{two.8}) whereas it was given
by eq. (\ref{two.11}) for a proton with structure.  In the present
case, we need to sum over all possible hadronic states.  Since now
there is no spinor available to describe the hadronic state, unlike the
earlier case of elastic scattering, it is not possible to repeat the
steps which led us to eq. (\ref{two.11}).  Recall, however, the
calculation leading to eq. (\ref{two.5}), which gave the cross-section
for an electron scattering off a point Dirac proton.  Then we had,
\be
\label{three.9}
|{\cal M}^2|  \sim e^4 \; L_{\mu\nu} H'_{\mu\nu}
      \quad {\rm where} \quad
L_{\mu\nu} = {1 \over 2e^2} \sum_{S,S'} J^e_\mu J^{e^\dagger}_\mu,
{}~~~H'_{\mu\nu} = {1 \over 2e^2} \sum_{S_p,S'_p} J^p_\mu J^{\dagger^p}_\mu.
\ee
$S,S'(S_p,S'_p)$ denote the spins of the initial and final state electron
(proton).  In the case of a non-point-like proton all we had to do was to
replace $J^p_\mu$ of eq. (\ref{two.8}) by $J^{p,\ext}_\mu$ of eq.
(\ref{two.11}).  Comparing figs. \ref{fig4} and \ref{fig6}, we see that the
lepton end in both the elastic and inelastic case is the same.  Hence we
can once again write $|{\cal M}|^2$ in a form analogous to eq.
(\ref{three.9}). However, in this case we know nothing about the tensor
involving hadronic variables.

To get more insight into the structure of the hadronic tensor for the case
of inclusive scattering, it might be instructive to look at a few steps
that led us to eq. (\ref{two.5}).  The differential cross-section for the
elastic process was written as:
\bea
d\sigma^{e\ell} &=& {1 \over 2M_p} \;{1 \over  2E_0} {d^3p \over (2\pi)^3} \;
{1 \over 2E} {d^3p' \over (2\pi)^3} {1 \over 2W} \; {e^4 \over q^4} \times
\nonumber \\[2mm]
& & \left[{1\over2} \sum_{S,S'} \bar u_e (\vec p,S') \gamma_\mu u_e (\vec
p_0,S) \bar u_e(\vec p_0,S) \gamma_\nu u_e(\vec p,S')\right] \nonumber \times
\\[2mm]
& & \left[{1\over2} \sum_{S_p,S'_p} \bar u_p (\vec p_1,S_p) \gamma_\mu
u_p(\vec p~^\prime,S'_p) \bar u_p(\vec p~^\prime, S'_p) \gamma_\nu
u_p(\vec p_1,S_p) \right] \times \nonumber \\
& & (2\pi)^4 \delta^4(P_1 + q - P_4) \;\; .
\label{three.10}
\eea
The first factor in the square bracket is $L_{\mu\nu}$ and the second is
$H'_{\mu\nu}$ of eq. (\ref{three.9}).  The above can be rewritten as,
\bea
d\sigma^{e\ell} &=& {1 \over 2M_p} {1 \over 2E_0} {d^3p \over (2\pi)^3
\; 2E} \; {d^3p' \over (2\pi)^3  2W} {e^4 \over q^4} \;
L_{\mu\nu} \times \nonumber \\[2mm]
& & {1\over2} \sum_{S_p,S'_p}  \langle P_1,S_p |\tilde J^\dagger_\mu|
P_4,S'_p\rangle \langle P_4,S'_p |\tilde J_\nu| P_1,S_p \rangle \times
\nonumber \\[2mm] & &  (2\pi)^4 \delta^4 (P_1 + q - P_4) \nonumber \\
{\rm with}&& \nonumber \\
&\langle & P_4,S'_p |\tilde J_\mu| P_1,S_p\rangle = \bar u_p (\vec p',S'_p)
\gamma_\mu u_p (\vec p_1,S_p) \;\; .
\label{three.11}
\eea
Here $\tilde J_\mu$ represents the electromagnetic current operator from
which certain constants have been removed.  The above equation can be used
to write the expression for the double differential cross-section for
elastic scattering as,
\be
{d^2 \sigma^{e\ell} \over d\Omega dE} = {\alpha^2 \over q^4} \; {E
\over E_0} \; L_{\mu\nu} H^{\mu\nu} \;\; ,
\ee
where
\bea
H_{\mu\nu} &=& {1 \over 4\pi M_p} {1 \over 2} \sum_{S_p,S'_p} \int {d^3 p'
\over (2\pi)^3} \; (2\pi)^4\; \delta^4 (P_1 + q - P_4) \;
\nonumber \\[2mm]  & & \langle P_1,S_p |\tilde
J^\dagger_\mu|P_4,S'_p\rangle \langle S'_p P_4 |\tilde J_\nu|P_1,S_p \rangle
\;\; .
\label{three.12}
\eea
For the case of inclusive scattering we can generalise the expression for
$H_{\mu \nu}$ given by eq. (\ref{three.12}) as:
\bea
H_{\mu\nu} &=& {1\over2} \sum_{S_p} \sum_X \int {1 \over 4\pi} {1 \over
M_p} \prod^n_{i=1} {d^3 \ell_i \over 2E_i (2\pi)^3} \; (2\pi)^4
\delta^4 \left(P_1 + q - \sum_{i=1}^n \ell_i\right) \nonumber \\[2mm]
& & \langle P_1,S_p |\tilde J^\dagger_\mu| X \rangle
\langle X|\tilde J_\nu| P_1,S_p \rangle \;\; ,
\label{three.13}
\eea
where $|X\rangle$ denotes a state containing $n$ particles with four
momenta $\ell_1,\ell_2,\cdots,\ell_n$ and we sum over all such states.
The differential cross-section $d\sigma^{ine\ell}$ for this case of
inclusive scattering is then given by (using eqs. (\ref{three.10}) to
(\ref{three.12}))
\be
d\sigma^{ine\ell} = {1 \over 2M_p} \; {1 \over 2E_0} \; {d^3p \over
(2\pi)^3} {1 \over 2E} \; {e^4 \over q^4} \; 4\pi M_p \;
L_{\mu\nu} H_{\mu\nu}  \;\; .
\label{three.14}
\ee
All we know about $H_{\mu\nu}$ given by eq. (\ref{three.13}) are its
symmetry properties.  Gauge invariance requires $\partial_\mu \tilde J_\mu
= 0$.  In momentum space, this translates into
\be
q_\mu H_{\mu\nu} = H_{\mu\nu} q_\nu = 0 \;\; .
\label{three.15}
\ee
We begin by writing $H_{\mu\nu}$ as the most general tensor
consistent with gauge invariance and parity invariance that can be
constructed out of the linearly independent four-momenta available
at the hadronic end; $q_\mu$ and $P_{1\mu}$.  Also note that
$L_{\mu\nu}$ is symmetric in $\mu \leftrightarrow \nu$.  Hence only  the
symmetric part of $H_{\mu\nu}$ will be relevant.  The most general form
for $H_{\mu\nu}$ can therefore be
written as
\bea
H_{\mu\nu} &=& W_1 (q^2,\nu)\delta_{\mu\nu} + W_2 (q^2,\nu) {P_{1\mu}
P_{1\nu} \over M^2_p} \nonumber \\[2mm]
& & + W_4 (q^2,\nu) {q_\mu q_\nu \over M^2_p} + W_5 (q^2,\nu)
\left({P_{1\mu} q_\nu + q_\mu P_{1\nu} \over M^2_p}\right) \;\; ,
\label{three.16}
\eea
where $W_i(q^2,\nu)$ $(i=1,2,4,5)$ are arbitrary functions of $q^2$ and
$\nu$.  The requirement of gauge invariance of eq. (\ref{three.15}) gives
\bea
q_{\nu} \biggl [ W_1 (q^2,\nu) &+& {q^2 \over M^2_p} W_4 (q^2,\nu)
+ {P_1 \cdot q\over M^2_p} W_5 (q^2,\nu) \biggr ]  \nonumber\\
&+& P_{1\nu} \biggl [W_5 (q^2,\nu) {q^2 \over M^2_p} + {P_1 \cdot q \over
M^2_p} W_2 (q^2,\nu) \biggr ] = 0 \nonumber
\eea
Since $q_\nu$ and $P_{1\nu}$ are two linearly independent fourvectors and
the terms in the square brackets multiplying them are Lorentz scalars, it
follows that each of the brackets must be identically zero.  Hence we get,
\bea
& &W_5 (q^2,\nu)\; =  - {P_1 \cdot q \over q^2} W_2 (q^2,\nu) \;\; , \nonumber
\\[2mm]
& & W_1 (q^2,\nu) + {q^2 \over M^2_p} W_4 (q^2,\nu) + {P_1 \cdot q
\over M^2_p} W_5 (q^2,\nu) = 0 \;\; .
\label{three.17}
\eea
Using this, the expression for $H_{\mu\nu}$ becomes,
\be
H_{\mu\nu} = W_1 (q^2,\nu) \left[\delta_{\mu\nu} - {q_\mu q_\nu \over
q^2}\right]  + {W_2 (q^2,\nu) \over M^2_p}
\left[P_{1\mu} - {P_1 \cdot q \over q^2} q_\mu\right] \left[P_{1\nu} -
{P_1 \cdot q \over q^2} q_\nu\right] \;\; .
\label{three.18}
\ee
Thus we see that in effect the most general expression for the hadronic
tensor involves only two arbitrary functions.  $L_{\mu\nu}$ of eq.
(\ref{three.14}) is the same as that for the elastic scattering and is
given by
\be
L_{\mu\nu} = - {1 \over 2} Tr\left[P\!\!\!/ \gamma_\mu P\!\!\!/_0
\gamma_\nu\right] =\; -2\left[P \cdot P_0 \; \delta_{\mu\nu}
- P_\mu P_{0\nu} - P_\nu P_{0\nu}\right] \;\; .
\label{three.19}
\ee
To calculate $d\sigma^{\rm inel}$ of eq. (\ref{three.14}) we need to know
$L_{\mu \nu} H_{\mu\nu}$.  Using eqs. (\ref{three.18}) and
(\ref{three.19}), we get
\be
L_{\mu\nu}H_{\mu\nu} =\; -4P \cdot P_0 \; W_1 (q^2,\nu) + {W_2 (q^2,\nu) \over
M^2_p}\; (2M^2_p P \cdot P_0 + 4P \cdot P_1 P_1 \cdot P_0) \;\; .
\label{three.20}
\ee
With our choice of normalisation for $W_i (q^2,\nu)$, (cf. eqs.
(\ref{three.14}) and (\ref{three.16}),
we have
\be
{d^2 \sigma^{\rm inel,em} \over d\Omega dE} = {\alpha^2 \over q^4} \;
{E \over E_0} \; L_{\mu\nu} H_{\mu\nu} \;\; .
\label{three.21}
\ee
The superscript `em' denotes here the nature of interaction involved in
the scattering process, viz. electromagnetic interaction.  In the
laboratory $P_1 \cdot P_0 = -M_p E_0$, $P \cdot P_1 = -M_p (E_0 - \nu)$,
and $P \cdot P_0 = -q^2/2 = -2EE_0 \sin^2 \theta/2$.  Using these
relations and eq. (\ref{three.20}), we get
\be
{d^2 \sigma^{\rm inel,em} \over d\Omega dE} = {4\alpha^2 \over q^4} E^2
\cos^2 {\theta \over 2} \left[W_2 (q^2,\nu) + 2W_1 (q^2,\nu) \tan^2
{\theta \over 2}\right] \;\; .
\label{three.22}
\ee
Thus the cross-section for the inelastic scattering process $e^-p
\rightarrow e^- X$ is parametrised in terms of two functions $W_1
(q^2,\nu)$ and $W_2 (q^2,\nu)$.  Since these functions contain all the
information about the proton structure as revealed to an electromagnetic
probe, these are called the structure functions.  The differential
cross-section of eq. (\ref{three.22}) can be equivalently written in terms of
the pair of variables $(q^2,\nu)$ or the pair of dimensionless variables
$(x,y)$ introduced earlier.  We can show that
\bea
{d^2\sigma \over dx dy} = 2ME_0 x {d^2\sigma \over dx dq^2} = 2ME^2_0 y
{d^2\sigma \over dq^2 d\nu} &=& 2M_p \pi {E_0 \over E} \; y {d^2\sigma
\over d\Omega dE} \nonumber \\[2mm]
&=& M_p {E_0 \over E} \; y {d^2\sigma \over d\cos \theta dE}  \;\; \; .
\label{three.23}
\eea

Before we go on to discuss the subject of experimental measurements of
functions $W_1 (q^2,\nu)$, $W_2(q^2,\nu)$ and the information they yield
about the proton structure, it is instructive to see to what the
general functions $W_1,W_2$ reduce to for the special cases of elastic
scattering of an electron off a pointlike ``Dirac'' proton (eq.
(\ref{two.5})) and a proton with structure (eq. (\ref{two.16})).  The
cross-section of eq. (\ref{two.5}) can be rewritten as a double differential
cross-section by using the identity,
\be
{d\sigma^{e\ell} \over d\Omega} = \int dE \; \delta \left(E - {E_0
\over 1 + {2E_0 \over M_p} \sin^2 {\theta \over 2}}\right)
{d\sigma^{e\ell} \over d\Omega}\;\; .
\ee
The above equation follows from realising that for elastic scattering the
energy of the scattered electron is fixed via eq. (\ref{one.18p}),  once the
angle is fixed.  Using the definitions of $\nu$ and $q^2$, as well as
properties of $\delta$-function, we can rewrite the above expression for
$ d\sigma^{e\ell} / d\Omega$ as
\be
{d\sigma^{e\ell} \over d\Omega} = \int dE~\delta\left(-\nu + {q^2 \over
2M_p}\right) \;  \left(1 + {2E_0 \over M_p}
\sin^2 {\theta \over 2}\right)\;{d\sigma^{e\ell} \over d\Omega} \;\; .
\ee
{}From this equation it is obvious that the double differential
cross-section for the elastic case is
\be
{d^2\sigma^{e\ell} \over d\Omega dE} = \;
\delta\left(-\nu + {q^2 \over 2M_p}\right) \; \left(1 + {2E_0 \over
M_p} \sin^2 {\theta \over 2}\right)\;{d\sigma^{e\ell} \over d\Omega}
\;\; .
\label{three.24}
\ee
Formally, eq. (\ref{three.24}) can also be derived by using,
\be
\int {d^3 p' \over 2W} {\delta^4 (P_1 + q - P_4) \over (2\pi)^3} = {1
\over 2M_p} \delta\left(\nu - {q^2 \over 2M_p}\right) \;\; .
\ee
This is merely a restatement of the relation between $E,E_0$ and $\sin^2
\theta/2$ given by eq. (\ref{one.18p}).

Using eqs. (\ref{three.24}) and (\ref{two.5}) we can therefore write,
\be
{d^2\sigma^{ep \rightarrow ep} \over d\Omega dE}\Bigg|_{
{\rm `` Dirac} {\scriptstyle ''}} =
\delta\left(\nu - {q^2 \over 2M_p}\right)  \left[1 + {q^2 \over 2M^2_p}
\tan^2 {\theta \over 2}\right]\;\left({d\sigma \over
d\Omega}\right)_{\rm Mott} .
\ee
Using the expression for $\left(d\sigma / d\Omega\right)_{\rm
Mott}$ we therefore get,
\be
{d^2\sigma^{ep \rightarrow ep} \over d\Omega dE}\Bigg|_{{\rm
``Dirac}{\scriptstyle ''}} =\;
\delta\left(\nu - {q^2 \over 2M_p}\right)
{4\alpha^2 E^2 \over q^4} \cos^2 {\theta \over 2} \left[1 + {q^2 \over
2M^2_p} \tan^2 {\theta \over 2}\right]   .
\label{three.25}
\ee
Similarly, we can see from eq. (\ref{two.5p}) that for the imaginary case
of a spinless, pointlike proton  we will get,
\be
{d^2\sigma^{ep \rightarrow ep} \over d\Omega dE}\Bigg|_{\rm
spinless~point~proton} =\; \delta\left(\nu - {q^2 \over 2M_p}\right)
{4\alpha^2 E^2 \over q^4} \; \cos^2 {\theta \over 2} \;\; .
\label{three.26}
\ee
Using eq. (\ref{two.16}) we get similarly,
\be
{d^2 \sigma \over d\Omega dE}\Big|_{\rm ext.} = \;
\delta\left(\nu - {q^2 \over 2M_p}\right)\; {4\alpha^2 \over q^4}\;
E^2 \cos^2 {\theta\over2} \left[A^p(q^2) + B^p(q^2) \tan^2 {\theta \over 2}
\right] \;\; ,
\label{three.27p}
\ee
where $A^p(q^2)$ and $B^p(q^2)$ are given by eq. (\ref{three.27}).
Comparing eqs. (\ref{three.25}) -- (\ref{three.27p}) with eq.
(\ref{three.22}) we get,
\bea
W^{{\rm point},p}_2 (q^2,\nu) &=& \delta\left({q^2 \over 2M_p} -
\nu\right) = {1\over\nu} \delta(1 - x) \;\; ; \nonumber \\[2mm]
W^{{\rm point},p}_1 (q^2,\nu) &=& {q^2 \over 4M^2_p} \delta\left({q^2
\over 2M_p} - \nu\right) = {1 \over 2M_p} \; {q^2 \over 2M_p \nu}
\delta (1 - x) \;\; ,
\label{three.28}
\eea
while
\bea
W^{{\rm scalar},p}_2 &=& {1 \over \nu} \; \delta(1 - x)
\;\; ; \nonumber \\[2mm]
W^{{\rm scalar},p}_1 &=& 0 \;\; ,
\label{three.29}
\eea
and
\bea
W^{e\ell}_2 (q^2,\nu) &=& {1 \over \nu} A^p (q^2,\nu) \delta(1 - x)
\;\; ; \nonumber \\[2mm]
W^{e\ell}_1 (q^2,\nu) &=& \frac{B^p (q^2,\nu)} {2 \nu} \delta(1-x) =
\frac {1} {2M_p} \; \frac{q^2}{2 M_p \nu} {G^p_M}^2 (q^2)\;
\delta(1-x) \;\; .
\label{three.30}
\eea
The functions $G^p_E (q^2)$ and $G^p_M (q^2)$ have the very
steep $q^2$ dependence given by eq. (\ref{two.19}).  This means that the
structure functions $W^{e\ell}_{1,2} (q^2,\nu)$ fall off very steeply with
increasing $q^2$.  More interesting is the observation that the structure
functions $W^{\rm point}_{1,2}$ for a point scatterer do not depend upon
the variables $q^2$ and $\nu$ separately, but are functions only of the
combination $x = q^2/2M_p \nu$.  This is of course obvious.  In the case
of a point scatterer, the structure functions should depend only on
dimensionless variables as there is no intrinsic length scale associated
with the scatterer.  Thus if the scatterer has a finite size then the
structure functions for the elastic scattering $W^{e\ell}_1,~W^{e\ell}_2$ ,
fall off as a power of $q^2$ whereas for a pointlike scatterer they
depend on $q^2$ only through the combination $q^2/2M_p\nu$. Table
\ref{structure}  summarises the behaviour of the structure function
for different types of scatterers.
\begin{table}
\caption{
The structure functions $ M_p W_1(x,Q^2) $ and $\nu W_2(x,Q^2)$
for elastic scattering from different type of scatters.}
\medskip
\begin{tabular}{|c|c|c|}
\hline
& &  \\
{\rm Scatterer} & $ M_p W_1(x,Q^2)$ & $\nu W_2(x,Q^2)$ \\
& & \\
\hline
& & \\
{\rm pointlike spin 1/2 proton} & $ \frac{q^2}{4M_p \nu}\; \delta(1-x)$ &
$\delta(1-x)$ \\
& & \\
\hline
& & \\
{\rm pointlike scalar proton}& 0 &$\delta (1-x)$\\
& & \\
\hline
& & \\
spin 1/2 proton with structure & $\frac{q^2}{4M_p \nu}
(G_M^p)^2\; \delta (1-x)$&
$ \frac{\bigg( (G_E^p)^2 + \frac{q^2}{4 M_p^2}
(G_M^p)^2 \bigg)}{\bigg(1 + \frac{q^2}{4 M_p^2} \bigg)} \; \delta (1-x)$  \\
& & \\
\hline
\end{tabular}
\label{structure}
\end{table}

%
\subsection{Scaling of structure functions and partons}
The discussions of the earlier section tell us that the $q^2$ and $\nu$
dependence of the structure functions $W_{1,2} (q^2,\nu)$ depends on the
nature of scatterer.  Hence it is worthwhile asking, what do the
experimental measurements of $W_{2,1}^{{\rm inel},p} (q^2,\nu)$ look like?
With increasing $q^2$, the inelastic scattering begins to dominate the
elastic process for $q^2 \gsim 2$ -- 3 GeV$^2$ (recall eq. (\ref{two.19})).
For quasi-inelastic scattering, the resonance excitation corresponds to
$M_X = M^\star$ in eq. (\ref{three.4}).  Hence $\nu$ is fixed once $q^2$
is.  This means that ${d^2\sigma / dq^2 d\nu}$ will be a
$\delta$-function in $\nu$ for a fixed $q^2$.  Experimental measurements
of ${d^2 \sigma / dq^2 d\nu}$ do indeed show peaks in $\nu$ at a fixed
$q^2$.  Fig. \ref{fig8} taken from \cite{smith} shows this.  With
increasing $q^2$, the proton-resonance transition form factors also
show a power law fall off with $q^2$, just like $G_M^p (q^2)$.  At
still higher values of $q^2$, continuum production takes over.  The
interesting observation of {\it Bjorken scaling} was the fact that
for this inelastic scattering the structure functions $W_{1,2}^{\rm
inel} (q^2,\nu)$ do not fall off with increasing $q^2$ but they are
found to become independent of $q^2$ (for a fixed value of $q^2/2M_p
\nu$) as both $q^2$ \underbar{and} $\nu$  become large, {\it
i.e.},  they are functions of $x = q^2/2M_p\nu$ alone.  This is
illustrated for some data on $\nu W_2^{\rm inel} (q^2,\nu)$ in fig.
\ref{fig9} taken from the first of Ref. \cite{DIS}.  Thus the
experimental observation is,
\bea
\nu W_2^{\rm inel} (q^2,\nu) ;
& \stackrel{\hbox{scaling}}
{\!\a} & F_2^{ep}(x) \; \; ;\nonumber \\[-0.2cm]
& {\scriptstyle\nu\to\infty,\,q^2\to\infty} & \nonumber \\
M_p W_1^{\rm inel} (q^2,\nu)
& \stackrel{\hbox{scaling}}
{\!\a} & F^{ep}_1 (x) \;\; . \nonumber \\[-0.2cm]
& {\scriptstyle\nu\to\infty,\,q^2\to\infty} &
\label{three.31}
\eea
This phenomenon of `scaling' of structure functions was interpreted by
Bjorken \cite{bjpaschos} as an indication of the existence of pointlike
scatterers inside the proton.  This interpretation can be understood by
recalling eq. (\ref{three.28}).  Thus the observed scaling of the DIS
cross-sections indicates that  inelastic electron-proton scattering can
be understood in terms of {\it incoherent}, elastic scattering of
electron off the individual, pointlike constituents of the proton termed
`partons' \cite{feynman}.  The charged partons which take
part in the electromagnetic scattering are termed quarks.  Whether these
quarks are to be identified with the quarks whose existence is inferred
from spectroscopic studies is best discussed later on in the context
of the parton model.  Since there is no scale associated with these pointlike
objects,  eq. (\ref{three.28}) indicates that the individual
elastic cross-sections must scale.  We will show later, in a detailed
discussion of the parton model, that the variable $x$ can be interpreted
as the fraction of proton momentum that the parton carries.  Hence the
intuitive picture is as shown in  fig. \ref{fig10}.

In the scaling limit of eq. (\ref{three.31}), using eq. (\ref{three.23}),
eq. (\ref{three.22}) becomes
\[
{d^2 \sigma^{{\rm inel},em} \over dx~dy} = {4\pi \alpha^2  \over Sx^2y^2}
\left[ xy^2 F_1^{ep} (x) + \left(1 - y - {M_p \over 2E_0} xy\right)
F_2^{ep} (x)\right] \;\; .
\]
$S$ is the square of the total c.m. energy of the scattering process
$\simeq 2 M_p E_0$.  The scaling limit corresponds to high energies
of the incident electron, hence particle masses can be neglected and
we get,
\be
{d^2 \sigma^{{\rm inel},em} \over dx~dy} = {4\pi \alpha^2 \over Sx^2y^2}
\left[xy^2 F_1^{ep} (x) + \left(1 - y\right) F_2^{ep} (x)\right] \;\; .
\label{three.32}
\ee
Here $F_1^{ep} (x)$ and $F_2^{ep} (x)$ are the electromagnetic structure
functions of the proton.

It is worth noting here that  according to eq. (\ref{three.29}),
$W_1^{\rm scalar} = 0$. Hence if the pointlike constituents inside the proton
are scalars, the term proportional to $\tan^2 (\theta / 2) $ in
eq. (\ref{three.22}) (or equivalently the term proportional to $xy^2$ in
eq. (\ref{three.32})) will be absent.  Note also that all the discussions
will be completely unchanged if one were to use $\mu^-$ beams instead of
$e^-$ beams.  The structure functions are characterised by the target and
the type of interactions used as a probe, in the present case a proton.
So we have
$$
\begin{array}{l}
F_1^{ep} (x) = F_1^{\mu p} (x) \;\; ; \\[2mm]
F_2^{ep} (x) = F_2^{\mu p} (x) \;\; .
\end{array}
$$
%

\subsection{Neutrino Deep Inelastic Scattering}
So far, in our discussions the scattering process that we considered was
$e^{-}$--$p$ scattering.  This probes the structure of proton via
electromagnetic interactions and, as shown before, is characterised by two
independent, arbitrary functions $W_i (q^2,\nu)~(i=1,2)$ which are to be
determined experimentally.  One can also probe the structure of the proton
via weak interactions in the charged current reaction
\be
\nu_\ell   + p \rightarrow \ell^- + X
\label{three.32'}
\ee
and the neutral current reaction,
\be
\nu_\ell + p \rightarrow \nu_\ell + X \;\; .
\label{three.33}
\ee
Here we have started the discussion directly with the inclusive,
inelastic process.  The analogues of the form factors $G_M^p$, $G^p_E$
and $G^n_M$ for the elastic processes involving neutrinos exist.
Symmetries of strong interactions relate the electromagnetic
form--factors as measured in the $e^{-}$--$p$ or $\mu^{-}$--$p$
scattering and the weak form--factors as measured in $\nu_\ell\; p$
scattering.  The charged current weak form--factors have been measured
and indeed played a very important role in confirming these features
of strong interaction symmetries.  Here, however, we concentrate only
on the DIS process shown in fig. \ref{fig11}.

Analogous to the earlier discussions (cf. eqs. (\ref{three.10}) --
(\ref{three.15})) for the charged current reaction of eq.
(\ref{three.32'}), we can write
\be
d\sigma^{{\rm inel},\nu} = {1 \over 4M_p E_0} {d^3p \over (2\pi)^2 2E}
\left({g \over \sqrt{2}}\right)^2 \left({1 \over q^2 - M^2_W}\right)^2
4\pi M_p L'_{\mu\nu} H^{\prime\prime}_{\mu\nu} ,
\label{three.34}
\ee
where $L'_{\mu\nu}$ is the leptonic tensor evaluated using the weak
current $J^{\rm weak}_\mu (\ell,\nu_\ell)$ instead of the electromagnetic
current used in eq. (\ref{three.9}), $H^{\prime\prime}_{\mu\nu}$ is the
hadronic tensor (again analogue of eq. (\ref{three.13}) but replacing the
electromagnetic current by the weak current), $g$ is the weak coupling of
$\ell,\nu_\ell$ to the weak gauge boson $W$ and $M_W$ is the mass of the
$W$ boson.  The matrix element of the weak current is given
by \cite{halzen}
\be
\langle P_3,S_\ell |\hat J_\mu^{\rm weak}|P_2,S_{nu_\ell}\rangle \equiv
J_\mu^{\rm weak} (\ell,\nu_\ell) \equiv  {i g \over \sqrt{2}} \bar
u_\ell ({S_\ell},P_3) \gamma_\mu (1 + \gamma_5)
 u_{\nu_\ell} (S_{\nu_\ell},P_2) \;\; .
\label{three.35}
\ee
Hence
\bea
L'_{\mu\nu} &=& {1\over2} \sum_{S,S'} J_\mu^{\rm weak} J_\nu^{\dagger
{}~{\rm weak}} \nonumber \\[2mm]
&=& {1 \over 2} \; {g^2 \over 2} \; \sum_{S,S'} \left[\bar
u_\ell (S',P_3) \gamma_\mu (1 + \gamma_5) u_{\nu_\ell} (S,P_2)\right] \times
\nonumber \\[2mm]
& & ~~~~~~~~~~~~~~~~~  \left[\bar u_{\nu_\ell} (S,P_2) \gamma_4 (1 +
\gamma_5) \gamma_\nu \gamma_4 u_\ell (S',P_3)\right] \nonumber \\[2mm]
&=& - {1\over2} g^2 ~{\rm Tr}\left[\gamma_\mu (1 + \gamma_5) (-i
P\!\!\!/_2) \gamma_\nu (1 + \gamma_5) (-i P\!\!\!/_3)\right] \nonumber
\\[2mm] &=& {g^2 \over 2} {\rm Tr}\left[P\!\!\!/_3 \gamma_\mu P\!\!\!/_2
\gamma_\nu (1 + \gamma_5)\right] \nonumber \\[2mm]
&=& {g^2 \over 2} \; 4\left[P_{3\mu} P_{2\nu} + P_{3\nu} P_{2\mu} -
\delta_{\mu\nu} P_2 \cdot P_3 - \epsilon_{\mu\nu\alpha'\beta'}
P_{3\alpha'} P_{2\beta'}\right] \;\; . \nonumber
\eea
Here, as before, masses of the leptons are neglected.  We note now
that $L'_{\mu\nu}$ is no longer symmetric under an exchange $\mu
\leftrightarrow \nu$.  Hence the most general expression for
$H^{\prime\prime}_{\mu\nu}$ must involve antisymmetric tensors too.
A tensor decomposition of $H^{\prime\prime}_{\mu\nu}$ now involves
\underbar{six} linearly independent tensors that one can construct out
of the four momenta $P_1$ and $q$, and hence six arbitrary functions of
$q^2$ and $\nu$. This is to be contrasted with the electromagnetic case
where one needed only two arbitrary functions due to the its symmetric
nature and the requirement of gauge invariance. The most general
expression for $H^{\prime\prime}_{\mu\nu}$ in this case can be written as
\bea
H^{\prime\prime}_{\mu\nu} &=& W'_1 (q^2,\nu)\delta_{\mu\nu} + W'_2
(q^2,\nu) {P_{1\mu} P_{1\nu} \over M^2_p} \nonumber \\[2mm]
& & + W'_3 (q^2,\nu) {P_{1\alpha} q_\beta \over 2M^2_p}
\epsilon_{\mu\nu\alpha\beta} + W'_4 (q^2,\nu) {q_\mu q_\nu \over M^2_p}
\nonumber \\[2mm]
& & + W'_5 (q^2,\nu) \left({P_{1\mu} q_\nu + q_\mu P_{1\nu} \over
M^2_p}\right) + W'_6 (q^2,\nu) \left({P_{1\mu} q_\nu - q_\mu P_{1\nu}
\over M^2_p}\right)  \;\; .
\label{three.37}
\eea
Thus, in general, the DIS process $\nu_\ell \; + p \rightarrow \ell^-
+ X $
requires six structure functions.  However, it should be noted at this
point that the cross-section involves contraction of
$H^{\prime\prime}_{\mu\nu}$ with $L'_{\mu\nu}$, and contributions from
$W'_4$, $W'_5$ and $W'_6$ to the cross-section can be seen to be
proportional to the lepton masses and can therefore be dropped.  This is
demonstrated for the term containing $W'_4 (q^2,\nu)$.  Consider
\[
Y = W'_4 (q^2,\nu) {q_\mu q_\nu \over M^2_p} \; L'_{\mu\nu} = {W'_4
(q^2,\nu) \over M^2_p} \left[2(P_3 \cdot q) (P_2 \cdot q) - q^2(P_2
\cdot P_3)\right] \;\; .
\]
Using $-P_3 \cdot q = P_2 \cdot q = q^2 / 2 $; $P_2 \cdot P_3 = -
q^2 / 2 - m^2_e$ we get,
\bea
Y &=& {W'_4 (q^2,\nu) \over M^2_p} \left[2 \; {q^2 \over 2} \;
\left(- {q^2 \over 2}\right) - q^2\left(-{q^2 \over 2} -
m^2_e\right)\right] = {W'_4 (q^2,\nu) \over M^2_p} m^2_e q^2 \simeq
0(m^2_e) \nonumber
\eea
Similarly terms proportional to $W'_5$ and $W'_6$ in
$H^{\prime\prime}_{\mu\nu} L'_{\mu\nu}$ can be shown to be small.

The contraction  of the first two symmetric terms in eq. (\ref{three.37})
with the symmetric terms in $L'_{\mu\nu}$ gives results similar to the
electromagnetic case.  The contraction between the symmetric and
antisymmetric terms will obviously yield zero.  The contraction of the
antisymmetric term in $H^{\prime\prime}_{\mu\nu}$ with the one in
$L'_{\mu\nu}$ gives,
\[
X = - 4\;{g^2 \over 2} \; {W'_3 (q^2,\nu) \over 2M^2_p} \; P_{1\alpha}
q_\beta \epsilon_{\mu\nu\alpha\beta} \epsilon_{\mu\nu\alpha'\beta'}
P_{3\alpha'} P_{2\beta'} \;\; .
\]
Using, $
\epsilon_{\mu\nu\alpha\beta} \epsilon_{\mu\nu{\alpha'}{\beta'}} =
2 \; [\delta_{\alpha{\alpha'}} \delta_{\beta{\beta'}} -
\delta_{\alpha{\beta'}} \delta_{\beta{\alpha'}}] \; $,
we get
\[
X = - 4 \;{W_3' (q^2,\nu) \over 2M^2_p} \; {g^2 \over 2} \; 2 \; [P_1
\cdot P_3 P_2 \cdot q - P_1 \cdot P_2 P_3 \cdot q] .
\]
Using $P_1 \cdot P_3 = -M_p E$, $P_1 \cdot P_2 = -M_p E_0$ and the
expressions for $P_2 \cdot q$, $P_3 \cdot q$ as well as $q^2$ quoted
earlier, we get
\[
X = 8 EE_0 \sin^2 \theta/2 \;
{g^2 \over 2} \; {W'_3 (q^2,\nu) \over M_p} (E + E_0) \;\; .
\]
Neglecting the lepton masses, in the limit $q^2 \ll M^2_W$, the general
expression for the inelastic, inclusive, differential cross-section for
$\nu$ reactions becomes,
\bea
{d^2 \sigma^{\rm inel} \over d\Omega dE} (\nu_\ell\; p \rightarrow \ell^- X)
&=& {G^2_F \over 2\pi^2} E^2 \; \Bigg[2W'_1 (q^2,\nu) \sin^2 {\theta
\over 2} + W'_2 (q^2,\nu) \cos^2 {\theta \over 2} \nonumber \\[2mm]
& & + W'_3 (q^2,\nu) {E_0 + E \over M_p} \sin^2 {\theta \over 2}\Bigg]
\;\; ,
\label{three.38}
\eea
where $G_F/\sqrt{2} = g^2/8 M_W^2$.
The maximum $\nu$-beam energy that has been reached in current
experiments is $E_0 = 500$ GeV.  From eq. (\ref{three.8ppp}) this means
that the maximum $q^2$ that can be reached in these experiments is
$\approx 1000$ GeV$^2$.  Since $M_W \simeq {\cal O} ~(1000 ~$ GeV) , the
approximation $q^2 \ll M^2_W $ is valid even at these highest attainable
values of $q^2$.  The charged current weak structure functions at higher
values of $q^2$ are accessible only in the study of the reaction
\[
e^- + p \rightarrow \nu + X
\]
at the HERA collider where $q^2$ can reach upto $10^5$ GeV$^2$.

Again, in the Bjorken limit  of large $q^2$ and $\nu$
the structure functions are found to scale just as in  eq. (\ref{three.31}),
{\it i.e.,} $M_p W'_1 (q^2,\nu)
\rightarrow F_1^{\nu \rm p} (x)$, $\nu W'_2 (q^2,\nu) \rightarrow
F_2^{\nu \rm p} (x)$ and $\nu W'_3 (q^2,\nu) \rightarrow F_3^{\nu \rm p} (x)$,
giving
\bea
{d^2\sigma^{\rm inel} \over dx dy} (\nu_\ell\; p \rightarrow \ell^- X) &=&
{G^2_F \over 2\pi} \; S \; \Big[xy^2 F_1^{\nu p} (x) + F_2^{\nu p}
(x) (1 -y) \nonumber \\[2mm]
& & + F_3^{\nu p} (x) xy\left(1 - {y \over 2}\right)\Big] \;\; .
\label{three.39}
\eea
$S$ in the above equation stands for the square of the centre of mass
energy given by $M^2_p + 2M_p E_0 \simeq 2M_p E_0$.  Note that the result
of eq. (\ref{three.39}) is obtained from eq. (\ref{three.38}) in the limit
of vanishing particle masses just as in the case of eq. (\ref{three.32}).
Here we have three structure functions  as opposed to the
electromagnetic case.   The third structure function arises from the
term containing $\epsilon_{\mu\nu\alpha\beta}$ in the tensor $L'_{\mu
\nu} ~(H^{\prime\prime}_{\mu\nu})$.  This is  the parity violating
structure function.  If instead of neutrino scattering we were to
consider
\[
\bar\nu_\ell + p \rightarrow \ell^+ + X
\]
the corresponding term in $L'_{\mu\nu}$ will change sign.  Hence in the
cross-section the term containing $F_3$ will change sign.  Hence,
\[
F_3^{\bar\nu p} = -F_3^{\nu p}.
\]
This can be physically understood by realising that $\bar\nu (\nu)$ is
right (left) handed and the term containing $F_3$ essentially arises from
the $V - A$ interference term in $L'_{\mu\nu} H^{\prime\prime}_{\mu\nu}$,
which  changes sign as we go from a right handed $\bar\nu$ to a left
handed $\nu$.\footnote{The terminology of  left (right) handedness
of the $\nu$($ \bar \nu$) corresponds to its helicity being -1(1).}

At this stage of analysis, it is not at all clear that the functions $W'_1
(q^2,\nu)$, $W'_2 (q^2,\nu)$ and $W_3' (q^2,\nu)$ (or equivalently
$F_i^{\nu p} (x)$, $i = 1,3$) have anything to do with the electromagnetic
structure functions $W_i (q^2,\nu)$, $i = 1,2$ (or equivalently
$F_i^{ep} (x)$ or $ F_i^{\mu p} (x)$, $i=1,2)$.  One can make
predictions for relations between these only in the framework of the
parton model. As a matter of fact, experimental test of these
relations was an important step in establishing the parton model firmly.
This will become clearer as we go on to discuss DIS and the parton model
in the next section.

\subsection{Relationship between scaling of cross-section and
pointlike constituents}
The phenomenon of  scaling of the cross--sections  reflecting the
existence of pointlike scattering centers occurs at different energy
scales twice as we go from $e^{-} A$ to $e^{-} p$ elastic scattering to
$e^- p$ deep inelastic scattering.  This is illustrated by the data on
nuclear scattering very nicely.  Fig. \ref{fig12} taken from ref. \cite{close}
shows this schematically whereas fig. \ref{fig13} \cite{hofstadter}
shows  some of the actual data.  The ``large''ness or ``small''ness
of a particular $q^2$ value has always to be understood with respect
to the inverse size.  Recall, eg.,  that for $q^2 \ll (0.71)$  GeV$^2$
the $W^{\it el}_{1,2}$ of eq. (\ref{three.30}) will look just like
their counterpart for elastic scattering from a pointlike scatterer
of eq.  (\ref{three.28}).  In general therefore, for elastic
scattering we can write,
\[
(M_p W_1^{e\ell})\;  \nu W_2^{e\ell} \sim f_{(1)2} (x) \; g(q^2) \; \; .
\]
For $q^2$ values such that $qR_{\rm target} \ll 1$ ({\it i.e.},  $q^2 \ll
\Lambda^2_{\rm target}$, where $\Lambda_{\rm target} =$ inverse size of
the target $\sim (1 / R)$), $\nu W_2^{e\ell}$ $(M_p  W_1^{e\ell})$
will scale, {\it  i.e.}, will not show any extra $q^2$ dependence.  At these
values of $q^2$, $g(q^2) \approx 1$.  In the case of elastic scattering,
be it from a nucleon or a nucleus, the variables $q^2$ and $\nu$ always
satisfy the relation
\[
q^2 = 2M_{\rm target}\; \nu\;\;.
\]
This also indicates that the function $f_{1(2)} (x)$ must also have a
factor $\delta \left( {q^2 \over 2M_{\rm target}\nu} - 1\right) $ or
equivalently a factor $ = \delta (x-1)$, {\it i.e.}, $f_{1(2)} (x)$ will
thus have a peak at $x = q^2 / (2M_{\rm target} \; \nu) = 1$.  As
$q^2$ values increase, the function $g(q^2)$ starts differing from 1
and falls off with increasing $q^2$.  For $q^2R^2 = 1$, {\it i.e.},
$q^2 \approx \Lambda^2_{\rm target}$, form factors cause a measurable
suppression of the cross-section.  For $qR \gg 1$, the function
$g(q^2)$ falls off very steeply indeed and the elastic peak at $x=1$
disappears.

In fig. \ref{fig12}(a) the elastic peak at $x = {q^2 / 2M_c \nu} = 1$
at $q^2 = 0.01$ GeV$^2$ in $e C \rightarrow e C$ is shown.  The quasi
elastic excitation of the resonance $C^\star$ in $e C \rightarrow e C^\star$
appears as a peak at a lower $x$ values (cf. eq. (\ref{three.3})).  At this
value of $q^2$, we have $q^2 \gg \Lambda^2_C$ and the Carbon nucleus
appears like a point particle and elastic scattering dominates.
As $q^2$ increases further, at $q^2 = (0.1)$ GeV$^2$, we have the situation
\[
\Lambda^2_{\rm proton} \gg q^2 \gg \Lambda^2_C\;\; .
\]
In this regime the proton appears to be pointlike and the elastic peak for
$e C \rightarrow e C$ disappears.  The dominant process is no longer the
elastic process $e C \rightarrow e C$, but  the inelastic one which
is now an incoherent sum of scattering off the $N$ nucleons inside the
Carbon nucleus.  In principle the elastic scattering off a nucleon should
show up as a peak at $x = q^2 / 2 M_C\; \nu = 1/N$ (for
the elastic $e p$ scattering $q^2 = 2M_p \nu$; hence
$x = q^2 / 2M_C\; \nu = 1 / N$).  The Fermi motion of the proton
in the nucleus however changes the kinematics and smears the
$\delta$-function peak.  This is shown in fig. \ref{fig12}(b).  Now
\[
\nu W^C_2 \sim f_2^C (x) \; g^p(q^2) \;\; .
\]
But $g^p (q^2) \approx 1$ since $q^2 \ll \Lambda^2_{\rm  proton}$.  Thus $f^C_2
(x,Q^2)$ now scales.  This scaling thus reveals the existence of $N$ pointlike
nucleons in the nucleus $C$.  Fig. \ref{fig13} shows data on $e
\alpha$ scattering at two different $q^2$ values $q^2 \simeq 0.08$
GeV$^2$ and $q^2 \simeq 0.1$ GeV$^2$.  This figure illustrates the same
point,  only the value of N here is 4.

If we now consider a proton at rest, at higher values of $q^2$ ($\simeq
0.5$ GeV $^2$) and higher $\nu$ values, the elastic peak for $e p
\rightarrow e p$ will occur at $x = {1 \over N}$ and the quasi-elastic
excitations corresponding to ,say, $\Delta^{++}$ will show up as a peak at
still lower values of $x$.  This is the proton analogue of Fig.
\ref{fig12}(a) and which was discussed already in section 2; except for
the fact that the variable defined there had been in terms of proton mass
$\left(x_p = (q^2 / 2M_p \nu)\right)$.  Now the cross-section (more
precisely $\nu W_2^p$) again will consist of two factors: one an $x$
dependent kinematic function and a $q^2$ dependent form factor reflecting
the target size.  This is the violation of scaling.

Now if $q^2$ is further increased to $q^2 \gg 0.71$ GeV$^2$ (Fig.
\ref{fig12}(d) corresponds to $q^2 \simeq 5$ GeV$^2$), then the
rapid fall-off of the form factor, $g^p (q^2)$, with increasing $q^2$ will
cause the elastic peak to vanish as well as the inelastic $\Delta^{++}$ peak.
Now the incoherent scattering from the pointlike partons (quarks) $eq
\rightarrow eq$ will show up as a smeared peak in $x$ distribution.  The
$x$ value at which this peak appears should give information about the
number of partons off which the electrons get scattered.  The
structure function scales again, revealing the existence of pointlike
constituents inside the proton and this peak will be at $x \simeq 1
/3N $ indicating the existence of three valence quarks.


At still higher $q^2$ values, for the proton target, ($q^2
\simeq 200$ -- 400 GeV$^2$ for the $e p$ , $\mu  p$ and $\nu p$ experiments
discussed earlier or $q^2 \lsim 10^4$ -- $10^5$ GeV$^2$ at the $ep$ collider,
HERA) $q\bar q$ production begins and that increases the number of
pointlike constituents in the proton effectively.  This shifts the peak
in the structure function to lower and lower $x$ values.  These scaling
violations are better discussed in terms of QCD and have been studied
extensively in current DIS experiments\cite{virchaux}. However,
these will not be discussed here any further.

\setcounter{equation}{0}
\section{Parton Model}
%
\subsection{Formalism}
As discussed in the last chapter, the DIS $e^-p \rightarrow e^-X$
cross-section scales at large $q^2$ and large $\nu$ (in the Bjorken
limit).  The observed scaling is evidence that the DIS cross-section is
given by an incoherent sum of scattering of the electron against
individual partons inside the proton.  This interpretation is basically
the parton model.  Implicit in this model are two assumptions:
\begin{itemize}
\item
Interactions among the partons are negligible during the time of
interaction between the electron and the parton.  The higher the
energy, shorter is the interval of time of this interaction.  Before
the proof of asymptotic freedom \cite{gross} of QCD, this assumption
was justified only by the success of the parton model, but asymptotic
freedom provides now justification for this assumption.
\item
 The second assumption is that final state interactions can be
neglected.  If the struck parton receives a huge kick then it gets
removed from the parton and the final state interactions are hence
negligible (recall here fig. \ref{fig10}).
\end{itemize}
Fig. \ref{fig15} shows a schematic description of DIS in the parton model.
Let $\xi$ be the momentum fraction of the proton carried by the struck
parton and $e_q$ denote its electric charge in units of the proton charge.
Let $f_{i/p} (\xi)$ be the probability that the  parton $i$ carries
momentum fraction $\xi$ of the proton.  Momentum conservation implies
\be
\sum_i \int^1_0 d\xi\; f_{i/p}(\xi)\; \xi = 1\;\; .
\label{four.1}
\ee
Note here that the sum is over all types of partons, not just the charged
ones which the incident electron (or equivalently the virtual photon) sees.
In this model, one has neglected transverse momentum of the
partons.  This is justified by the experimental observation that  apart
from the struck parton which causes hadrons to emerge at large angles,
the remaining particles in the final state emerge at small angles w.r.t.
the beam direction.

Let us choose the $z$ axis to be the direction of motion of the proton and
hence of the parton.  The magnitude of the three momentum is $|\vec p_1|$
and $\xi|\vec p_1|$ respectively for the proton and the parton.  The four
momenta are given by $P_1 = (\vec 0, P_L,iE_1)$ and $P_q = (\vec 0,\xi
P_L, i\xi E_1)$, where $P_L = |\vec p_1|$.  This gives $P^2_1 = -M^2_p$
and $P^2_q = -\xi^2 M^2_p$.  It appears as if the partons have a
variable mass $\xi M_p$.  This clearly is not what we mean.  This is just
a reflection of the fact that the above kinematics is strictly correct
only in the limit where all masses can be neglected.  In this case the
kinematics given above simply corresponds to a collinear emission of a
massless particle from another massless particle.

The frame of reference in which the above kinematics is strictly valid is
called the infinite momentum frame.  In the infinite momentum frame, time
dilation slows down the rate at which partons interact with one another
and this time scale is now much bigger than the time taken by the current
({\it i.e.}, the electron or the virtual photon) to interact with the parton.
Hence the impulse approximation (assumption (i) above) is justified in
this frame.

After scattering the struck parton (quark) appears in the detectors as
a stream of hadrons.  The time scale of hadronisation $(\sim
10^{-23}~{\rm sec}.)$ is much bigger than the interaction time scale
$(\lsim 10^{-25}$ sec.) for $\nu \gsim {\cal O} $(10 GeV).  This
description of scattering as a two step process is the second basic
tenet of parton model.  It is clear from the above discussion that for
both these assumptions to be justified and the picture of
fig. \ref{fig15} to be true, $q^2,\nu$ and $W$ all need to be large.

Since the partons are pointlike objects, $m_q W_1^{\rm parton}
(q^2,\nu)$ and $\nu W_2^{\rm parton} (q^2,\nu)$ are given by eqs.
(\ref{three.28}), replacing $M_p \rightarrow m_q = \xi M_p$.  Hence we
have,
\bea
m_q W_1^{\rm parton} (q^2,\nu) = e^2_q {q^2 \over 4m_q} \delta\left(\nu -
{q^2 \over 2m_q}\right)\;\; ; \nonumber \\[2mm]
\nu W_2^{\rm parton} (q^2,\nu) = e^2_q \delta\left(1 - {q^2 \over
2m_q\nu}\right) \;\; .
\label{four.2}
\eea
If we now define
\be
\omega = {2M_p \nu \over q^2} = {-2P_1 \cdot q \over q^2} = {1 \over
x}\;\; ,
\label{four.3}
\ee
then we get
$$
M_p W_1^{\rm parton} (q^2,\nu) \equiv e^2_q \; {q^2 \over 4m_q \nu \xi}
\delta\left(1 - {1 \over \xi\omega}\right) e^2_q = {q^2 \over 4M_p \nu
\xi^2} \delta\left({-x \over \xi} + 1\right)\;\; .
$$
The right hand side of this equation is clearly a function of $x$ alone,
and we can write (with an analogous discussion for $\nu W_2^{\rm parton}
(q^2,\nu)$),
\[
F_1^{\rm parton} (x) = M_p W_1^{\rm parton} (q^2,\nu) = {e^2_q \over 2}
{q^2 \over 2M_p \nu \xi^2} \delta\left(1 - {x \over \xi}\right) \;\; ;
\]
$$
F_2^{\rm parton} (x) = \nu W_2^{\rm parton} (q^2,\nu) = \delta\left(1 - {x
\over \xi}\right) e^2_q \;\; .
\label{four.4}
$$
Since in the limit of large $q^2,\nu$ and $W$, the total cross-section for
$ep$ scattering (and hence $M_p W_1^{\rm parton}$, $\nu W_2^{\rm parton}$)
are given by an incoherent addition over all the charged partons (quarks)
we get in the scaling limit,
\bea
F_2^{ep} (x) &=& \sum_q \int^1_0 d\xi F_2^{\rm parton} (\xi)\; f_{q/p} (\xi)
\nonumber \\[2mm]
&=& \sum_q e^2_q \int^1_0 d\xi \delta\left(1 - {x \over \xi}\right) f_{q/p}
(\xi) \nonumber \\[2mm]
&=& \sum_q e^2_q\; f_{q/p} (x)\; x \;\; . \nonumber
\eea
Similarly,
\bea
F^{ep}_1 (x) &=& {1\over 2} \sum_q \int^1_0 d\xi e^2_q {q^2 \over 2M_p \nu
\xi} \delta (\xi - x) f_{q/p} (\xi) \nonumber \\[2mm]
&=& {1\over2} \sum_q e^2_q \;f_{q/p} (x) \;\; . \nonumber
\eea
It should be noted here that the summation is over  particles as well
as antiparticles.

Thus we find that the scaling variable $x$ which we defined earlier can be
identified with the momentum fraction $\xi$ that the parton $q$ carries.
The structure functions appearing in eq. (\ref{three.31}) (which are the
scaling limits of the arbitrary functions $W_i (q^2,\nu), ~i=1,2$ which
appeared in the tensor decomposition of eq. (\ref{three.18})) are related
to the probability of finding a parton of charge $e_q$ (in units of proton
charge) with momentum fraction $x$ of the proton.  There is yet another
way of seeing the same thing.  In the infinite momentum frame all masses
are negligible.  The final four momentum of the struck parton is given by
$(\xi P_1 + q)$.  Hence we have

$$
(\xi P_1 + q)^2 \simeq 0 \; \; ,
$$
which gives us,
\be
\xi = {-q^2 \over 2P_1 \cdot q} \;\; .
\label{four.5}
\ee
This is same as the variable $x$ defined in eq. (\ref{three.7a}).  The
above expressions for $F_i^{ep} (x)$, $i = 1,2$ also imply
\be
F_2^{ep} (x) = 2x F_1^{ep} (x)\;\; ,
\label{four.6}
\ee
which is known as the Callan-Gross relation.
Using eq. (\ref{four.6}) we get from eq. (\ref{three.32})
\bea
{d^2 \sigma^{inel} \over dx dy} (e^-p \rightarrow e^- X) &=& {2\pi \alpha^2
 \over Sx^2y^2} \left[\sum_q e^2_q x f_{q/p} (x)\right] \left[1 +
(1-y)^2\right] \nonumber \\[2mm]
&=& {4\pi \alpha^2  \over Sx^2 y^2} F_2^{ep} (x) \left[{1\over2} + {(1-y)^2
\over 2}\right] \;\; .
\label{four.8}
\eea
In the parton model picture which we have developed above, now the
structure function $F_2^{ep} (x)$ is expressed in terms of the probability
density functions $f_{q/p} (x)$.  Since this probability should be
independent of the probe that is used to extract it from the data, one
expects that structure functions $F_i^{\nu p} (x) ~(i=1,2,3)$,
measured in $\nu p$ DIS, must also be some combinations of the same functions
$f_{q/p}(x)$ that are extracted from the data on electromagnetic DIS processes,
the specific form being decided by the nature of $\nu q$ interactions.
As a matter of fact, identifying the charged partons in the proton with
the constituent quarks of $SU(3)$ flavour, one can make definite
predictions for the weak structure functions of the parton as well as
electromagnetic and weak structure functions of other targets such as
neutron or nuclei and relate them to each other.  An experimental
verification of these relations in the $\nu$ DIS experiment
\cite{neu} played a very important role in establishing the parton
model on a firm footing.

An alternative definition of eq. (\ref{four.8}) can be given as follows.
Consider the cross-section ${d\sigma / d\Omega}$ for scattering of an
$e^-$ from a pointlike object of charge $e_q$ (in units of proton charge).
The corresponding expression is given by eq. (\ref{two.5}).  For the case
of elastic scattering, we know that $y$ is given by (recall
eq. (\ref{three.8})) ,

\[
y = {\nu \over E_0} = {q^2 \over S} = {2E_0/M_p \sin^2 {\theta \over 2}
\over 1 + {2E_0 \over M_p} \sin^2 {\theta \over 2}} \;\;\; .
\]
Hence
\[
{d\sigma^{el} \over dy} = {d\sigma^{el} \over d\Omega} \; \left({2\pi
E_0 M_p \over E^2}\right) = {S \over E^2} {d\sigma^{el} \over d\Omega}
\;\;\; ,
\]
where $S$ is the square of the cm energy.  In the present case we have to
consider scattering of the $e^-$ from a charged parton carrying a fraction $x$
of the four-momentum $P_1$ of the proton.  The square of the cm
energy $\hat S$ of the elastic electron-parton scattering is given by,
\be
\hat S = -(xP_1 + P_2)^2 = -2x P_1 \cdot P_2 = 2x M_p E_0 = xS \;\; .
\label{four.9}
\ee
Using the expression for $d\sigma^{el} / d\Omega$ given by eq.
(\ref{two.5}) but now for c.m. energy $\sqrt{\hat S}$ as given by eq.
(\ref{four.9}), we have (again neglecting particle masses),
\bea
{d\sigma^{el} \over dy} (e^- q \rightarrow e^- q) &=& {4\pi\alpha^2 \over
q^4} \; {\hat S \over E^2} e^2_q \; E^2 \left({E \over E_0}\right)
\left[\cos^2 {\theta \over 2} + {q^2 \over 2M^2_p} \sin^2 {\theta \over
2}\right] \nonumber \\[2mm]
&\simeq& {4\pi \alpha^2 e^2_q \over q^4} S \; x \left[{1 + (1-y)^2
\over 2}\right] \;\; .
\label{four.10}
\eea
If $f_{q/p} (x) dx$ is the probability of finding a parton with momentum
fraction of proton between $x$ and $x + dx$, we have
\[
{d\sigma \over dy} (e^- p \rightarrow e^- p) = {4\pi\alpha^2 \over q^4}
\left[\sum_q e^2_q \int^1_0 f_{q/p} (x) x ~dx\right] \left[{1 + (1-y)^2 \over
2}\right] \;\; .
\]
Hence we get
\be
{d\sigma \over dx~dy} (e^-p \rightarrow e^-p) = {4\pi \alpha^2 S \over
q^4} \left[\sum_q e^2_q f_{q/p} (x) x\right] \left[{1 + (1-y)^2 \over
2}\right] \;\; .
\label{four.11}
\ee
Note that above equation is the same as eq. (\ref{four.8}) or eq.
(\ref{three.32}) where,
\be
F^{ep}_2 (x) = 2x F^{ep}_1 (x) = \sum_q e^2_q \; x \; f_{q/p} (x) \;\; .
\label{four.11'}
\ee
This alternative derivation of eq. (\ref{four.8}) indicates that, to derive
the corresponding expressions for the double differential DIS
cross-section for $\nu p $ or $ \bar\nu p$ processes, we need to know
$d\sigma^{el} / dy (\nu{_\ell} \; q \rightarrow
\ell \; q')$, $d\sigma^{el} / dy (\bar \nu_{\ell}
\; q \rightarrow \bar\ell \; q')$ etc.  Due to the parity violating
nature of weak interactions, the angular distribution (and hence $y$
distribution) of $\nu q$ and $\nu \bar q$ scattering are different in
nature.  This is in contrast to the situation in the case of
electromagnetic scattering.  The differential distributions $
d\sigma^{el} / dy $ for the elementary scattering process can be
obtained after a simple calculation to be
\bea
{d\sigma^{el} \over dy} (\nu_{\ell}\; q) & = & {d\sigma^{el} \over dy}
(\bar\nu_\ell\; \bar q) = {G_F^2\; S \over \pi} \;\; ; \nonumber \\[2mm]
{d\sigma^{el} \over dy} (\bar\nu_\ell\; q)& = &{d\sigma^{el} \over dy}
(\nu_\ell\; \bar q) = {G_F^2\; S \over \pi} (1 - y)^2 \;\; . \nonumber
\eea
The second equation above indicates the impossibility of $\nu_{\ell}\; \bar q\;
(\bar \nu_{\ell}\; q)$ scattering in the backward direction ($\theta = \pi$
corresponds to $y = 1$).  This can be easily understood from fig.
\ref{fig16P}.  This shows that the backward scattering in this case will
correspond to $|\Delta J_Z| = 2$ which is not possible.  We can use the
above expressions for $d\sigma^{el} / dy (\nu_{\ell}\;
q)$ etc. in a manner similar to the one used in ariving at eq.
(\ref{four.11}).  In this case we get for the charged current DIS
cross-section,
\bea
{d^2\sigma \over dx dy} (\nu_{\ell}\; p \rightarrow \ell^- n) &=& {4G^2_F\; S
\over 2\pi} \left[2x \sum_q f_{q/p} (x) + 2x \sum_q f_{\bar q/p} (x) (1 -
y)^2\right] \nonumber \\[2mm]
&=& {G_F^2\; S \over 2\pi} \Bigg[2x \sum_q \left(f_{q/p}(x) + f_{\bar q/p}
(x)\right) \left({1 + (1-y)^2 \over 2}\right) \nonumber \\[2mm]
& & + 2x \sum_q \left(f_{q/p} (x) - f_{\bar q/p} (x)\right) \left({1 - (1-y)^2
\over 2}\right)\Bigg] \;\; . \nonumber
\eea
Of course it is understood that the sum is to be taken over those quarks
or antiquarks which can take part in allowed transitions.  The first term
in the bracket is parity conserving whereas the second one violates
parity.  This equation is the same as eq. (\ref{three.39}) if we identify
\be
F_2^{\nu p} (x) = 2x F_1^{\nu p} (x) = 2x\left[\sum_q f_{q/p} (x) +
f_{\bar q/p} (x)\right] \;\; ;
\label{four.12}
\ee
\be
x F_3^{\nu p} (x) = 2x\left[\sum_q f_{q/p} (x) - f_{\bar q/p}
(x)\right] \;\; .
\label{four.13}
\ee
%
\subsection{Structure functions for proton, neutron and isoscalar targets}
In this section we will write the form for the electromagnetic and weak
structure functions expected in the parton model, if one identifies the
charged parton with the Gell-Mann-Zweig constituent quarks.  In the
$SU(3)_f$ picture, e.g., the proton contains two $u$-quarks with charge
${2\over3} e_p$ and one $d$-quark with charge $-{1\over3} e_p$ .  Hence one
expects that $F_2^{ep}$ is given by,
\be
F_2^{ep} = {4 \over 9} x u^p (x) + {1\over9} xd^p (x)\;\; .
\label{four.14}
\ee
Here $u^p (x) = f_{u/p} (x)$ and so on.  Isospin invariance would imply,
$$
u^n (x) = d^p (x); ~d^n (x) = u^p (x)\;\; .
$$
Hence we expect,
\be
F_2^{en} (x) = {4\over9} d^p (x) + {1\over9} u^p (x) \;\; .
\label{four.15}
\ee
However, this presupposes the picture that the proton (neutron) has
$2(1)$ $u$-quarks and $1(2)$ $d$-quarks.  However, all the
conclusions, verified by experiments, about static properties of
proton/neutron will remain unchanged, if in addition to these partons
one had large number of $u\bar u$, $d\bar d$, $s\bar s$ pairs forming
an $SU(3)$ singlet, which have been radiated from those ``valence''
(large momentum) partons.  One would expect the proton/neutron to
contain such radiated partons in the parton model picture.  Since
these arise from radiation, these will have carry less momentum than
the parent quarks, they will have a softer momentum distribution than
the ``valence'' quarks.  These were called ``wee'' or ``sea'' partons
by Feynman.  Then we will have, assuming $SU(3)$ symmetry for the sea
({\it i.e.}, the sea quark density is the same for all three types of
quarks $u,d$ and $s$),
\bea
u^p (x)& = &u^p_V (x) + u^p_S (x) ,\nonumber \\[2mm]
d^p (x)& = &d^p_V (x) + d^p_S (x) ,\nonumber\\[2mm]
u^p_S(x)& =& \bar u_S^p (x) = d_S^p (x) = \bar d_S^p (x) = S^p_S (x) =
\bar S^p_S (x) = K(x) \;\; .
\label{four.16}
\eea
In this case eq. (\ref{four.14}) will modify to
\bea
F_2^{ep} (x) &=& x\Bigg[{4\over9} (u^p_V (x) + u^p_S (x)) + {1\over9}
(d_V^p (x) + d_S^p (x)) + {4\over9} \bar u^p_S (x) \nonumber \\[2mm]
& & + {1\over9} \bar d_S^p (x) + {2\over9} \bar S^p_S (x)\Bigg]
\nonumber \\[2mm]
&=& x\left[{4\over9} u^p_V (x) + {1\over9} d_V^p (x) + {4\over3}
K(x)\right] \;\;  .
\label{four.17}
\eea
Again isospin invariance gives us
\be
F_2^{en} (x) = x\left[{4\over9} d_V^p (x) + {1\over9} u^p_V (x) +
{4\over3} K(x)\right] \;\; .
\label{four.18}
\ee
In writing eqs. (\ref{four.17}) and (\ref{four.18}) we have assumed,
in addition to the flavour symmetry of the sea, also absence of
heavier charm and bottom quarks in the sea.  At higher energies, even
these can be radiated.  In that case the factor before $K(x)$ will
change.  Using eqs. (\ref{four.17}) and (\ref{four.18}) we get,
\be
{F_2^{en} (x) \over F_2^{ep} (x)} = {u^p_V (x) + 4d_V^p (x) +
{4\over3} K(x) \over 4u^p_V (x) + d_V^p (x) + {4\over3} K(x)} \;\;\; .
\label{four.19}
\ee

Can we compute these structure functions $u^p_V (x)$, $d_V^p (x)$ and
$K(x)$?  The answer is no, not without a model.  However, we can try
to measure them experimentally.  The experimentally measured DIS
cross-sections for $ep \; (\mu p)$ reactions will yield $F_2^{ep}
(F_2^{\mu p})$.  How does one extract the specific parton densities
from it?  Before one goes into this question, let us see how one can
obtain some qualitative information about $u^p_V (x)$, $d_V^p (x)$ and
$K(x)$.  Consider the ratio of eq. (\ref{four.19}).  If $K(x)$
dominates over $u^p_V (x)$, $d_V^p (x)$, this ratio will be 1.  If
either $u^p_V (x)$ or $d_V^p (x)$ dominates over all the other
densities  then this ratio will go  to $1/4$ or $4$
respectively.  Hence, under the assumption of valence dominance
we will get
\be
{1\over4} < {F_2^{en} (x) \over F_2^{ep} (x)} < 4
\label{four.20}
\ee
If, on the other hand, we have $K(x) = 0$ and $u^p_V (x) = 2d_V^p
(x)$ (as can be expected in a pure constituent picture), this ratio
will have the value $2/3$.

With these qualitative predictions we turn to the question as to what
the experiments say.
As was the case with the form factors, $F_2^{en} (F_2^{\mu n})$
is measured by combining the data on $ep$, $ed$ $(\mu p,\mu d)$
scattering.  Early data\cite{rmp} tell us:

\begin{enumerate}

\item[{(a)}] $K(x)$ dominates at small $x$ and the ratio is indeed
close to 1 at small $x$.

\item[{(b)}] At large $x (x \rightarrow 1)$, $F_2^{ep} (x) /
F_2^{en} (x) \rightarrow 4$.  This means that $u^p_V (x)$ dominates
over $d_V^p (x)$, as well as over $K(x)$ at large $x$.

\item[{(c)}] Observation (b) also tells us that the naive expectation
of $u^p_V (x) = 2d_V^p (x)$ is not fullfilled.

\end{enumerate}
Fig. \ref{fig17} shows the recent high statistics data of the
NMC collaboration on the ratio $F_2^{\mu n} / F_2^{\mu p}$
\cite{plot}.
What are the other theoretical constraints on these densities?  We
know that the net number of $u(d)$ quarks in a proton is $2(1)$ and
the net number of strange quarks in both is zero.  Hence we have,
\bea
\int^1_0 \left(u^p(x) - {\bar u}^p (x)\right)dx &=& 2 = \int^1_0 u^p_V (x)
dx\;\; , \nonumber \\
\int^1_0 \left(d\;^p (x) - {\bar d}^p (x)\right)dx & =& 1 = \int^1_0 d_V^p
(x) dx\;\;, \nonumber \\
\int^1_0 \left(S^p(x) - \bar S^p (x)\right)dx &=& 0 \;\; .
\label{four.21}
\eea
The first two constraints of eq. (\ref{four.21}) can also be obtained
by considering electromagnetic charge conservation, which implies, for
the proton and neutron respectively,
\bea
1 &=& \int^1_0 \left[ {2\over3} \left(u^p(x) - \bar u^p(x)\right) -
{1\over3} \left(d^p(x) - \bar d^p (x)\right)\right] dx \;\; , \nonumber \\
0 &=& \int^1_0  \left[{2\over3} \left(d^p (x) - \bar d^p (x)\right) -
{1\over3} \left(u^p (x) - \bar u^p (x)\right) \right] dx \;\; .
\label{four.22}
\eea
We have to use the fact that both the proton and neutron have zero
strangeness, in order to get the third of eq. (\ref{four.21}).  It
should be emphasized here that the actual determination of $u^p (x)$,
$d^p (x)$ and $K(x)$ has to be done by combining data on
electromagnetic structure functions as well as the weak structure
functions for different targets; $p,n$ and isoscalar nuclei and then
fitting a form to the densities $u^p_V (x)$, $d^p_V (x)$ and $K(x)$,
using the data, subject to the above sum rules.  Some of these details
will be discussed below.
%
%
\subsection{Properties of partons as determined from DIS}
%
\subsubsection{Spin of partons}
Let us start from the Callan-Gross relation of eq. (\ref{four.6}).
Note here that if the charged partons were scalars then, according to
eq. (\ref{three.29}), $F_1^{\rm parton} (x) = 0$.  Hence the ratio
$F_2^{ep} (x) /(2x F_1^{ep}) (x) \rightarrow \infty$
whereas eq. (\ref{four.6}) predicts the ratio to be 1 for spin 1/2
partons.  It is customary to define a longitudinal structure function
$F_L^{ep} (x)$ by
\be
F_L^{ep} (x) = F_2^{ep} (x) - 2x F_1^{ep} (x) \;\;\; .
\label{four.7}
\ee
Since $F_2^{ep} (x)$ and $F_1^{ep} (x)$ can be extracted from the
measured DIS cross-section, an experimental verification of the above
relation will imply that the pointlike constituents inside the proton
revealed in DIS have spin 1/2.  $F_L^{ep} (x)$ is termed the longitudinal
structure function as it is proportional (in the large $q^2$, $\nu$
limit) to the virtual photoabsorption cross-section for longitudinal
photons ({\it i.e.}, photons with helicity $\lambda = 0$).  In
fig. \ref{fig16}, we see the experimental data as the ratio
\be
R = {F_L^{ep} (x) \over 2x F_1^{ep} (x)}\;\; .
\label{four.7n}
\ee
In the large $q^2$, $\nu$ limit this is the
ratio of virtual photoabsorption cross-section for longitudinal and
transverse photons $(\lambda = \pm 1)$.  As noticed before, the
denominator is zero for scalar partons. This can be physically
understood as the impossibility of absorption of transverse photons
by a scalar target.  The data of fig. \ref{fig16} taken from
\cite{rmp} show clear evidence that the charged partons are spin 1/2
objects and not spin 0 objects. Deviation of this quantity from zero is yet
another `check' for QCD but again will not be discussed here further.

%
%
%
\subsubsection{Momentum carried by charged partons}
As can be seen from eqs. (\ref{four.17}) and (\ref{four.18}) the area
under the $F_2^{ep} (F_2^{en})$ vs. $ x $ curve will measure the
weighted sum of the momentum fractions carried by the charged partons
in the proton (neutron).  Hence
\be
\begin{array}{l}
\int^1_0 F_2^{ep} (x) dx = {4\over9} \epsilon_u + {1\over9} \epsilon_d
+ {1\over9} \epsilon_s \;\; , \\[2mm]
\int^1_0 F_2^{en} (x) dx = {1\over9} \epsilon_u + {4\over9} \epsilon_d
+ {1\over9} \epsilon_s \;\; ,
\label{four.25}
\end{array}
\ee
where $\epsilon_u,\epsilon_d$ and $\epsilon_s$ are the fractions of
the proton momentum carried by $(u + \bar u)$, $(d+\bar d)$ and $(s +
\bar s)$.  If we define $\delta = \epsilon_s / (\epsilon_u +
\epsilon_d) $ , we can write
\bea
\int^1_0 \left[F_2^{ep} (x) + F_2^{en} (x)\right]dx & =& {5\over9}
(\epsilon_u + \epsilon_d) + {2\over9} \delta (\epsilon_u + \epsilon_d)
\;\; ,
\nonumber \\[2mm]
{9(\delta + 1) \over 5 + 2\delta} \int^1_0 \left[F_2^{ep} (x)
+ F_2^{en} (x)\right]dx & =& \epsilon_u + \epsilon_s + \epsilon_d \;\; .
\label{four.26}
\eea
It is clear from eq. (\ref{four.26}) that a knowledge of $\delta$ is
necessary to determine $(\epsilon_u + \epsilon_s + \epsilon_d)$.  An
extraction of  quark densities from the DIS data using
$ep,en,\nu p,\nu n$ experiments and eq. (\ref{four.24}) shows that
$\delta \leq 0.06$.  Using the data on $(F_2^{ep} (x) + F_2^{en} (x))$,
this gives,
\be
\epsilon_u + \epsilon_d + \epsilon_s \simeq (0.54 - 0.56) \pm 0.04
\;\; .
\label{four.27}
\ee
Eq. (\ref{four.27}) above implies that some momentum of the proton
$(\sim 50$\%) is carried by partons which are neutral to the probe,
{\it i.e.}, partons which do not have electromagnetic or weak interactions.
These are called gluons which hold the quark-partons together in a
proton.
%
%
\subsubsection{Charge assignment of different partons}

To confirm the fractional charge
assignment used in arriving at, e.g., eqs. (\ref{four.14}),
(\ref{four.15}) one has to combine data on $F_2^{ep}, F_2^{en}$ with
the data on $F_2^{\nu p}, F_2^{\nu n}$.  For this we should obtain
expressions for $F_2^{\nu p}, F_2^{\nu n}$ analogous to eqs.
(\ref{four.14}), (\ref{four.15}).  To do this one has to note the
following :
\begin{enumerate}
\item[(i)]
A neutrino can scatter off only the charge $-
{1\over3} e_p$ quarks and charge $-{2\over3} e_p$ antiquarks, as it has
to have a $\ell^-$ in the final state.  Hence only $\nu_{\ell}\; d$,
$\nu_{\ell}\; s$ and $\nu_{\ell}\; \bar u$ processes take place.  (Recall, we
have at present neglected the heavier, charm and bottom, quark content
of the proton/neutron).
\item[(ii)]
At low energies, $\nu_{\ell}\; s \rightarrow \ell^- c$
transition can be neglected and $\nu_{\ell}\; s \rightarrow \ell^- u$
transition is suppressed in the limit of the
mixing angle in the $s$--$d$ sector (Cabibbo angle) $\theta_c \simeq 0$.
This is a good approximation as $\cos\theta_c = 0.98$.  Under these
approximations, the only transitions that  contribute to the DIS
cross-section for incident $\nu_\ell$  are $\nu_{\ell}\; d\; (\bar u)
\rightarrow \ell^- u\;(\bar d)$ and  $\bar\nu_{\ell}\; u\; (\bar d)
\rightarrow \ell^+ d\; (\bar u)$ for incident $\bar \nu_\ell $.
\end{enumerate}
Using eq. (\ref{four.13})
then we have, at low energies and with the above approximations,
\bea
F_2^{\nu p} (x) &=& 2x\left[d\;^p (x) + \bar u^p (x) \right] \;\; ,
\nonumber \\
F_2^{\nu n} (x) &=& 2x\left[u^p (x) + {\bar d}^{p} (x) \right] \;\; .
\label{four.23}
\eea
Using eqs. (\ref{four.17}), (\ref{four.18}) and (\ref{four.23}) we
get,
\be
{F_2^{ep} (x) + F_2^{en} (x) \over F_2^{\nu p} (x) + F_2^{\nu n} (x)}
= {5 \over 18} {\left[u^p (x) + d\;^p (x) + \bar u^p (x) + {\bar d}^p (x)
+ S^p (x) + \bar S^p (x)\right] \over \left[u^p (x) + d^p (x) + \bar
u^p (x) + {\bar d}^p (x)\right]} \;\;\; .
\label{four.24}
\ee
Excess of this ratio above 5/18 is a measure of the momentum carried
by the $s(\bar s)$ quarks. Production of charmed particles in DIS in
$\nu p$ reactions provides a direct measurement of this strange sea
content of the proton.  The factor of 5/18 in the eq.
(\ref{four.24})  above is the average squared charge of the $u,d$ quarks.

The electromagnetic and weak structure functions are usually measured
not only for light targets such as proton/deuterium, but more often
experiments are performed with heavier, nuclear targets so as to get
large cross-sections.  Normally one expects that for nuclear targets, the
cross-section from different nuclei will add incoherently.  For an
isoscalar target therefore, the structure function per nucleus
becomes,
$$
{1 \over A} F_2^{eA} = {1\over2} (F_2^{en} +  F_2^{ep})\;\; .
$$
With eqs. (\ref{four.17}) and (\ref{four.18}), we get
\bea
{1 \over A} F_2^{eA} &=& x\Bigg\{{5 \over 18} \left[u^p(x) + d^p(x)
+ \bar u^p(x) + {\bar d}^p(x) + S^p(x) + \bar S^p(x)\right] \nonumber \\[2mm]
& & ~~~~~~~~~~~~~~~~~~~~ - {1 \over3} \left[S^p(x) + \bar
S^p(x)\right]\Bigg\} \;\; . \nonumber
\eea
If we include also the charm sea contribution then we will have
\be
{1 \over A} F_2^{eA} \simeq {5 \over 18} \sum_q x q^p(x) \;\; .
\label{four.28}
\ee
Note again the factor of 5/18.  It is clear from this discussion that
the ratio ${\displaystyle {F_2^{eA} \over F_2^{\nu A}}}$ is the same
as the r.h.s. of eq. (\ref{four.24}).  A collection of some of the
data taken from the third of Ref.
\cite{neu}  presented in fig. \ref{fig18} shows experimental proof of
this relation.  This confirms the identification of the charged
partons with the Gell-Mann-Zweig\cite{qmodel} quarks.

Note also that a measurement of  ${1 \over A} \displaystyle \int^1_0
F_2^{eA} dx$ for an isoscalar target can directly give information
about the total momentum carried by the quark-partons.  The  recent
high $Q^2$ data on $\mu p/\mu d$ DIS scattering by the different
collaborations like EMC, BCDMS, NMC  and $\nu p/\nu Fe$ DIS data by the
CCFR collaboration \cite{virchaux} confirm  all the parton model
features quite beautifully.


\subsection{Interpretation of sea densities}
The existence of gluons gives a very simple understanding of the
sea-quark densities.  At some very low $q^2$-scale, the proton can be
looked upon as made up of only three valence quarks $u,u$ and $d$, all
carrying $( {1 \over 3} )^{\rm rd}$ of the proton momentum.  The probability
distributions $u^p_V (x)$ and $d_V^p (x)$ are just  $\delta (1 - x/3)$
with appropriate normalisation.  Emission of gluons, by bremsstrahlung,
by these quarks causes the probability distributions to
shift to lower $x$ values.  The emitted gluons give rise to $q\bar q$
pairs thus giving rise to sea quarks.  Since the process of
bremsstrahlung is naturally peaked at small values of momentum
fractions, it generates $q\bar q$ sea.  At higher and higher energies
the number of $q\bar q$ pairs produced goes on increasing.  Actually
this process will also give rise to scaling violations as it causes
$u^p (x)$, $d^p(x)$ to shift to lower $x$ values with increasing $q^2$
and  makes $F_2^{ep} (x)$ $q^2$ dependent.  However, this clearly
takes us out of the realm of the Quark-Parton-Model (QPM) and causes
corrections to the simple QPM picture.  These corrections (scaling
violations) actually played an important role in establishing the
nature of interactions among quarks and gluons.  But this will not be
discussed further here.

Another way of understanding more about the sea-quark densities is to
try to construct a quantity which is independent of sea quark
densities.  Consider the following combination:
\be
F_2^{ep} (x) - F_2^{en} (x) = {x \over 3} \left[u_V^p (x) - d_V^p
(x) \right] .
\label{four.29}
\ee
This difference does not involve the sea quark densities at all {\it
if one assumes isospin--symmetric sea densities.}  The
experimental data on $(F_2^{ep} (x) - F_2^{en} (x))$ \cite{rmp} show a
peak at $x = 1/3$.  Fig. \ref{fig19} shows (just to show the
increased accuracy of the newer data) the much more recent and
high statistics data taken from \cite{plot}. This clearly supports the
picture of three constituents of mass ${1\over3} M_p$ and also the
interpretation of sea quark pairs as arising due to bremsstrahlung from
valence quarks.
%
%
\subsection{Sum rules on parton densities}
Since there is, as yet, no theory which can compute parton densities
in a proton, all the knowledge about parton densities is to be
obtained from experiments, which can then be  used for testing
models of the proton structure functions.  Within the framework of the
quark--parton model, various relations, sum rules have been derived for
$F_2^{ep}$ and $F_2^{en}$ or combinations thereof.  Some of these sum
rules have already been written down in eqs. (\ref{four.21}) and
(\ref{four.22}).  From eq. (\ref{four.29}) we get
\be
\int^1_0 {\left(F_2^{ep} (x) - F_2^{en} (x)\right) \over x} dx =
{1\over3} \int^1_0 \left(u^p_V (x) -  d_V^p (x)\right) dx = {1\over3}
\;\; .
\label{four.30}
\ee
This is called  the Gottfried sum rule.  This sum rule is arrived at by
assuming $\bar u_s^p (x) = \bar d_s^p (x)$.  If that is not the case,
this will be violated.  Current data from $\mu$ DIS experiments
\cite{plot} show that the sum rule is violated, and the above integral
(obtained from extrapolation of the data for the region $x > 0.8 $
and $ x < 0.004$ ) is,
\[
\int^1_0 {F_2^{ep} (x) - F_2^{en} (x) \over x} dx = 0.258 \pm 0.017
\;\; .
\]
These data use measurements of the structure function at $q^2 = 4 $
GeV$^2$. These indicate a departure from the expected
value of $1/3$ and a breaking of the isospin symmetry of the sea
densities which has been assumed in arriving at eq.  (\ref{four.30}).

One can derive yet another sum rule using $F_2^{\nu p}$ and $F_2^{\nu
n}$.  We see from eq. (\ref{four.23}) that ,
\be
\int^1_0 {(F_2^{\nu n} - F_2^{\nu p})dx \over x} = 2 \int^1_0
\left(u^p(x) - d\;^p(x)\right)dx = 2 \;\; .
\label{four.31}
\ee
This sum rule is called the Adler sum rule.  Again it assumes isospin
symmetry for the sea-quark densities.  The sum rule is known to be
satisfied.  But the $\nu$ scattering data have intrinsically much
larger errors compared to the $\mu p(\mu  D)$ scattering data.  As a
result the neutrino data are not in a position to test the violation
of isospin symmetry of sea densities implied by the $\mu$ DIS data
from EMC.

Yet another sum rule that has been written down is for the parity
violating structure function $F_3^{\nu p}$. One can write down
expressions for $F_3^{\nu p}$ and $F_3^{\nu n}$ using eq.
(\ref{four.13}) and using the isospin invariance of the
neutron/proton parton densities. Using eq. (\ref{four.13}) we get,
$$
\int_0^1 (F_3^{\nu p}(x)  + F_3^{\nu n}(x)) dx = 3 \;\; .
$$
This is called the Gross--Llewellyn Smith sum rule. The earliest
$\nu$ experiments \cite{neu} verified this sum rule. Currently it has
been a focus of lot of discussions as the deviation of the r.h.s from
three can provide important information about and test of pQCD. The
currently measured value is  $2.50 \pm 0.018 (stat.) \pm 0.078 (syst.)$.
%
%
\subsection{Parton model in processes other than DIS}
DIS processes provided the first measurement of parton densities.  We
saw in the previous sections that a combination of the different (weak
and electromagnetic) structure functions with different targets
(proton, neutron, isoscalar nuclei) can be used to establish different
properties of the charged partons such as their spin, charge
assignment etc.  The individual parton densities can be extracted from
the data on $F_2$ only under certain assumptions and one usually fits
a form to these.  The fits are constrained by various sum rules which
are derived on general principles.  However, none of the DIS processes
can give information about the gluon densities.  In the framework of
perturbative QCD, outside the realm of quark-parton model, some
information about gluon densities is obtained by studying the scaling
violation ($q^2$ dependence) of the structure functions.  But no
direct information on the gluons is available from these  processes.  It is
also important to note that since there are, as yet, no theoretical
predictions for either quark/gluon densities, it is imperative to find
processes other than DIS to extract  combination of parton
densities  different from those measured in DIS, to supplement our knowledge.
This, in turn, can help to get a more complete picture of parton
densities.

A better knowledge of parton densities is not only essential to check
our ideas about perturbative QCD but it is also necessary to be
able to make  correct predictions for the cross-sections of
different physical processes expected at the high energy $e$--$p$
or hadron-hadron colliders.  High energy collisions of hadrons
(say $p$--$p$, $\pi$--$p$ or $\pi$--$A$) with each other can be
described in parton model as an incoherent sum of interactions
among partons.  This means that our predictions as to what is
likely to happen at high energies in collisions of these hadrons
will depend on our knowledge of these parton densities.

As an illustration of the above, consider the production of a vector
boson $W^+$ in $p\bar p$ collisions as shown in fig. \ref{fig20}.  The
production cross-section is given by
\be
d\sigma (p\bar p \rightarrow W^+ x) = \int^1_{M^2_W/ S} dx_1
\int^1_{M^2_W / Sx_1} dx_2 u^p (x_1) {\bar d}^{\bar p} (x_2)\;
d\hat\sigma (u\bar d \rightarrow W^+)\bigg|_{\hat s = Sx_1x_2} .
\label{four.32}
\ee
The important point here is that the densities $u^p$ or ${\bar
d}^{\bar p}$ in eq. (\ref{four.32}) are exactly the same as those
extracted from DIS. That the same probability functions are applicable
in all `hard' processes be it DIS or be it (say) $W^+$ production, is
an assumption in the QPM.  In the framework of perturbative QCD this
has been proved quite rigorously for a large number of hard
processes\cite{QCDper}.  So any hard processes in hadron-hadron
collision is computed as an incoherent sum over all the partons where
the individual $2 \rightarrow n$ parton subprocess is convoluted with
the parton distribution functions.  An agreement of measured
cross--section with the predictions made in the QPM provides
consistency checks on our knowledge of these parton densities.
Further, by singling out final states which are sensitive to a
specific parton in the initial state, we can better our knowledge of
parton densities.  For example, a comparison of $W^+$ production
discussed above with $W^-$, can be a good test of the isospin
symmetry of the sea.

This simple QPM picture has been justified theoretically in
perturbative QCD, has been used extensively and has strong
experimental support.  The high energy processes of interest to the
Nuclear physics community are the heavy ion collisions.  The plain QCD
contributions to different final states in these collisions, arising
from the partons in the nuclei, need to be computed correctly before
one can assess the observability of different signals of
Quark-Gluon-Plasma (QGP) formation such as $J/\psi$ suppression,
dilepton production or direct photon production \cite{satz}.  For this
one needs a good knowledge of parton densities inside the nuclei.
This is the topic of the next lecture.
%
%
\subsubsection{Jet Production}
The most common hard process in hadronic collisions is called jet
production.  This arises basically from the scattering of two partons
in the hadrons which produce light partons in the final state.  As
already said in eq. (\ref{four.32}), in QPM the total cross-section at
the hadronic level is obtained by convoluting the parton densities
with the cross--section of this basic $2 \rightarrow 2$ subprocess.
Since this scattering is between pointlike particles, it involves
large momentum transfers and the final state partons are thus produced
at large angles ({\it i.e.}, with large momenta transverse to the beam
direction, $(p_T)$) as opposed to the remaining partons in the two
hadrons which go along the beam direction.  There are in all 8 hard
scattering $2 \rightarrow 2$ subprocesses which will produce light
partons in the final state.
The $q$ or $g$ in the final state does not appear in the detector as
a free gluon or quark due to the phenomenon of `colour confinement'.
They appear in the detector as a `stream' or `jet' of particles
clustered around the original direction of the parton.  As said
earlier, the major feature of QPM is that these jets are predicted to
have large $p_T$ due to the hard scattering of the two partons.
Observation of this large $p_T$ jet production was one of early
confirmation of QPM in processes other than the DIS.

Quantitative information about the parton densities can be obtained
using jet production, however, only in the framework of perturbative
QCD.  The qualitative success of the QPM is formally justified only in
the context of perturbative QCD (pQCD) any way.  Use of jet-production
as a probe of parton densities is a little complicated as both quarks
and gluons in the initial state contribute and over most of the range
of $p_T$ neither quarks or gluons dominate.  In the framework of pQCD
one can show that \cite{PQCD} the large $p_T$ jet production in
hadronic collisions is given by,
\bea
{d\sigma (H_1 H_2 \rightarrow j_1 j_2) \over dp_T dy_1 dy_2} &=&
F^{H_1} (x_1) F^{H_2} (x_2) f (p_T,y_1,y_2) \nonumber \\
&=& x_1 x_2 \left[{4\over9} \sum_q \left(q^{H_1} (x_1) + \bar q^{H_1}
(x_2)\right) + g^{H_1} (x_1)\right] \times \nonumber \\
& &~~~~ \Bigg[{4\over9} \sum \bigg(q^{H_2}
(x_2) + \bar q^{H_2} (x_2)\bigg) + g^{H_2} (x_2)\Bigg] \times \nonumber \\
& &~~~~~f(p_T,y_1,y_2) \;\; ,
\label{four.32P}
\eea
where $y_i = {1\over2} \ell n\left({E_i + P^i_L \over E_i -
P^i_L}\right)$ is the rapidity of the $i$th jet in terms of its energy
$E_i$ and longitudinal momentum $P^i_L$.  The momentum fractions
$x_1(x_2)$ of the hadrons $H_1(H_2)$ carried by the partons are
related to rapidities $y_1,y_2$ and $p_T$ of the jets and
$f(p_T,y_1,y_2)$ is a function approximating the dominant part of all
the subprocess cross-sections ${d\hat\sigma / dp^2_T} (p_1p_2
\rightarrow p_3 p_4)$ where $p_i$ stand for partons.  Measurement of
large $p_T$-jet, triple differential cross-section of
eq. (\ref{four.32P}) by UA-1 collaboration in $p\bar p$ collisions
\cite{Alt} allowed extraction of $F^p(x)$ and its comparison with the
parton densities extracted from DIS, has tested the above picture
quite well.
%
%
\subsubsection{Drell-Yan process}
Another hard process which was studied in the early days of QPM and
does not involve pQCD in the subprocess (apart from its role as a
justification of QPM) is the  production of a massive $\mu^+\mu^-$ pair
with zero $p_T$, via $q\bar q$ fusion \cite{PQCD}.  The hard subprocess,
shown in fig. \ref{fig21}, is
\be
q\bar q \rightarrow \gamma^\star \rightarrow \mu^+ \mu^- \;\; .
\label{four.33}
\ee
The process at the level of hadrons is
\be
H_1 H_2 \rightarrow \mu^+ \mu^- X \;\; .
\label{four.34}
\ee

Let us take the special case when $H_1 = p$, $H_2 = \bar p$ .  Then the
total cm energy available for the subprocess (say) $q^p + \bar q^{\bar
p} \rightarrow \mu^+ \mu^-$ is given by,
\bea
m^2_{\mu^+\mu^-} = \hat s &=& (P_qp + P_{\bar q}\bar p)^2 = (x_1 P^p +
x_2 P^{\bar p})^2 \nonumber \\[2mm]
&\simeq& 2x_1 x_2 P^p \cdot P^{\bar p} \simeq x_1x_2 (P^p + P^{\bar
p})^2 \simeq x_1 x_2 S \;\; ,
\label{four.35}
\eea
where $S$ is the centre of mass energy of the $p\bar p$ system.  The
process shown in fig. \ref{fig20} is an electroweak variant of the
same.  The $\mu^+\mu^-$ pair so produced has no net momentum
transverse to the relative direction of motion of $p$ and $\bar p$, as
the transverse momenta of the initial state partons is negligible in
the QPM.

The cross-section for the Drell-Yan (DY) process obviously reflects
the quark and anti-quark content of both the beam and target.  Since
the cross-section for $\mu^+\mu^-$ pair production is proportional to
$e^2_q$, the information available from its study is essentially the
same as that obtained from a study of DIS.  In the QPM some simple
relations exist between DY cross-sections for a fixed target and
different beams, e.g., the valence quarks in $\pi^+$ are $u$ and $\bar
d$, while those in $\pi^-$ are $\bar u$ and $d$.  Since the valence
quarks in proton are $u$ and $d$, the dominant contribution to DY pair
production with $\pi^+ (\pi^-)$ beam and proton target comes from $d^p
\bar d^{\pi^+} (u^p \bar u^{\pi^-})$ annihilation.  For an isoscalar target
like Carbon nucleus, we know from our discussion in the earlier
section, $d^C = u^C$.  Hence we expect
\be
{\sigma (\pi^+ C) \over \sigma(\pi^- C)} \simeq {1\over4} .
\label{four.36}
\ee
This very basic prediction of QPM, which tests the ideas of valence
quarks, sea-quarks and the  quark-content of $\pi^+/\pi^-/p$ etc.
as given by the static quark-model, was tested quite adequately in the
early experiments and thus played an important role in establishing
the QPM.

Going outside the framework of plain QPM, in the context of
perturbative QCD, one can also compute production of massive
$\mu^+\mu^-$ pair, at large $p_T$, in hadron-hadron collisions.  It
takes place via the hard scattering process,
\be
q\bar q \rightarrow \gamma^\star g \rightarrow g \mu^+\mu^-; ~~qg
\rightarrow q\gamma^\star \rightarrow q\mu^+\mu^-.
\label{four.37}
\ee

As discussed for jet production, the final state light parton appears
as a jet in each case.  The $\gamma^\star$ (equivalently the
$\mu^+\mu^-$ pair) produced in the hard $2 \rightarrow 2$ scattering
subprocess carries a large $p_T$, which is balanced by the jet,
even though the initial state partons have zero $p_T$.  The process at
the hadron level is
\be
H_1 + H_2 \rightarrow \mu^+ + \mu^- + jet + X ,
\label{four.38}
\ee
where the $\mu^+\mu^-$ pair is produced with non-zero $p_T$.  It is clear
from eq. (\ref{four.37}) that the large $p_T$ DY $\mu^+\mu^-$ pair can
yield `direct' information about the gluon content of the hadron.

%
\subsubsection{Direct photon-production}
A related process which can also probe the gluon content of a hadron
is the `direct' photon production \cite{PQCD} via the subprocess
\be
q\bar q \rightarrow g\gamma, ~~~qg \rightarrow q\gamma .
\label{four.39}
\ee
The only difference from eq. (\ref{four.37}) is that the photon in the
final state is a `real' photon; {\it i.e.}, the invariant mass of the photon
here is zero.  The nomenclature `direct' (or prompt) has to do with
the fact that this photon is produced on the short time scale of the
hard scattering process as opposed to the photons which are produced
in the decay of the final state hadrons like $\pi^0$ and hence appear
on a longer (hadronisation) time scale.  Of course the latter are
produced in much more plentiful numbers as the corresponding
cross-sections are much higher.  Therefore, a study of `direct' or
`prompt' photons is experimentally quite challenging.  But
the effort is well worth it, as one can isolate certain kinematic
regions of the final state where the cross-section is dominated by the
gluon content of the colliding particles and hence the measurement
offers an almost `direct' probe of the gluon densities.
%
%
\subsubsection{Heavy quark production}
An even better probe of the gluon densities is provided by production
of heavy quark \cite{PQCD} (charm, bottom etc.) in hadronic collisions
via the subprocesses,
\be
gg \rightarrow Q\bar Q; ~~~q\bar q \rightarrow Q\bar Q\;\; ,
\label{four.40}
\ee
where $Q$ stands for $c/b$ quark.  This final state has a very
distinctive signature and can be separated from the very big
background of jet production (from the same initial states) with
comparative ease.  The corresponding process at the hadronic level is
\be
H_1 H_2 \rightarrow Q\bar Q X\;\; .
\label{four.41}
\ee
This heavy quark production is dominated by gluon densities due to
dynamical reasons that make the corresponding cross-sections larger.
The situation can be further improved (from the point of view the
determination of the gluon densities) by considering photoproduction
of heavy quarks \cite{PQCD} in the process
\be
\gamma h \rightarrow Q\bar Q X \;\; .
\label{four.42}
\ee
In this case the basic subprocess is
\be
\gamma g \rightarrow Q\bar Q X \;\; .
\label{four.43}
\ee
This is just an analogue of the corresponding process for the
hadronic production of heavy flavour,
where a
gluon is replaced by $\gamma$.  Thus a study of photoproduction
(production with incident photons) of heavy quarks can give pretty
good information about the gluon content of the target.

A process with smaller cross-section but a more distinctive signature
is the production of a bound $Q\bar Q$ pair (quarkonium) instead of a
free $Q\bar Q$ pair.  The processes which contribute to the
hadroproduction (production in hadronic collisions) and
photoproduction  of a quarkonium\cite{reya} are the same as those
in eqs. (\ref{four.40}) and (\ref{four.43}) respectively.

If we relax the restriction to the $2 \rightarrow 2$ subprocess
which is inherent to the the QPM and allow $2 \rightarrow 3$
subprocesses (which is justified by pQCD) then we can describe
large $p_T$, photo- and hadro-production of quarkonium in the
above picture ({\it i.e.}, convolution of parton densities with a
subprocess).  Some of the subprocesses are,
\be
q\bar q \rightarrow Q\bar Qg, ~~~gg \rightarrow Q\bar Qg, ~~~\gamma
g\rightarrow Q\bar Qg .
\label{four.44}
\ee

A process which is closely related to the above is the associated
production of a quarkonium with a photon in hadronic collisions, via
the subprocess, e.g.,
\be
gg \rightarrow \bar Q Q \gamma .
\label{four.45}
\ee
This was recently suggested as a probe of gluon densities.  The
cross--sections are quite a bit smaller than hadroproduction of a
quarkonium but, due to the associated photon, is much cleaner for
detection.  It is better than the `direct' or `prompt' photon in
association with jet (eq. (\ref{four.39})), again due to the ease of
discrimination against background.  It also has the advantage of being
`directly' proportional to gluon densities for certain spin-parities
of quarkonia.  In the next section we will discuss how these
processes can be used to glean information about the gluon densities
in the nucleus.

\setcounter{equation}{0}
\section{EMC effect}

%
\subsection{EMC effect : Experimental situation}
As alluded to in the earlier discussion, because of the high
energies involved in  DIS, it was expected that the nuclear
structure function $F_2^{eA}$ should simply be an incoherent sum
of the structure functions of the individual nucleons. This
picture, based on the impulse approximation, depicted in fig.
\ref{fig22} was implicitly assumed in all our discussions of the
parton model and various tests of the parton model using isoscalar
nuclei involved using this approximation. The success of these tests,
albeit qualitative, supports the assumption. However, experiments
which set out to test this quantitatively met with a surprise
\cite{oldemc}.  This experiment compared ${1 \over A} F_2^{\mu A}$
with $F_2^{\mu p}$. The original experiment studied the EMC ratio
($\rho^{EMC} (x)$) defined by
\be
\rho^{EMC}(x) = {1 \over A} {F_2^{\mu A}(x) \over F_2^{\mu p}(x)}\;\;,
\label{five.1}
\ee
as a function of the scaling variable $x$ for $A = {\hbox {Fe}}$. The
deviation of this ratio from unity is a measure of the failure
of the impulse picture. The initial data gave $\rho^{EMC} > 1$
for $x < 0.3 $ and a suppression for $0.3 < x < 0.8 $. Since then
a lot of DIS experiments verified this non-trivial nuclear
dependence of $F_2^{lA}$ ($\rho^{EMC}(x) \neq 1$) for a wide
range of nuclear targets, with a variety of lepton types ($e^- $
beams, $\mu$ beams, $\nu $ beams) and over a wide range of $q^2$
($4 < q^2 < 200 $ GeV$^2) $\cite{emcdata}.  The newer data by
NMC\cite{newreview} probed the effect to very {\it low values
of} $x$ upto $ x = 0.0035$.  Fig. \ref{fig23} shows a collection
of some of these data \cite{pdg,newreview,virchaux}.
The data show the following features :
\begin{enumerate}
\item There is no low $x (x < 0.2) $ enhancement of the
nuclear structure functions which was present in the original
EMC data.
\item There is a definite `shadowing' effect, {\it i.e.},
$\rho^{EMC} < 1$ for $x < 0.05$ even at large values of $q^2$.
The suppression of the nuclear structure function $F_2^{lA} $ w.r.t.
the nucleon structure function $F_2^{lp}$ rises with $x$ for $x <
0.2$. The `shadowing' means that the DIS cross--section per nucleon
is reduced as compared to that for a free nucleon, due to the
presence of the other nucleons in the nucleus. This shadowing effect
depends only weakly  on  $q^2$.
\item For $0.05-0.1 < x < 0.2\; $ $\rho ^{EMC} (x) $ goes slightly
above unity and then falls below 1 for $0.3 < x < (0.8-0.9) $. This
depletion of ${1 \over A} F_2^{lA} $ w.r.t $F_2^{lp} $ is $\sim
10-15 \%$.
\item For values of $x > 0.8$ the EMC ratio goes above unity.
\item The  effect does not show any $q^2 $ dependence.
\item Nor does the effect  show strong dependence on the
mass number $A$ of the target.
\item An experimental measurement of the ratio $ R
= F_L(x) / (2 x F_1(x)) $ does not show any appreciable
nuclear dependence.
\end{enumerate}
Due to the intuitive appeal of the impulse approximation,
observation of this nontrivial nuclear dependence of $F_2^{lA} $
was almost unexpected by theorists except for a
suggestion\cite{SAKRZ} of a possible enhancement of the nuclear
sea--densities. For the same reason, the observation of the EMC effect also
gave rise to a large amount of theoretical activity and a large number
of models for nuclear structure functions. Considering that we do not
as yet have a credible model (let alone a theory) even of the nucleon
structure function, it is clear that all the models proposed to explain
the nuclear structure functions do so only by giving a recipe to
calculate the nuclear structure function in terms of that of the
nucleon. Different models differ in the theoretical ideas about the
effect of the nuclear environment on the parton densities. All these
models of course involve parameters some of which are estimated and
some are usually fitted to reproduce the observed EMC effect. Since the
DIS experiments probe only the quark--parton densities directly, it is
not surprising that all the models agree on the form of the nuclear
quark--parton densities. However, the nuclear gluon densities which are
unconstrained by the DIS data are {\it predictions} of  these different
models and generally differ greatly from model to model. The different
models differ radically in the physics phenomenon they invoke to
explain the EMC effect. Hence, to arrive at the correct theoretical
understanding of the EMC effect, it is essential to be able to
distinguish between the various models. This can be done effectively if
one can extract nuclear gluon densities  and hence measure the gluonic
EMC ratio
\be
\rho_g = {1 \over A} {g^A(x) \over g^p(x)}\;\; .
\label{five.1p}
\ee

There is yet another reason which makes such a determination
imperative.  This has to do with the signals of the
Quark--Gluon--Plasma (QGP) mentioned in the earlier sections. To assess
the observability of any `hard'  signal of QGP formation in
Heavy ion collisions it is absolutely essential to understand the
contributions to the `hard' final state under question,  coming from a
combination of the nuclear dependence of the parton densities and
perturabtive QCD. This of course requires a good knowledge of $\rho_g$
defined above.

The discussions of the last section outlined various hard processes
other than the DIS which can be used to glean information about the
gluonic EMC ratio. Hence, a study of correlation between the
nontrivial nuclear dependence of the structure functions (the EMC
effect) and the $A$--dependence of the different hard processes such as
large $p_T$ jet--production, DY $\mu^+ \mu^-$ pair
production(including large $p_T$ DY), electro- and photo-production of
quarkonia as well as their production in hadronic collisions etc. can
help shed light on the EMC effect. It is worthwhile to ask at this
point whether there exists any evidence of a nontrivial nuclear
dependence for the abovementioned hard processes, before we turn to a
discussion of such correlations. Below we first discuss some of the
models that have been suggested to explain the EMC effect, summarize
the experimental evidence of nontrivial nuclear effects in hard
processes other than the DIS and then examine the implications of
these data for various models of the EMC effect.


\subsection{EMC effect: Theoretical Models}
As mentioned in the earlier sub--section, observation of the till
then unexpected nuclear dependence of the structure functions gave
rise to a large variety of models\cite{newreview} for the EMC effect.
Almost all the models address primarily the region $0.2 < x < 0.8$.
The small $x$ `shadowing' region is interesting and has been recently
the focus of theoretical discussions\cite{shadowing} but will not be
discussed here.  Broadly speaking the models can be divided into
different classes:
%

\subsubsection{Nuclear Physics based models}
 Models based on `conventional' nuclear physics try to explain the
depletion of $F_2^{A}$ in the valence region as being due to the
virtual pions present in the nucleus (as a result of the nuclear
force). The pions can carry a momentum fraction up to $M_\pi /M_p$ and
hence will cause a depletion of quarks in the valence region and also
a low $x$ enhancement. This idea
\cite{llsmith} almost always will give rise to a  enhancement
of the anti--quark content of the nuclear structure function as the
pions contain more valence anti--quarks. The nuclear structure
function is given by
\be
F_2^A(x,q^2) = \int_{x}^{1} dy f_N(y) F_2^N(x/y,q^2) +
               \int_{x}^{1} dy f_{\pi}(y) F_2^{\pi}(x/y,q^2)  \;\; ,
\label{five.3}
\ee
where $f_N$ and $f_{\pi}$ denote the nucleon and the pion distribution
functions in the nucleus.  Free parameters of this model are the
average number of pions in the nucleus and the momentum fraction $\eta$
carried by the pions. Of course the form of the quark--distributions
in the free proton and pion, {\it i.e.}, $F_2^{A} {\rm and} F_2^{\pi} $, are
also an input to the model. The original idea was extended further by
a large number of authors \cite{newreview} including also the effect of
$\Delta 's $ in the nucleus.

Attempts\cite{akulin} were made to calculate  the two abovementioned parameters
in nuclear physics framework by using the measured values of the
nuclear separation energies. There was much excitement initially   as
the original calculations seemed to yield values of $\eta$ and
the average number of pions required by the best fit to the data on
EMC effect. However, since then the issue has been revisited by a lot
of nuclear physicists\cite{newreview} and there is no clear consensus
about the exact size of the nuclear binding contribution to the EMC
effect; but it is fair to say that it can, at the most, account for
only  10-15\% of the EMC effect\cite{brown}. Since the model always predicts an
enhancement of the anti--quark content and also in the low $x$
region,  the experimental information on the nuclear dependence of
hard processes other than the DIS constrain these types of model
rather strongly, as we will see later.

%

\subsubsection{Cluster models}
Several authors conjectured that the nontrivial nuclear dependence of
the structure function can be understood in terms of nucleon clusters
inside the nucleus. The idea is that clusters of nucleons deconfine in
the nucleus and share quarks with each other. So the quarks
belonging to these clusters now occupy the volume of the cluster
rather than that of a single nucleon. Idea of such clusters was
initially suggested
{
to explain the deep inelastic data
from ${^3}He$ in the $q^2$ range $1 < q^2  < 4 {\rm GeV}^2$.
The probabilities of  N--quark cluster formation can be computed
theoretically. A feature of the cluster models is the existence of
large momentum partons in the nucleus. For an N--quark cluster, the
variable $x = q^2/2 M_p \nu $ can take values up to $N/3$. Thus the
fractional momentum $x$ carried by the quark can be greater than
unity in this case.  The cluster models(which involve 6 or 9--quark
clusters) invoke only a partial deconfinement of the quarks in the
nucleus. Furthermore the quark distribution functions in the
free--nucleon and in the N--quark clusters are both input functions.
There exist also another class of models where  the quarks from the
nucleons are postulated to deconfine to the whole nucleus and these
deconfined partons are assumed to form a gas. These `gas' models
then compute the parton distributions of the `deconfined' quarks in
terms of the parameters of the model. Below two representative models
of this kind are discussed.

\vspace{0.3cm}
\noindent{\bf 1) The Gas Model} \\

\noindent The parton densities of a nucleus with atomic number
$A$ are defined in the Gas model\cite{gas} as a sum of two
components.

\begin{equation}
f_{i/A}(x) = (1 - \omega) {\tilde f}_{i/N}(x) + {1\over A}
\sum^A_{r=1} \omega^r (1-\omega)^{A-r} f_{i,r}^{\rm gas} (x;\mu,T)
\;\; .
\label{gasmodel}
\end{equation}
The first term, occurring with a weight $(1 - \omega)$, is that for a
free nucleon parton density after corrections for its Fermi motion inside
the nucleus.  The second component is written in terms of thermal
distributions of momenta at a temperature $T$,
leading to the following functions $f_{i,r}^{\rm gas} (x;\mu,T)$:
\begin{equation}
f_{q,r}^{\rm gas} (x;\mu,T) = {2r_0^3 T^3 \over \pi} \Bigg[\phi ~\ell n(1 +
ze^{-\phi}) + {1\over2} \ell n^2(1 + ze^{-\phi}) + Li_2\bigg({z \over
z+e^\phi}\bigg)\Bigg]
\label{gasq}
\end{equation}
and
\begin{equation}
f_{g,r}^{\rm gas} (x;\mu,T) = {2r_0^3 T^3 \over \pi} \bigg[Li_2(e^{-\phi}) -
\phi ~\ell n(1 - e^{-\phi})\bigg] \;\; .
\label{gasg}
\end{equation}
Here $r_0 = 1.2$ fm, $\phi = Mx/2T$, $z = {\rm exp}(-\mu/T)$, where $M$
is the nucleon mass and $Li_2(x)$ is the Euler dilogarithm function.
Using the constraint on the total baryon number of the nucleus to
eliminate the chemical potential $\mu$, one has two model parameters,
$T$ and $\omega$, for each nucleus.

Using the CDHS parametrisations\cite{cdhs},
\begin{eqnarray}
\label{cdhs}
 F_2^{\nu p}(x)\;&=&\;1.1(1+3.7x)(1-x)^{3.9} ,\nonumber\\[2mm]
 x\sigma_p(x)\;&=&\;0.17(1-x)^{8.54} \; \; ,\\[2mm]
 xf_{g/p}(x)\;&=&\;2.62(1+3.5x)(1-x)^{5.9} \;\; ,\nonumber
\end{eqnarray}
for $F_2^p(x)$ (here $\sigma_p(x)$ is the total sea density) and
the data on $\rho(x)= F_2^A(x)/A F_2^p(x)$
these parameters have been fixed\cite{gas} for many nuclei, including the
ones used for the E772 experiment.  The corresponding $\rho_g (x)$
is then predicted uniquely. They are  summarized  in Table \ref{tabgas}.

It may be mentioned here that, using a different set of structure
functions for proton instead of eq. (\ref{cdhs}) necessitates a
time-consuming re-analysis of the EMC-data to obtain $T$ and $\omega$.
For this reason we have not used more recent parametrisations for the
proton structure function and we are also constrained to use {\em
different} proton structure functions in different models.

\vspace{0.3cm}

{\bf 2) The six-quark cluster model}\\

\noindent The six-quark cluster model\cite{bag} is also a
representative of the two-component models for EMC effect.
In this model, it is assumed that when two nucleons get closer
to each other than a certain critical radius they merge together to
form a six-quark cluster.  By assuming the probability to form higher
clusters to be negligible, the remaining model inputs are the probability of
forming such a cluster and the form of the parton distributions in the
3-quark and 6-quark clusters.  The latter are chosen using the
quark-counting rules and the constraints of i)
normalisation of valence densities (an N-quark cluster has N valence
quarks), and ii) conservation of momentum.  It is further assumed that the
average momentum carried by the sea partons is the same
for the three- and six-quark clusters and
that it is $\sim 0.2$ of that of the gluons.  The forms of
nuclear densities per nucleon in this model are,
\begin{equation}
f_{i/A} (x) = (1 - \epsilon) f_{i,3} (x)
+~ {\epsilon \over 2}~f_{i,6} \left({x \over 2}\right)~~,~~
\end{equation}
where $\epsilon$ is the probability to find a six quark cluster which
increases with $A$ \cite{pirner} and the subscripts denote the cluster
size.  The values of $\epsilon$  used  are given in Table \ref{tabgas}.
\begin{table}
\caption{
Model parameters for nuclei used in E772 experiment for gas model(T,$\omega$)
and six quark cluster model($\epsilon$).}
\vspace{0.5cm}
\begin{center}
\begin{tabular}{|c|c|c|c|}
\hline
{\rm A} & {\rm T (MeV)} & $\omega$ & $\epsilon$ \\
\hline
& & & \\
12 & 54 & 0.069 & 0.112 \\
& & & \\
\hline
& & & \\
40 & 47 & 0.057 & 0.170 \\
& & & \\
\hline
& & & \\
56 & 45 & 0.117 & 0.186 \\
& & & \\
\hline
& & & \\
184 & 42 & 0.132 & 0.230 \\
& & & \\
\hline
\end{tabular}
\end{center}
\label{tabgas}
\end{table}
Specific choices\cite{bag} for the valence density $V(x) = f_{u_V}(x)
+ f_{d_V}(x)$, sea density $S(x)$ and the gluon $G(x)$ for an $N$
quark cluster ($N = 3, 6)$ from Ref.\cite{bag,sulassila} are given by,
\begin{eqnarray} \label{bagden}
x V_N (x)&=& {1 \over B\left(1/2,~2N -2 \right) } \;
N x^{0.5} (1-x)^{2N-3}   \nonumber\\[2mm]
x S_N (x)&=&{N-1 \over 2(4N - 3)}~ (a_N + 1) ~(1 -x)^{a_N} \\[2mm]
x f_{g,N}(x) \equiv x G_N(x)&=&{5(N - 1) \over 2(4N-3)} ~ (c_N +1)
{}~(1-x)^{c_N}~~,~~\nonumber
\end{eqnarray}
with $a_3 = 9,~ a_6 = 11, ~~ c_3 = 7$, and $c_6 = 10$\cite{sulassila}.
Here $B$ is the usual Euler Beta function and
$S_N(x)$ represents the sum of sea quark densities over all flavours.
With a further assumption of
$\bar f_{\bar s,N} (x) = {1 \over 2}~f_{\bar u,N} (x) = {1 \over 2}
{}~f_{\bar d,N} (x)$,
the $\bar u$ distribution for the $N$-quark cluster is given by
$f_{\bar u,N} (x) = {1 \over 5}~ S_N (x)$.
The six quark--cluster model described is an updated version
(in their choice of the parameters $\epsilon $ and the input
proton densities) of the original six quark--bag model\cite{oldbag}.


\subsubsection{Rescaling models}
A large class of models\cite{resc,GGB,G} try to model the effect of
the nuclear medium in terms of the different length scales associated
with the nucleon and the nucleus. The precise fashion in which it is
done varies from model to model.  In the $q^2$ rescaling models the
nuclear parton densities at a scale $q^2$ are obtained from the parton
densities in a proton at the same $q^2$ by evolving them to a scale
$\xi_A q^2$, {\it i.e.}, the nuclear parton density per nucleon
$f_{i/A} (r, q^2)$ is given by
\begin{equation}
f_{i/A}(x, q^2) = f_{i/p} (x, \xi_A q^2) \;\; .
\end{equation}
Here  the results for the rescaled nuclear densities as obtained in Refs.
\cite{GGB,G} are shown  where $\xi_A = A^{2/3}$ where  the starting
nucleon parton densities are taken\cite{G} to be a parametrisation of
the EMC Deuterium data at $q^2 = 20~{\rm GeV}^2$.  It should be added
here that the rescaling models along with the nuclear physics based
models always tend to enhance the nuclear structure function at small
values of $x$.

There exist also hybrid models \cite{hybrid} which combine the ideas
of both, the rescaling and the cluster models. In these types of
models the nuclear parton densities(per nucleon) are given by,
\begin{equation}
f_{i/A}(x,q^2) = f_{i/p} (x/\alpha_{A},\xi_A q^2) \;\; .
\end{equation}
The two parameters are introduced to model the change in the scale in
the nuclear case as well as the the possibility of cluster formation.
The two parameters are then fitted to reproduce the data on
$\rho_{EMC}$. Of course the fitted values of the parameters depend on
the choice of the parametrisation for the parton densities in the
proton.  The values we obtained\cite{rosou} are $$
\alpha_A = 0.012,\;\; \xi_A = A^{0.4}.
$$

%

\subsection{Comparisons of different model predictions for the gluon density}
In fig. \ref{fig24} are shown the fits to the data
obtained in, e.g., the rescaling and the hybrid models. This
makes the point that the fits to the data on $\rho_{EMC} $ in
different  models are all of the same quality and all have
similar quark parton distributions. Fig. \ref{fig25} shows the
expectations for the gluonic EMC ratio $\rho_g$ of eq.
(\ref{five.1p}), for some of the models of the EMC effect
discussed above. It should be noted here that the different fits
to the data use different parametrisations for the proton
densities and hence it is more meaningful to compare the
predictions for $\rho_g$ for  the different models of the EMC
effect rather than the  absolute gluon densities.  The figure
shows clearly that the differences in the predictions of the
various models are indeed sizable.


\subsection{$A$ dependence of the hard processes}
\subsubsection{Experimental situation}
Even before the EMC effect was discovered \cite{oldemc}, there
existed a few experiments which reported an anomalous nuclear
enhancement of cross--sections for large--$p_T$ particle/jet
production \cite{oldjet} and DY $\mu^+ \mu^-$ pair production
\cite{CIP} with nuclear targets.  The experiments parametrised the
cross--section for nuclear targets with a beam $B$ as,
\be
{1 \over A} \sigma^{BA}  = A^{\alpha -1} \sigma^{Bp}\;\; .
\label{five.2}
\ee
Similarly the ratio of the differential cross--sections, e.g.,
$\displaystyle {d\sigma \over dp_T}$ is parameterised in terms of
$\alpha (p_T)$.  Again for no nontrivial nuclear dependence we must
have $\alpha = 1$. A deviation of $\alpha$ from unity signals an
anomalous nuclear effect. It should be noted here that due to the
heavy nuclear targets that are used, a small deviation from unity
for $\alpha$ means a rather large difference between the (per
nucleon) cross--section with the nuclear and the free nucleon target.
The initial experiments \cite{oldjet} reported indeed very large
values of $\alpha$ increasing with the $p_T$ values reaching 1.8 at
the highest $p_T = 6$ GeV. The rise was seen for both p and $\pi^+$
beams. However the jet--like character of these data were
questionable. In case of the non--jet like data, a large nuclear
enhancement can also be caused by final state multiscattering effects,
which can conceivably be larger for nuclear targets as opposed to the
nucleon target. Recently there has been more data on the nuclear
dependence of large $p_T$ jet production \cite{newjet}. The data
show the following features:
\begin{enumerate}
\item The data do show an anomalous nuclear enhancement,
{\it i.e.}, $\alpha$ values bigger than unity. The enhancement
increases with the transverse energy $E_T$, which is essentially a
measure of the tranverse momentum $p_T$.
\item The `jettier' events give smaller values of $\alpha$
than the non-jet-like events for the same value of $E_T$.
\item The `jets' seen carry a large fraction of the available c.m.
energy, $\sqrt{S}$.
\end{enumerate}
The second feature above is consistent with the much larger values of
$\alpha$ quoted by the earlier experiments. However, in the case of
the newer data, due to better characterisation of the jet-like
nature, multiscattering cannot be invoked to explain the anomalous
nuclear enhancement and hence has to be interpreted as a reflection
of the enhancement of $F_2^{eA}$ over $F_2^{ep}$ in certain $x$ regions.
Recently, some data on large $p_T$ particle production
has become available which shows only a modest rise of the
cross--sections with the mass number $A$.

High statistics data on the $A$ dependence of the $J/\psi$ and
$\Upsilon$ production \cite{kasta,fnal1,fnal2} as well as on the
large $p_T$ DY ($\mu^+\mu^-$ pair) production \cite{fnal3} have
become recently available.  Experimental data indicate a nuclear
suppression of the total cross--section.  Differential
cross--sections are studied in two variables: $p_T$ and $x_F$. $x_F$
is given by $ {2 P_L / \sqrt{S}}$, where $P_L$ is
the longitudinal momentum of the quarkonium or the $\mu^+ \mu^-$ pair
and $\sqrt{S}$ is the total c.m. energy. $\alpha(p_T)$ and
$\alpha(x_F)$ show a modest $p_T$ and $x_F$ dependence respectively.
For the large $p_T$ DY pairs the $\alpha$ values are mostly in the
vicinity of unity as opposed to the very large nuclear enhancement
reported by earlier experiments \cite{CIP}.

As discussed in the earlier section, yet another interesting probe of
the gluon  densities is the photo- and electro-production of
quarkonia. The early  FNAL and SLAC data\cite{suppjpsi} reported
suppression of the lepto-/photo-production cross--sections of the
$J/\psi$ as opposed to the EMC data\cite{EMCjpsi} which reported an
enhancement. The situation was clarified by the latest NMC
experiments\cite{NMCjpsi} where a modest $A$--dependence of the
differential distributions of the $J/\psi$ production has been reported.

Thus the experimental studies of the nuclear dependence of the
different hard processes show quite different behaviours, viz.,
\begin{enumerate}
\item  Large $p_T$ jet production shows considerable nuclear enhancement.
\item  Hadronic quarkonium production shows a suppression for nuclear
targets.
\item The DY $\mu^+ \mu^-$ pair production and large $p_T$ particle
production show a very modest (almost nil) nuclear dependence.
\end{enumerate}
{}From our discussion in the earlier section we know that the dominant
subprocesses in each of the above hard processes involve different
initial state partons. Also as a result of the different kinematical
conditions the different experiments probe parton densities at
different values of $x$. Hence a demand that a given model  of nuclear
parton densities explain all these data on the nuclear dependence of
the hard processes consistently can indeed constrain these models
considerably and help us discriminate among them.


\subsubsection{Comparison of the model predictions with data}
In this section we present some of the model predictions for the
$A$--dependence of the different hard scattering processes and compare
it with the abovementioned data.  It should be added here that for
consistency, one has to use different proton densities while
calculating predictions of the different models of the EMC effect for
the nuclear dependence of various hard scattering processes.


\vspace{0.3cm}
{\bf 1)  A dependence of jet production}\\

\noindent The first process we consider is  large $p_T$
particle and jet production with nuclear targets. As mentioned
in the description of the data, the interesting features of the
data are that the high--$p_T$ jets seen carry a rather large
fraction of the total c.m. energy, $\sqrt{S}$. The large value of
$\alpha$ implies therefore that the nuclear gluon distributions
are somehow harder compared with that in a nucleon. This points
towards cumulative or cluster effects which predict an extension
of the nuclear parton densities beyond $x_{bj} > 1.0$ as opposed
to the rescaling models. It can be proved on quite general
grounds \cite{rosouprd}, using simply the experimental
information on the signature of $d \rho^{EMC} /dx$, that
$d\alpha / dp_T$ will always be less than 0 unless the nuclear
structure functions extend into the region beyond $x_{bj} = 1$,
the so called cumulative region. These qualitative expectations
are indeed borne out by a comparison of the model predictions
with the data. Table \ref{tabalph} shows the data  along with the
predictions of different models for the EMC effect.

\begin{table}
\caption{
Expected values of $\alpha$ in different models compared with the data
\protect\cite{newjet}.}
\vspace{0.5cm}
\begin{center}
\begin{tabular}{|c|c|c|c|c|c|}
\hline
& {\rm Data} &{\rm Gas}&{\rm Six--Quark}&{\rm Hybrid}&{\rm Rescaling}  \\
{\rm Config.}& Ref. \cite{newjet} &Ref. \cite{gas}& Ref. \cite{oldbag}&
Ref. \cite{hybrid}& Ref. \cite{resc}  \\
\hline
$E_T > 15 $ {\rm GeV} &  &  &  &  & \\
$P > 0.8 $ & $1.14 \pm 0.02 $& 1.16 & 1.10 & 1.03  & 0.98  \\
{\rm (Jet--like)}  &  &  &  &  & \\
{\rm (with P cut)} &  &  &  &  & \\
\hline
$E_T > 18 $ {\rm GeV} & & & & & \\
{\rm Jet--like} & $ 1.45 \pm 0.01 $ & 1.47  & 1.16  & 1.06 & 0.98\\
{\rm  without P cut} & & & & & \\
\hline
\end{tabular}
\end{center}
\label{tabalph}
\end{table}

A cut on the planarity P helps to choose jet--like events for events
with lower $E_T$. For larger values of $E_T$ the jet--like nature of
the events is clear and no such cut is required.  As one can see, the
rescaling model fails to give an enhancement of the jet
cross--sections. Furthermore, the predicted value of $\sigma^A$ fails
even to show the $A^\alpha$ behaviour.  The cluster models
\cite{gas,oldbag,hybrid} all give values of $\alpha$ bigger than 1.
Stronger the cumulative effects higher are the values of $\alpha$
predicted. Here it is worth pointing out that we had used the older
version of the six--quark bag model.  We see  that  these
data already seem to prefer cluster models over the rescaling type
models of the EMC effect.

\vspace{0.3cm}
{\bf 2) Nuclear dependence of the large $p_T$ quarkonia  and
$\mu^+ \mu^-$ pair production}\\

\noindent Next we discuss a comparison of the data on
hadroproduction of large $p_T$ quarkonia ($J/\psi \; {\rm and }
\; \Upsilon$) and $\mu^+\mu^-$ pairs with model predictions. We
choose here  FNAL data \cite{fnal1,fnal2,fnal3} for its high
statistics, although comparisons with the earlier data with pion
beams\cite{kasta} do exist\cite{sriro,souraj}.

The E772 experiment has provided data for the ratio
\begin{equation}
R^{J/\psi}(p_T) = {d\sigma(pA \rightarrow J/\psi X) \over dp_T} \biggm/
                   A {d\sigma(pp \rightarrow J/\psi X)\over dp_T}
\label{rjpsiexp}
\end{equation}
with an $x_F$-cut of $0.15 \le x_F \le 0.65$ on the $J/\psi$'s, while
for the $\Upsilon$-production cross sections, they chose to present only
$\alpha(p_T)$, where
\begin{equation}
 {d\sigma(pA \rightarrow \Upsilon X) \over dp_T} = A^{\alpha(p_T)}
 {d\sigma(pp \rightarrow \Upsilon X) \over dp_T}
\label{rupsexp}
\end{equation}
with a corresponding $x_F$-cut for $\Upsilon$ of $-0.2 \le x_F \le 0.6$.
The nuclei used were carbon, calcium, iron and tungsten.  One can compute
each of the individual $p_T$-distributions in eqs.\,
(\ref{rjpsiexp}-\ref{rupsexp}) incorporating these $x_F$-cuts. However
one needs to use a specific model for the quarkonium formation.
Fig. \ref{jpsisld} exhibits the results for a specific model of
hadronisation of the quarkonium,  for all the four nuclei along with
the corresponding data from the E772 collaboration. One sees that for
the lighter nuclei both the two-component models, namely, the gas model
and the six-quark cluster model, describe the data rather well.  For
the tungsten nucleus, however, {\em none} of the models seems to be
in agreement with data.

Fig. \ref{upsalph} shows a comparison of the model predictions for
$\alpha$ values for $\Upsilon$-production with the E772-data.  One sees
a similar general agreement for the gas model and the six-quark cluster
model as for $J/\psi$-production at moderate values of $p_T$.  At the
largest $p_T$, however, the E772-data rise too sharply compared to any
model and could possibly indicate that these models tuned to earlier
large $x$-data have to be better tuned to perform well in the small
$x$-region.

The discrepancy at large $p_T$ values for the  $\Upsilon$-data and
for the  tungsten target for the $J/\Psi$ production do
expose the inadequacy of all the three models of the EMC effect and
the corresponding parametrisation of the nuclear parton densities
considered here but the general agreement in other cases tells us that
the data can indeed be described in terms of the structure function
effects in general and the data are accurate enough to allow
discrimination between different models of the nuclear structure functions.

The proton--induced dimuon pair production was studied over a wide
range of $x_F$ and $p_T$ values.  The data on the ratio of the {\em
integrated} dimuon yield for different nuclei were compared with
theoretical predictions, obtained by using the $q\bar q$ annihilation
process, for various models of the EMC effect.  It seemed\cite{fnal3}
to rule out the 6-quark cluster model\cite{bag}.  However, a later
comparison\cite{sulassila} with an improved version of the model,
showed that this model too can be consistent
with the information on the ratio of the integrated dimuon yields.

Experimental information \cite{fnal3} is also available for the ratio
\begin{equation} \label{rdyexpt}
R^{DY} = {{d\sigma^{DY} \over dp_T} ~ (p A \rightarrow \mu^+\mu^-
X) \biggm/ {d\sigma^{DY} \over dp_T} ~ (p p \rightarrow \mu^+\mu^- X)}
\;\; ,
\end{equation}
where ${d\sigma^{DY}/dp_T}$ is the differential $DY$
cross section integrated over the continuum region (avoiding the
resonances) $4 < M_{\mu^+ \mu^-} < 9~{\rm GeV}$ and $M_{\mu^+\mu^-}
\geq 11~ {\rm GeV}$, with $x_F > 0$.

Fig. \ref{dimu} exhibits the results of a computation for the four
different nuclei and the three different models with the corresponding
data.  Again we see that, similar to the case of resonance production,
the general trends of the data are well described by the model
predictions for the gas model and the 6-quark cluster model.

Thus in conclusion we see that already the available experimental
information on quarkonia, dimuon and large $p_T$ jet production
indicate that two component models of the EMC effect seem to be
preferred by the data.  Further experiments with direct photon
production with nuclear targets \cite{sousri} or  associated production of
$J/\Psi$ and $\gamma$\cite{srirajro} will help in this direction even
more. But what is important to note that it is possible now to tune the
nuclear gluon densities in different EMC models using the data already
available. This should go a long way in helping us understand the
physical origin of the EMC effect as well as help us estimate the QCD
backgrounds to  the `hard' signals of QGP formation even better.

\newpage
\begin{figure}[hp]
\begin{center}
 Figure Captions
\end{center}
\end{figure}
\begin{figure}[hp]
\caption{Electromagnetic scattering process $eA \to eA $ along
with the four momenta assignment for the various particles that
are involved.}
\label{fig1}
\end{figure}
\vspace{-0.5cm}
\begin{figure}[hp]
\caption{Kinematics of the elastic scattering process $eA \to eA
$ in the laboratory frame.}
\label{fig2}
\end{figure}
\vspace{-0.5cm}
\begin{figure}[hp]
\caption{Scattering of an electron $e^-$ from the nuclear charge
distribution.~~~~~~~~~~~~~}
\label{fig3}
\end{figure}
\vspace{-0.5cm}
\begin{figure}[hp]
\caption{Feynman diagram for the elastic scattering process $e p
\to e p $ for a pointlike proton.}
\label{fig4}
\end{figure}
\vspace{-0.5cm}
\begin{figure}[hp]
\caption{The form--factor $G^p_M(q^2)/\mu{_p}$  as measured in SLAC
experiments as a function of $q^2$ (data taken from Ref.
\protect\cite{kirk}). The solid line is the curve $
\left( 1 + q^2/0.71({\rm GeV}^2)\right)^{-2}$.}
\label{fig5}
\end{figure}
\vspace{-0.5cm}
\begin{figure}[hp]
\caption{Kinematics of the inelastic scattering process $ep \to eX$.~~~~~~~
{}~~~~~~~~~~~~~~~~~~~~~~~~~~~}
\label{fig6}
\vspace{-0.5cm}
\end{figure}
\begin{figure}[hp]
\caption{Allowed region in the $q^2 - \nu $ plane for the
elastic, quasi--elastic and inclusive inelastic scattering.}
\label{fig7}
\end{figure}
\vspace{-0.5cm}
\begin{figure}[hp]
\caption{$\tilde W_2 = M^2_p \; W_2(\nu, Q^2)/\pi \alpha$
 as a function of $q^2 {\rm and}\;\nu$ (Figure taken from Ref.
\protect\cite{smith}).  The details of
the source of the data etc.  to be found in Ref. \protect\cite{smith}.}
\label{fig8}
\end{figure}
\vspace{-0.5cm}
\begin{figure}[hp]
\caption{Data on $\nu W_2^{ep}$ as a function of $q^2$ for
$\omega = {1 / x} = 1/4$ (taken from Ref. \protect\cite{DIS}).
Different data correspond to different scattering angles
as indicated in the figure.}
\label{fig9}
\end{figure}
\vspace{-0.5cm}
\begin{figure}[hp]
\caption{Parton model picture of the DIS scattering of an $e^-$
off a proton target.~~~~~~~}
\label{fig10}
\end{figure}
\vspace{-0.5cm}
\begin{figure}[hp]
\caption{DIS scattering of a $\nu$ off a proton target.~~~~~~~~~~~~~~~~
{}~~~~~~~~~~~~~~~~~~~~~~~~~~~~}
\label{fig11}
\end{figure}
\vspace{-0.5cm}
\begin{figure}[hp]
\caption{Schematic drawing of $F_2$ as a function of x for
different values of $q^2$ in the range $ 0.01 - 200 \; \rm
{GeV} ^2 $ (taken from Ref. \protect\cite{close}).}
\label{fig12}
\end{figure}
\clearpage
\begin{figure}[hp]
\caption{Data on $e-\alpha $ scattering  at $q^2 = 0.08 \;\rm
{GeV}^2$ (taken from Ref. \protect\cite{hofstadter}). A is the
elastic peak for $\alpha$ particle while the elastic proton peak
is shown by the dashed line which corresponds to $x =
q^2/2m_{\alpha}\; \nu \simeq  0.25$. The portion BCDE indicates
the momentum distribution of the nucleons in the $\alpha$ particle.}
\label{fig13}
\end{figure}
\vspace{-0.5cm}
\begin{figure}[hp]
\caption{Parton model.~~~~~~~~~~~~~~~~~~~~~~~~~~~~~~~~~~~~~~~~~~~~~~
{}~~~~~~~~~~~~~~~~~~~~~~~~~~~~~~~~~~~~~~~~~}
\label{fig15}
\end{figure}
\vspace{-0.5cm}
\begin{figure}[hp]
\caption{Impossibility of the backward scattering for the
helicity preserving interaction.~~~~~~~~~~~~~~~~~~~~~~~~~~~~~}
\label{fig16P}
\end{figure}
\vspace{-0.5cm}
\begin{figure}[hp]
\caption{Data on ${{F_2^{\mu n} (x) } \over {F_2^{\mu p} (x) }}$ as a
function of $x$ (data taken from Ref. \protect\cite{plot}).~~~~~~~~~~~~~~}
\label{fig17}
\end{figure}
\vspace{-0.5cm}
\begin{figure}[hp]
\caption{The ratio R of eq. (\protect\ref{four.7n}) as a
function of $q^2$ (taken from Ref.
\protect\cite{rmp}).~~~~~~~~~~~~~~~~~~~~~~~~~~~~~~~~~}
\label{fig16}
\end{figure}
\vspace{-0.5cm}
\begin{figure}[hp]
\caption{Ratio of $F_2^{\nu Fe}$ from the CCFR and CDHSW
experiment to $F_2^{\mu A}$ from different $\mu$ DIS
experiments. The $\mu $ structure functions are normalised so
thata ratio of unity means mean square charge of $5\over 18$
(figure taken from Ref. \protect\cite{neu}). See
Ref.\protect\cite{neu} for more details.}
\label{fig18}
\end{figure}
\vspace{-0.5cm}
\begin{figure}[hp]
\caption{Data on $F_2^{\mu p} - F_2^{\mu n}$ from NMC as a function
of $x$ (data from Ref. \protect\cite{plot}).~~~~~~~}
\label{fig19}
\end{figure}
\vspace{-0.5cm}
\begin{figure}[hp]
\caption{Parton model picture of the $W^+$ production in $p \bar
p $ collisions.~~~~~~~~~~~~~~~~~~~~~~~~~}
\label{fig20}
\end{figure}
\vspace{-0.5cm}
\begin{figure}[hp]
\caption{DY process of production of a $\mu^+ \mu^-$ pair
production in hadronic collisions.~~~~~~~~~~~~~~~~~~~~}
\label{fig21}
\end{figure}
\vspace{-0.5cm}
\begin{figure}[hp]
\caption{Impulse approximation for the nuclear DIS scattering.~~~~~~~~~~~~~
{}~~~~~~~~~~~~}
\label{fig22}
\end{figure}
\vspace{-0.5cm}
\begin{figure}[hp]
\caption{Compilation of the data on EMC effect (from Ref.
\protect\cite{pdg}). Details of the data available in Ref.
\protect\cite{pdg}}
\label{fig23}
\end{figure}
\vspace{-0.5cm}
\begin{figure}[hp]
\caption{Fits to the data on $\rho_{EMC}$ in some  models of the
EMC effect \protect\cite{rosou}.~~~~~~~~~}
\label{fig24}
\end{figure}
\vspace{-0.5cm}
\begin{figure}[hp]
\label{fig25}
\caption{The ratio $\rho_g$ of eq. \protect\ref{five.1p} as a function
of $x$ for some models of the EMC effect \protect\cite{srirajro}.}
\end{figure}
\clearpage
\begin{figure}[thp]
\caption{E772 data (inverted filled triangles) on the ratio
$R^{J/\psi}$ of Eq. \protect\ref{rjpsiexp} compared with the
predictions  for the gas model (squares), six-quark cluster
model (circles) and the rescaling model (open triangles) of the EMC
effect \protect\cite{roraj}.}
\label{jpsisld}
\end{figure}
\vspace{-0.5cm}
\begin{figure}[thp]
\caption{E772 data on $\alpha (p_T)$ of eq.
\protect\ref{rupsexp} compared with predictions of the three
different models of the EMC effect mentioned in the text.
 Notation is same as in Fig. \protect\ref{jpsisld} \protect\cite{roraj}.}
\label{upsalph}
\end{figure}
\vspace{-0.5cm}
\begin{figure}[thp]
\caption{E772 data on the ratio $R^{DY}$ of Eq. \protect\ref{rdyexpt}
compared with predictions of the three models of the EMC effect.
Notation is same as in Fig. \protect\ref{jpsisld} \protect\cite{roraj}.}
\label{dimu}
\end{figure}

\appendix
\newpage
\section {Notations used}
%

\noindent If $A$ and $B$ are two four vectors given by $A \equiv
(\vec{a},ia_o)$, $B \equiv (\vec{b},ib_o)$ then,
$$
A \cdot B = \vec a \cdot \vec b - a_0  b_0 = \vec a \cdot \vec b + a_4 b_4.
$$
The $\gamma $ matrix algebra is given as
\bea
\left\{ \gamma_\mu,\gamma_\nu \right\} & = & 2\; \delta_{\mu \nu} \nonumber\\
 \gamma_\mu^{{}^{ \dag}} & = & \gamma_\mu \nonumber\\
\gamma_\mu^2  & = & 1 \nonumber \\
\gamma_5 = \gamma_1 \gamma_2 \gamma_3 \gamma_4 & = &{1 \over 4!}
\epsilon_{\mu \nu \rho \sigma} \gamma_\mu \gamma_\nu \gamma_\rho \gamma_\sigma
\nonumber\\
\left\{ \gamma_5,\gamma_\nu \right\} & = & 0 \;\; ; \;\; \mu = 1,2,3,4
\nonumber\\
 \gamma_5^{{}^{\dag}} &=& \gamma_5\nonumber \\
\sigma_{\mu \nu} & = & {1 \over 2i} \left[ \gamma_\mu, \gamma_\nu\right]=
-i\; \gamma_\mu \gamma_\nu (\mu \not = \nu) \nonumber
\eea
The Dirac equation for a particle of mass $m$  is given in this metric by
$$
(\gamma_\mu \partial_\mu + m) \Psi = 0.
$$
In momentum space this becomes:
$$
(i\not\!\!P + m) u(P) = 0\;\; {\rm where }\;\; \not\!\!P = \gamma_\mu P_\mu
$$
P stands for the four--momentum of the particle and $u(P)$ is the
free particle spinor.

In this metric we have
$$
\sum_s u_\alpha(P,s) \bar u_\beta(P,s) = (-i \not\!\!P + m)_{\alpha \beta}.
$$
The trace theorems in this metric are given by:
\bea
Tr (\gamma_\mu \gamma_\nu) & = & 4\; \delta_{\mu \nu} \nonumber\\
Tr \left[\gamma_\mu \gamma_\nu \gamma_\rho \gamma_\sigma\right] & =&
4\; \left[\delta_{\mu\nu} \delta_{\rho \sigma} + \delta_{\mu\sigma}
\delta_{\nu \rho} - \delta_{\mu \rho} \delta_{\nu \sigma} \right]\nonumber\\
Tr \left[ \gamma_\mu \gamma_\nu \gamma_\rho \gamma_\sigma \gamma_5 \right]
& = &4\; \epsilon_{\mu \nu \rho \sigma} \nonumber \\
Tr \left[\gamma_1 \cdots \gamma_n\right] & = & 0\;\; ( n\;\; {\rm odd} )
\nonumber \\
Tr \left[\gamma_5 \gamma_\mu \cdots \right] & = & 0\;\; {\rm if}\;\;
\gamma_\mu \cdots\;\;\;{\rm has}\;\; {\rm less } \;\; {\rm than}\;\; 4\;\;
\gamma\;\; {\rm matrices}\nonumber
\eea
One more relation required in calculation of $(|{\cal M}|^2) $ is
$$
\gamma_4 {\left[ (\gamma \cdot A) (\gamma \cdot B).... (\gamma
\cdot L) \right]}^{{}^{\dag}} \gamma_4 = (-\gamma.L) \cdots (-\gamma \cdot A);
$$
where $A,B \cdots L$ stand for four vectors representing four momenta.

\newpage
%
\section{ Problems }
\begin{itemize}
\item[1] What is the energy of the probing $e^-$ beam that will be
required to probe the structure at a distance scale (say)
$10^{-17}~m$?
\item[2] Calculate the form factors $F(Q^2) $ for the following
charge distributions: 1) $\rho (R) = \delta^3(\vec R)$, 2) $\rho
(R) = {m^2\over 4 \pi} exp(-m R)/R $ and
3)$ {m^3 \over 8 \pi} exp(-m R)$
where $ m  > 0$.
\item[3] Show by explicit calculation that the factor
$\cos^2(\theta /2) $ in eq. (\ref{one.27}) corresponds to the
helicity non-flip amplitude.
\item[4] Calculate the expression for the differential
cross--section $d \sigma \over d \Omega $ for the electromagnetic
scattering of an electron incident on a target which is
\begin{enumerate}
\item[1)] spin 0, pointlike particle
\item[2)] spin 1/2, pointlike particle
\item[3)] spin 1/2 particle which is not pointlike.
\end{enumerate}
The expression for the cross--section is given by eq. (\ref{two.4})
The matrix element is given by eq. (\ref{two.3}) where the
electromagnetic current of the electron is given by eq.
(\ref{two.8}), and the electromagnetic current of the target in cases
2 and 3 are given by eqs. (\ref{two.8}) and (\ref{two.11})
respectively, whereas the electromagnetic current for the case 1) of
a spinless, pointlike proton is given  by $$ J_\mu^p = i q_p (P_4 +
P_1)_\mu $$ where all the  notations are as given in the lecture.
The answers are given by eqs. (\ref{two.5p}),(\ref{two.5}) and
(\ref{two.14}).
\item[5] Show that considerations of gauge invariance and
Lorentz invariance restrict the form of $H^{\prime \prime}$ as given
by eq. (\ref{three.37}), by following steps analogous to those used in
deriving eq. (\ref{three.18}). Using this, show that the expression for
the differential cross--section is given by eq. (\ref{three.38}).
\item[6] Calculate the expressions for $F_2^{\nu  p}$ and $
F_3^{\nu p} $ given that,
\begin{eqalignno}
\frac{d \sigma}{dy} ( \nu_{\ell}\; q \rightarrow \ell\ q') & =
\frac{G_F^2 s}{\pi} ,\nonumber\\
\frac{d \sigma}{dy} (\nu_{\ell}\; \bar q \rightarrow \ell\ \bar q`)
&= {G_F^2 s \over \pi}  (1-y)^2 .\nonumber
\end{eqalignno}
Assume that the Callan-Gross relation given by eq. (\ref{four.6})
is satisfied for the Neutrino structure functions $F_2^{\nu\;p}$ and
$F_1^{\nu\;p}$ and use the expression for the double differential
cross--section for the inelastic $\nu p$ scattering given by eq.
(\ref{three.39}).
\item[7] Show that in parton model, for an isoscalar nuclear target
$$
{F_2^{\ell A} \over F_2^{\nu A}} = {5 \over 18} + {1 \over  18}
\left[ \frac{4\; K(x)} {u_v(x) + d_v(x) + 4\;K(x)} \right],
$$
where $K(x)$ is the SU(3) symmetric sea densities defined in eq.
(\ref{four.16}).
\item[8] Derive eq. (\ref{four.26}).
\end{itemize}


\newpage
\begin{thebibliography}{120}
\bibitem{qmodel} M. Gell-Mann, Phys. Lett. {\bf 8},214 (1964).
\bibitem{DIS} For a summary of the early DIS results see, e.g.,
J.I. Friedman and H.K. Kendall, Ann. Rev. Nucl. Part. Sci. {\bf 22},
203 (1972) and references therein;\\
G.B. West, Phys. Rep. {\bf 18C}, 264 (1975) and references therein.
\bibitem{bjpaschos} J.D. Bjorken  and E.A. Paschos, Phys. Rev.
{\bf 185}, 1975 (1969).
\bibitem{feynman}
R.P. Feynman, Phys. Rev. Lett. {\bf 23}, 1415 (1969).
\bibitem{gross} D.J. Gross and F. Wilczek, Phys. Rev. Lett.
{\bf 30}, 1343 (1973); Phys. Rev. {\bf D8}, 3633 (1973);\\
H.D. Politzer, Phys. Rev. Lett. {\bf 30}, 1346 (1973).
\bibitem{QCD} M. Gell-Mann, Acta Phys. Austriaca, Suppl.{\bf
IIX}, 733(1972);\\
M. Gell-Mann and H. Fritzsch,  $XVI^{th}$ {\it Int. Conf. on
High Energy Physics}, Batavia, {\bf II}, 135 (1972);\\
M. Gell-Mann, H. Fritzsch and H. Leutweyler,  Phys. Lett. {\bf
B47},365 91973);\\
For a recent review  giving a persepective of QCD see, e.g.,  F. Wilczeck,
Proceedings of workshop {\it QCD 20 years later}, {\bf I}, 16, ed. P. Zerwas
(Aachen 1992).
\bibitem{QCDper} For a review of how parton model can be
understood in the context of pQCD see, e.g., \\
G. Altarelli, Phys. Rep. {\bf 81C}, 1 (1982);\\
E. Reya, Phys. Rep. {\bf 69C}, 195 (1981).
\bibitem{oldemc} EMC Collaboration, J.J. Aubert et al., Phys. Lett {\bf
B~123}, 275 (1983).
\bibitem{newreview} For a recent review of the data on and models of the EMC
effect see, e.g.,  M. Arneodo, Nuclear Effects in Structure Functions,
CERN-PPE/92-113.
\bibitem{myreview} R.M. Godbole, in: Frontiers in particle physics,
ed. Z. Ajduk, S. Pokorski and A.K. Wroblewski (World Scientific,
Singapore, 1990) p.483.
\bibitem{(g-2)cal} T. Kinoshita and W.B. lindquist, Phys. Rev. Lett. {\bf
47},1573 (1981).
\bibitem{pdg} Review of particle properties, part II, Phys. Rev.
{\bf 45} (1992).
\bibitem{rutherford} E. Rutherford, Philos. Mag. {\bf 21}, 669 (1911).
\bibitem{hofstadter} R. Hofstadter, Rev. Mod. Phys. {\bf 28}, 214
(1956) and references therein.
\bibitem{kirk} P. N. Kirk et al., Phys. Rev. {\bf D8}, 63 (1973).
\bibitem{smith} Field Theory in  Particle Physics, B. de Wit and
J. Smith, (North Holland).
\bibitem{halzen} For a description of standard model and details
see e.g. : Quarks and Leptons, F.~Halzen and A. D.  Martin, (John
Wiley, New York, USA)
\bibitem{close} For a review  of the early developments in parton model
see ,e.g., An Introduction to Quarks and Partons, F.E. Close, (Academic
Press, London.)
\bibitem{virchaux} For a recent summary of the DIS data see,
e.g, M. Virchaux, Proceedings of the workshop {\it QCD 20 years later}
ed. P. Zerwas,  (Aachen, 1992).
\bibitem{neu}
For an early review see, e.g., C.H. Llewellyn Smith, Phys. Rep.
{\bf 3 C}, 261(1972);\\
For a more recent review of neutrino DIS see, e.g, Sanjib  Mishra and
F. Scuilli, Ann. Rev. Nucl. Part. Sci., {\bf 39}, 259 (1988).
\bibitem{rmp} R.E. Taylor, Rev. Mod. Phys. {\bf 63}, 573 (1991);\\
H. W. Kendall, Rev. Mod. Phys. {\bf 63}, 597 (1991);\\
J.I. Friedman, Rev. Mod. Phys. {\bf 63}, 615 (1991).

\bibitem{plot}  NMC, M. Arneodo et al, CERN-PPE/93-117, subm. to
Rapid. Comm. in Phys. Rev.

\bibitem{satz} For a review see, e.g., H. Satz, in proceedings of
the workshop {\it QCD 20 years later}, ed. P. Zerwas, (Aachen 1992).
\bibitem{PQCD} For a good summary of various subprocess cross--sections
as calculated in perturbative QCD see, e.g., Applications of
Perturbative QCD, R. D. Field,(Addison-Wesley, Redwood City, USA).
\bibitem{Alt} For a good summary of early tests of
perturbative QCD at the $p \bar p$ collider experiments see, e.g.,
G. Altarelli, Ann. Rev. of Nucl. and Part. Sci. {\bf 39}, 357 (1989).
\bibitem{reya} For an early  discussion of electro- and
photo-production of quarkonia see, e.g., the second of reference
\protect\cite{QCDper}.
\bibitem{emcdata} For a compilation of the data on EMC effect see,
e.g., Ref. \protect\cite{pdg}.
\bibitem{SAKRZ}
A. Krzywicki, Phys. Rev. {\bf D 14}, 152 (1976);\\
R.M. Godbole and K.V.L. Sarma, Phys. Rev. {\bf D25},120 (1982).
\bibitem{shadowing}
N.N. Nikolaev and B.G. Zakharov, Phys. Lett. {\bf B260}, 414 (1991);
\bibitem{llsmith} C.H. Llewellyn Smith, Phys. Lett. {\bf B128},
107 (1983).
\bibitem{akulin} S.V. Akulinichev et al, Phys. Rev. Lett. {\bf
55}, 2239 (1985).
\bibitem{brown} G.L. Li, K.F. Liu and G.E. Brown, Phys. Lett.
{\bf B213}, 531 (1988).
\bibitem{gas} S. Gupta and K. V. L. Sarma, Z. Phys. {\bf C29}, 329 (1985).
\bibitem{cdhs} CDHS Collaboration, J. G. H. de Groot et al., Z. Phys.
{\bf C17}, 283 (1983).
\bibitem{bag} K.E. Lassila and U.P. Sukhatme, Phys. Lett. {\bf B209},
343 (1988); U.P. Sukhatme, G. Wilk and K.E. Lassila, Z. Phys. {\bf
C53}, 439 (1992).
\bibitem{pirner} M. Sato, S. Coon, H. Pirner and J. Vary, Phys. Rev.
{\bf C33}, 1062 (1986).
\bibitem{sulassila} K.F. Lassila, U.P. Sukhatme, A. Harindranath and J.
Vary, Phys. Rev. {\bf C44}, 1188 (1991).
\bibitem{oldbag} C.E. Carlson and T.J. Havens, Phys. Rev. Lett.
{\bf 51}, 261 (1983).
\bibitem{resc} F.E. Close, R.G. Roberts and G.G. Ross, Phys. Lett.
{\bf B129}, 346 (1993).
\bibitem{GGB} S. Gupta, S. Banerjee and R.M. Godbole, Z. Phys. {\bf
C28}, 483 (1985).
\bibitem{G}  S. Gupta, Pramana {\bf 24}, 443 (1985).
\bibitem{hybrid} J. D. Deus, M. Pimenta and J. Varela, Phys.
Rev. {\bf D30}, 697 (1984).
\bibitem{rosou} R.M. Godbole and S. Gupta, Phys. Lett. {\bf B228}, 129 (1989).
\bibitem{oldjet} C. Bromberg et al., Phys. Rev. Lett. {\bf 42},
1202 (1979).
\bibitem{CIP} K. J. Anderson et al., Phys. Rev. Lett. {\bf 42},
844 (1979)
\bibitem{newjet} H.E. Miettinen et al., Phys. Lett. {\bf B207},
202 (1988).
\bibitem{kasta} S. Kastanevas et al., Phys. Rev. Lett. {\bf 60}, (2121)
(1988).
\bibitem{fnal1} D.M. Alde et al., Phys. Rev. Lett. {\bf 66},
133 (1991).
\bibitem{fnal2} D.M. Alde et al., Phys. Rev. Lett. {\bf 66},
2285 (1991).
\bibitem{fnal3} D.M. Alde et al., Phys. Rev. Lett. {\bf 64},
2479 (1990).
\bibitem{suppjpsi}
M. D. Sokoloff et al, Phys. Rev. Lett. {\bf 57}, 3003 (1986);\\
R.L. Anderson et al., Phys. Rev. Lett. {\bf 38}, 263 (1977).
\bibitem{EMCjpsi} J.J. Aubert et al.,  Phys. Lett. {\bf B152}, 1985.
\bibitem{NMCjpsi}
D. Allasia et al., Phys. Lett. {\bf B258}, 493 (1991);\\
P. Amaudruz et al., Nucl. Phys. {\bf B371}, 553 (1992).
\bibitem{rosouprd} S. Gupta and R.M. Godbole, Phys. Rev. {\bf D33},
(1986) 3453.
\bibitem {sriro} R.M. Godbole and Sridhar K., Z. Phys. {\bf C51},
417 (1991).
\bibitem {souraj} R.V. Gavai and S. Gupta, Z. Phys. {\bf C49},
663 (1991).
\bibitem{sousri} S. Gupta and Sridhar K., Phys. Lett. {\bf B197},
259 (1987); Sridhar K., Z. Phys. {\bf C55}, 401 (1992).
\bibitem {srirajro} R.V. Gavai, R.M. Godbole and Sridhar K., Phys.
Lett. {\bf B299}, 157 (1993).
\bibitem{roraj} R.V. Gavai and R.M. Godbole, Nucl. Phys. {\bf A566}
(1994) 375 C, Proceedings of QM 93, B\"orlange, Sweden, June, 1993;
BU-TH-93/5 (submitted for publication).
\end{thebibliography}
\end{document}